\documentclass[journal,draftclsnofoot,onecolumn]{IEEEtran}
\usepackage{graphicx}
\usepackage{amssymb}
\usepackage{subfigure}
\usepackage{caption}
\usepackage{color}
\renewcommand{\theequation}{\arabic{section}.\arabic{equation}}
\newtheorem{theorem}{Theorem}

\newtheorem{lemma}{Lemma}

\newtheorem{remark}{Remark}

\begin{document}

\title{Finite State Markov Wiretap Channel with Delayed Feedback}

\author{Bin~Dai,
        Zheng~Ma,
        and~Yuan~Luo
\thanks{B. Dai is with the
School of Information Science and Technology,
Southwest JiaoTong University, Chengdu 610031, China, and with
the State Key Laboratory of Integrated Services Networks, Xidian University, Xi$'$an, Shaanxi 710071, China,
e-mail: daibin@home.swjtu.edu.cn.}
\thanks{
Z. Ma is with the
School of Information Science and Technology,
Southwest JiaoTong University, Chengdu 610031, China,
e-mail: zma@home.swjtu.edu.cn.}
\thanks{Y. Luo is with the computer science and engineering department, Shanghai Jiao Tong University, Shanghai 200240, China,
Email: luoyuan@cs.sjtu.edu.cn.}
}

\maketitle

\begin{abstract}

\textcolor[rgb]{1.00,0.00,0.00}{The finite state Markov channel (FSMC), where the channel transition probability is controlled by
a state undergoing a Markov process, is a useful model for
the mobile wireless communication channel. In this paper, we investigate the security issue in the mobile wireless communication systems by considering the FSMC with an eavesdropper, which we call the finite state Markov wiretap channel (FSM-WC).
We assume that the state is perfectly known by the legitimate receiver and the eavesdropper, and through a noiseless feedback channel,
the legitimate receiver sends his received channel output and the state back to the transmitter after some time delay.
Inner and outer bounds on the capacity-equivocation regions of the
FSM-WC with delayed state feedback and with or without delayed channel output feedback are provided in this paper, and we show that these bounds meet
if the eavesdropper's received symbol is a degraded version of the legitimate receiver's.
The above results are further explained via degraded Gaussian and Gaussian fading examples.}

\end{abstract}


\section{Introduction \label{secI}}

\subsection*{A. The finite state Markov channel}

\textcolor[rgb]{1.00,0.00,0.00}{The finite state Markov channel (FSMC) is a discrete channel, and its transition probability depends on a channel state which
takes values in a finite set and undergoes a Markov process.
Wang et al. \cite{wang} and Zhang et al. \cite{zhang} first found that the FSMC is
a useful model for characterizing the time-varying fading channels, and the capacity of the FSMC was studied by \cite{god}.
Here note that the capacity provided in \cite{god} is a multi-letter characterization, and it is difficult to calculate. A single-letter
characterization of the capacity of the FSMC remains open.}

\begin{figure}[htb]
\centering
\includegraphics[scale=0.6]{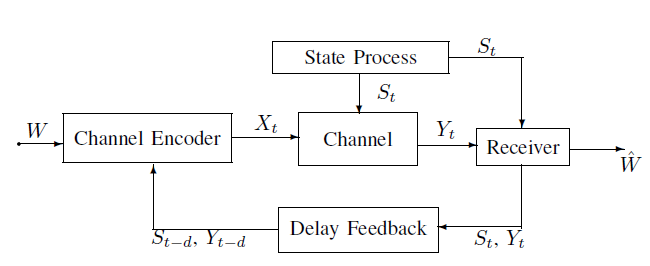}
\caption{The FSMC with delayed feedback}
\label{f1}
\end{figure}

\textcolor[rgb]{1.00,0.00,0.00}{It is known to all that for a point-to-point discrete memoryless channel (DMC), feeding back the channel output of the receiver to the transmitter via
another noiseless channel
does not increase the channel capacity \cite{coverx}.
However, Cover et al. showed that the capacity regions of several multi-user channels, such as multiple-access channel (MAC) and relay channel, can
be enhanced by feeding back the receiver's channel output to the transmitter over a noiseless channel, see \cite{coverz} and \cite{CG1}.
Then, it is natural to ask: does the receiver's channel output feedback help
to enhance the capacity of the FSMC? Viswanathan \cite{vis} answered this question by considering
a practical mobile wireless communication scenario, where the channel state is perfectly obtained by the receiver, and
the receiver noiselessly feeds back the state and his own channel output to the transmitter after some time delay.
Viswanathan \cite{vis} showed that 
this communication scenario can be modeled as the FSMC with delayed feedback, see Figure \ref{f1}.
The capacity of the model of Figure \ref{f1} is totally determined in \cite{vis}, and unlike the works of \cite{coverz} and \cite{CG1},
the capacity results in \cite{vis} imply that feeding back the receiver's channel output to the transmitter over a noiseless channel does not increase
the capacity of FSMC with only delayed state feedback.
Other related works on the FSMC with or without feedback are investigated in \cite{bash}-\cite{god2}.}

\subsection*{B. The wiretap channel}
Wyner, in his landmark paper on the wiretap channel \cite{Wy}, first investigated the
information-theoretic security in practical communication systems. In Wyner's wiretap channel model, a
transmitter sends a private message to a legitimate receiver via a discrete memoryless main channel,
and an eavesdropper eavesdrops the output of the main channel via a discrete memoryless wiretap channel.
We say that the perfect secrecy is achieved if no information about the private message is leaked to the eavesdropper.
The secrecy capacity, which \textcolor[rgb]{1.00,0.00,0.00}{is} the maximum reliable transmission rate with perfect secrecy constraint,
was
characterized by Wyner \cite{Wy}. After Wyner determined the secrecy capacity of the discrete memoryless wiretap channel model,
Leung-Yan-Cheong and Hellman \cite{CH} investigated the Gaussian wiretap channel (GWC),
where the noise of the main channel and the wiretap channel \textcolor[rgb]{1.00,0.00,0.00}{is} Gaussian distributed.
It \textcolor[rgb]{1.00,0.00,0.00}{is} shown in \cite{CH} that the secrecy capacity of the GWC \textcolor[rgb]{1.00,0.00,0.00}{is} obtained by subtracting
the capacity of the overall wiretap channel \footnote{Here the overall wiretap channel is a cascade of the main channel and the wiretap channel}
from the capacity of the main channel.
Wyner's work was generalized by Csisz$\acute{a}$r and K\"{o}rner \cite{CK}, where common and private messages \textcolor[rgb]{1.00,0.00,0.00}{are}
sent through a discrete memoryless general broadcast channel \footnote{Here note that Wyner's wiretap channel model is a kind of degraded broadcast channel}.
The common message \textcolor[rgb]{1.00,0.00,0.00}{is} assumed to be decoded correctly by both the
legitimate receiver and the eavesdropper, while the private message \textcolor[rgb]{1.00,0.00,0.00}{is} only allowed to be obtained by the
legitimate receiver. The secrecy capacity region of this generalized model was characterized in \cite{CK}, and later,
Liang et al. \cite{LPS} characterized the secrecy capacity region for the Gaussian case of Csisz$\acute{a}$r and K\"{o}rner's model \cite{CK}.
The work of \cite{Wy} and \cite{CK} lays the foundation of the information-theoretic security in communication systems.
Using the approach of \cite{Wy} and \cite{CK}, the security problems in multi-user communication channels, such as
broadcast channel, multiple-access channel, relay channel, and interference
channel, have been widely studied, see \cite{LMSY}-\cite{EU1}.

Recently, the wiretap channel with states has received much attention, see
\cite{MVL}-\cite{dai1}. These works focus on the scenario that the states are identical independent distributed (i.i.d.),
and to the best of the authors' knowledge, only Bloch et al. \cite{san1} and Sankarasubramaniam et al. \cite{san}
investigated the wiretap channel with memory states, where
a stochastic algorithm for computing the multi-letter form secrecy capacity of this model was provided.
A single-letter characterization for the secrecy capacity of \cite{san1} and \cite{san} is still open.

\subsection*{C. Contributions of This Paper and Organization}

In practical mobile wireless communication networks, security is a critical issue when people intend to transmit private information,
such as the credit card transactions and the banking related data communications. The secure transmission of these private messages in the practical
mobile wireless communication networks motivates us to study the finite-state Markov wiretap channel with delayed feedback,
see the following Figure \ref{f2}.
In Figure \ref{f2}, the transition probability of the channel at each time instant depends on a state which undergoes a
finite-state Markov process. At time $i$,
the receiver \footnote{Throughout this paper, the ``receiver'' is used as a shorthand for ``legitimate receiver''}
receives the channel output $Y_{i}$ and the state $S_{i}$, and sends them back to the transmitter after a delay time $d$
via a noiseless feedback channel.
The channel encoder, at time $i$,
generates the channel input according to the transmitted message $W$ and the delayed feedback $Y_{i-d}$ and $S_{i-d}$.
Moreover, at time $i$, an eavesdropper receives the channel output $Z_{i}$ and the state $S_{i}$, and he wishes to obtain the transmitted message $W$.
The delay time $d$ is perfectly known by
the receiver, the eavesdropper and the transmitter.
The main results of the model of Figure \ref{f2} are listed as follows.
\begin{itemize}
\item First, for the model of Figure \ref{f2} with only delayed state $S_{i-d}$ feedback,
we provide inner and outer bounds on the capacity-equivocation region, and we find that these bounds meet if
the eavesdropper's received symbol $Z_{i}$ is a degraded version of the receiver's $Y_{i}$.

\item Second, inner and outer bounds
on the capacity-equivocation region are provided for the model of Figure \ref{f2} with both delayed state $S_{i-d}$ and
delayed output $Y_{i-d}$ feedback. We also find that these bounds meet if
$Z_{i}$ is a degraded version of $Y_{i}$. Moreover, unlike the fact that the delayed receiver's channel output feedback does not
increase the capacity of the FSMC with only delayed state feedback \cite{vis},
we find that for the degraded case,
this delayed channel output feedback $Y_{i-d}$ helps to enhance the capacity-equivocation region of the FSM-WC with only delayed state feedback, i.e.,
sending back the receiver's channel output to the transmitter may help to enhance the security of the practical mobile wireless communication systems.

\item The above results are further explained via degraded Gaussian and Gaussian fading examples.

\end{itemize}

\begin{figure}[htb]
\centering
\includegraphics[scale=0.6]{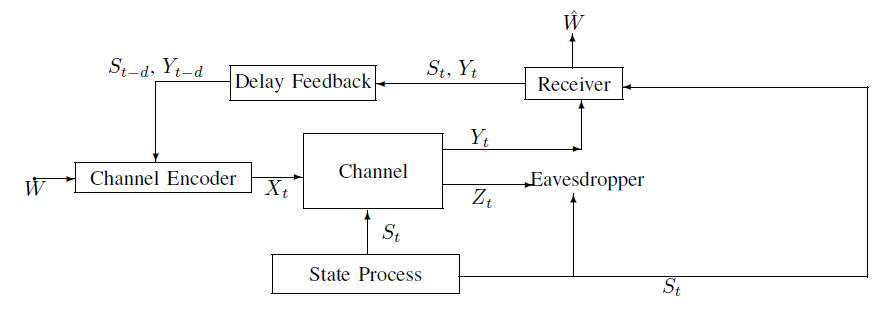}
\caption{The FSM-WC with delayed feedback}
\label{f2}
\end{figure}

The rest of this paper is organized as follows. In
Section \ref{secII}, we show the definitions, notations and the main results of the model of Figure \ref{f2}.
Degraded Gaussian and Gaussian fading examples of the model of Figure \ref{f2} are provided in Section \ref{secIII}.
Final conclusions are presented in Section \ref{secIV}.

\section{Basic Notations, Definitions and the Main Result of the Model of Figure \ref{f2}}\label{secII}

\emph{Basic notations:} We use the notation $p_{V}(v)$ to denote
the probability mass function $Pr\{V=v\}$, where $V$ (capital letter) denotes the random variable, $v$ (lower case letter)
denotes the real value of the random variable $V$. Denote the alphabet in which the random variable $V$ takes values by $\mathcal{V}$ (calligraphic letter).
Similarly, let $U^{N}$ be a random vector $(U_{1},...,U_{N})$, and $u^{N}$ be a vector value
$(u_{1},...,u_{N})$. In the rest of this paper, the log function is taken to the base 2.

\emph{Definitions of the model of Figure \ref{f2}:}

\begin{itemize}

\item The channel is a finite-state Markov channel (FSMC), where the channel state $S$ takes values in a finite alphabet
$\mathcal{S}=\{s_{1},s_{2},...,s_{k}\}$.
At the $i$-th time ($1\leq i\leq N$), the transition probability of the channel depends on the state $s_{i}$,
the input $x_{i}$ and the outputs $y_{i}$, $z_{i}$, and is given by $P_{Y,Z|X,S}(y_{i},z_{i}|x_{i},s_{i})$. The $i$-th time outputs of the channel $Y_{i}$
and $Z_{i}$
are assumed to depend only on $X_{i}$ and $S_{i}$, and thus we have
\begin{equation}\label{e202}
P_{Y^{N},Z^{N}|X^{N},S^{N}}(y^{N},z^{N}|x^{N},s^{N})=\prod_{i=1}^{N}P_{Y,Z|X,S}(y_{i},z_{i}|x_{i},s_{i}).
\end{equation}

\item
The state process $\{S_{i}\}$ is assumed to be a stationary irreducible aperiodic ergodic Markov chain.
The state process is independent of the transmitted messages, and it is independent of
the channel input and outputs given the previous states, i.e.,
\begin{equation}\label{e203}
Pr\{S_{i}=s_{i}|X^{i}=x^{i},Y^{i}=y^{i},S^{i-1}=s^{i-1}\}=Pr\{S_{i}=s_{i}|S_{i-1}=s_{i-1}\}.
\end{equation}
Here note that (\ref{e203}) also implies that
\begin{equation}\label{e203.1}
Pr\{S_{i}=s_{i}|X^{i}=x^{i},Y^{i}=y^{i},S^{i-d}=s^{i-d}\}=Pr\{S_{i}=s_{i}|S_{i-d}=s_{i-d}\},
\end{equation}
where $1\leq d\leq i-1$.
Denote the $1$-step transition probability matrix by $K$, and denote the steady state probability
of $\{S_{i}\}$ by $\pi$. Let the random variables $S_{i}$ and $S_{i-d}$ be the channel states at time $i$ and $i-d$, respectively.
The joint distribution of $(S_{i},S_{i-d})$ is given by
\begin{equation}\label{e204}
\pi_{d}(S_{i}=s_{l}, S_{i-d}=s_{j})=\pi(s_{j})K^{d}(s_{j},s_{l}),
\end{equation}
where $s_{l}$ is the $l$-th element of $\mathcal{S}$, $s_{j}$ is the $j$-th element of $\mathcal{S}$,
and $K^{d}(s_{j},s_{l})$ is the $(j,l)$-th element of the $d$-step transition
probability matrix $K^{d}$ of the Markov process\textcolor[rgb]{1.00,0.00,0.00}{.}

\item
Let $W$, uniformly distributed over the finite alphabet $\mathcal{W}=\{1,2,...,M\}$, be the message sent by the transmitter.
Here note that $W$ is independent of the state process $\{S_{i}\}$ ($1\leq i\leq N$) and $H(W)=\log M$.
For the model of Figure \ref{f2} without receiver's channel output feedback,
the $i$-th time channel input $X_{i}$ is given by
\begin{equation}\label{e206.xx1}
X_{i}=
\left\{
\begin{array}{ll}
f_{i}(W), & 1\leq i\leq d\\
f_{i}(W,S^{i-d}), & d+1\leq i\leq N\textcolor[rgb]{1.00,0.00,0.00}{,}
\end{array}
\right.
\end{equation}
and for the model of Figure \ref{f2} with receiver's channel output feedback,
$X_{i}$ is given by
\begin{equation}\label{e206}
X_{i}=
\left\{
\begin{array}{ll}
f_{i}(W), & 1\leq i\leq d\\
f_{i}(W,S^{i-d},Y^{i-d}), & d+1\leq i\leq N\textcolor[rgb]{1.00,0.00,0.00}{.}
\end{array}
\right.
\end{equation}
Here note that the $i$-th time channel encoder $f_{i}$ is a stochastic encoder.

\item
The channel decoder is a mapping
\begin{equation}\label{e207}
\psi: \,\, \mathcal{Y}^{N}\times \mathcal{S}^{N}\rightarrow \{1,2,...,M\},
\end{equation}
with inputs $Y^{N}$, $S^{N}$ and output $\hat{W}$.
The average probability of error $P_{e}$ is denoted by
\begin{equation}\label{e208}
P_{e}=\frac{1}{M}\sum_{\textcolor[rgb]{1.00,0.00,0.00}{j}=1}^{M}\sum_{s^{N}}P_{S^{N}}(s^{n})Pr\{\psi(y^{N},s^{N})\neq \textcolor[rgb]{1.00,0.00,0.00}{j}
|\textcolor[rgb]{1.00,0.00,0.00}{j} \,\,\mbox{was sent}\}.
\end{equation}

\item
Since the state is also known by the eavesdropper,
the eavesdropper's equivocation to the message $W$ is defined as
\begin{equation}\label{e210}
\Delta=\frac{1}{N}H(W|Z^{N},S^{N}).
\end{equation}

\item A \textcolor[rgb]{1.00,0.00,0.00}{rate-equivocation} pair $(R, R_{e})$ (where $R, R_{e}>0$) is called
achievable if, for any $\epsilon>0$, there exists a channel
encoder-decoder $(N, \Delta, P_{e})$ such that
\begin{eqnarray}\label{e211}
&&\frac{\log M}{N}\geq R-\epsilon, \,\,\,\, \Delta\geq R_{e}-\epsilon, \,\,\,\,P_{e}\leq \epsilon.
\end{eqnarray}
\end{itemize}
The capacity-equivocation region is a set composed of all achievable $(R, R_{e})$ pairs. Here the capacity-equivocation region of
the model of Figure \ref{f2} with only delayed state feedback
is denoted by $\mathcal{R}$, and $\mathcal{R}^{f}$ denotes the
capacity-equivocation region of the model of Figure \ref{f2} with delayed state and
receiver's channel output feedback.
In the remainder of this section, the bounds on the capacity-equivocation region $\mathcal{R}$ are given in Theorem \ref{T3} and Theorem \ref{T3.1}, and the
bounds on $\mathcal{R}^{f}$ are given in Theorem \ref{T1} and Theorem \ref{T1.1}, see the followings.

\emph{\textbf{Main results on $\mathcal{R}$:}}
\begin{theorem}\label{T3}
An inner bound $\mathcal{R}^{in}$ on $\mathcal{R}$ is given by
\begin{eqnarray*}
&&\mathcal{R}^{in}=\{(R, R_{e}): 0\leq R_{e}\leq R,\\
&&R\leq I(V;Y|S,\tilde{S}),\\
&&R_{e}\leq I(V;Y|U,S,\tilde{S})-I(V;Z|U,S,\tilde{S})\},
\end{eqnarray*}
where the joint probability $P_{UVS\tilde{S}XYZ}(u,v,s,\tilde{s},x,y,z)$
satisfies
\begin{eqnarray}\label{dota1}
&&P_{UVS\tilde{S}XYZ}(u,v,s,\tilde{s},x,y,z)\nonumber\\
&&=P_{YZ|XS}(y,z|x,s)P_{X|UV\tilde{S}}(x|u,v,\tilde{s})P_{V|U\tilde{S}}(v|u,\tilde{s})\cdot \nonumber\\
&&P_{U|\tilde{S}}(u|\tilde{s})K^{d}(\tilde{s},s)P_{\tilde{S}}(\tilde{s}),
\end{eqnarray}
and $U$ may be assumed to be a (deterministic) function of $V$. \textcolor[rgb]{1.00,0.00,0.00}{Here note that in $\mathcal{R}^{in}$, if $I(V;Y|U,S,\tilde{S})-I(V;Z|U,S,\tilde{S})<0$,
$R_{e}=0$.}
\end{theorem}

\begin{IEEEproof}
\textcolor[rgb]{1.00,0.00,0.00}{The inner bound $\mathcal{R}^{in}$ is achieved by the
following key steps:}
\begin{itemize}
\item \textcolor[rgb]{1.00,0.00,0.00}{First, combining the rate splitting technique used in \cite{CK} with the
multiplexing coding scheme used in \cite{vis}, we divide
the transmitted message $W$ into a common message $W_{c}=(W_{c,1},...,W_{c,k})$ and a confidential message $W_{p}=(W_{p,1},...,W_{p,k})$,
where $k$ is the cardinality of $\mathcal{S}$, and $W_{c,\textcolor[rgb]{1.00,0.00,0.00}{\tilde{s}}}$ 
(or $W_{p,\textcolor[rgb]{1.00,0.00,0.00}{\tilde{s}}}$) ($1\leq \textcolor[rgb]{1.00,0.00,0.00}{\tilde{s}}\leq k$) is 
the $\textcolor[rgb]{1.00,0.00,0.00}{\tilde{s}}$-th sub-message of $W_{c}$ (or $W_{p}$).
Further divide the sub-message $W_{p,\textcolor[rgb]{1.00,0.00,0.00}{\tilde{s}}}$ into two part, i.e., 
$W_{p,\textcolor[rgb]{1.00,0.00,0.00}{\tilde{s}}}=(W_{p,\textcolor[rgb]{1.00,0.00,0.00}{\tilde{s}},1},W_{p,\textcolor[rgb]{1.00,0.00,0.00}{\tilde{s}},2})$. 
Here note that the index $\textcolor[rgb]{1.00,0.00,0.00}{\tilde{s}}$ is 
the specific value of the delayed state $S_{i-d}$, which is represented by $\tilde{S}$.}

\item \textcolor[rgb]{1.00,0.00,0.00}{Similar to the superposition coding strategy used in \cite{CK}, the sub-message 
$W_{c,\textcolor[rgb]{1.00,0.00,0.00}{\tilde{s}}}$ ($1\leq \textcolor[rgb]{1.00,0.00,0.00}{\tilde{s}}\leq k$) is encoded as the cloud
center $U^{N_{\textcolor[rgb]{1.00,0.00,0.00}{\tilde{s}}}}$ (here $N_{\textcolor[rgb]{1.00,0.00,0.00}{\tilde{s}}}$ 
is the codeword length for $W_{c,\textcolor[rgb]{1.00,0.00,0.00}{\tilde{s}}}$ and $W_{p,\textcolor[rgb]{1.00,0.00,0.00}{\tilde{s}}}$),
and the message pair $(W_{c,\textcolor[rgb]{1.00,0.00,0.00}{\tilde{s}}}, W_{p,\textcolor[rgb]{1.00,0.00,0.00}{\tilde{s}}})$ 
is encoded as the satellite codeword $V^{N_{\textcolor[rgb]{1.00,0.00,0.00}{\tilde{s}}}}$. Here note that
the random binning coding strategy used in
\cite{CK} is also introduced into the construction of $V^{N_{\textcolor[rgb]{1.00,0.00,0.00}{\tilde{s}}}}$, i.e., 
there are three indexes in $V^{N_{\textcolor[rgb]{1.00,0.00,0.00}{\tilde{s}}}}$, the first index is chosen according to
the common message $W_{c,\textcolor[rgb]{1.00,0.00,0.00}{\tilde{s}}}$,
the second index is chosen according to $W_{p,\textcolor[rgb]{1.00,0.00,0.00}{\tilde{s}},1}$, and the third index
is randomly chosen from a bin with index $W_{p,\textcolor[rgb]{1.00,0.00,0.00}{\tilde{s}},2}$.}

\item \textcolor[rgb]{1.00,0.00,0.00}{Note that the state $S$ and the delayed state $S_{i-d}$ (represented by $\tilde{S}$) are known by all parties.
Then along the lines of the proof of \cite{CK},
for the sub-messages $W_{c,\tilde{s}}$ and $W_{p,\tilde{s}}$, we can obtain the following region $\mathcal{R}^{in}_{\tilde{s}}$
\begin{eqnarray*}
&&\mathcal{R}^{in}_{\tilde{s}}=\{(R_{\tilde{s}}, R_{e,\tilde{s}}): 0\leq R_{\tilde{s}}=R_{c,\tilde{s}}+R_{p,\tilde{s}},\\
&&0\leq R_{c,\tilde{s}}\leq \min\{I(U;Y|S,\tilde{S}=\tilde{s}), I(U;Z|S,\tilde{S}=\tilde{s})\},\\
&&0\leq R_{p,\tilde{s}}\leq I(V;Y|U,S,\tilde{S}=\tilde{s}),\\
&&0\leq R_{e,\tilde{s}}\leq R_{p,\tilde{s}},\\
&&R_{e,\tilde{s}}\leq I(V;Y|U,S,\tilde{S}=\tilde{s})-I(V;Z|U,S,\tilde{S}=\tilde{s})\},
\end{eqnarray*}
where $R_{c,\tilde{s}}$, $R_{p,\tilde{s}}$ and $R_{\tilde{s}}$ are the rates of the sub-messages $W_{c,\tilde{s}}$ , $W_{p,\tilde{s}}$ 
and $W_{\tilde{s}}=(W_{c,\tilde{s}},W_{p,\tilde{s}})$, respectively, and
$R_{e,\tilde{s}}$ is the equivocation rate of the sub-message $W_{p,\tilde{s}}$. 
Here note that in $\mathcal{R}^{in}_{\tilde{s}}$, if $I(V;Y|U,S,\tilde{S}=\tilde{s})-I(V;Z|U,S,\tilde{S}=\tilde{s})<0$, 
$R_{e,\tilde{s}}=0$.}

\item \textcolor[rgb]{1.00,0.00,0.00}{Finally, using Fourier-Motzkin elimination (see e.g., \cite{lall}) to eliminate $R_{c,\tilde{s}}$ and $R_{p,\tilde{s}}$
from $\mathcal{R}^{in}_{\tilde{s}}$, and
multiplexing all the sub-messages, the region $\mathcal{R}^{in}$ is obtained.}

\end{itemize}
The details of the proof
are in Appendix \ref{appen1}.
\end{IEEEproof}

\begin{theorem}\label{T3.1}
An outer bound $\mathcal{R}^{out}$ on $\mathcal{R}$ is given by
\begin{eqnarray*}
&&\mathcal{R}^{out}=\{(R, R_{e}): 0\leq R_{e}\leq R,\\
&&R\leq I(V;Y|S,\tilde{S}),\\
&&R_{e}\leq I(V;Y|U,S,\tilde{S})-I(V;Z|U,S,\tilde{S})\},
\end{eqnarray*}
where the joint probability $P_{UVS\tilde{S}XYZ}(u,v,s,\tilde{s},x,y,z)$
satisfies
\begin{eqnarray}\label{dota2}
&&P_{UVS\tilde{S}XYZ}(u,v,s,\tilde{s},x,y,z)\nonumber\\
&&=P_{YZ|XS}(y,z|x,s)P_{XVUS\tilde{S}}(x,v,u,s,\tilde{s}).
\end{eqnarray}
\textcolor[rgb]{1.00,0.00,0.00}{Here note that in $\mathcal{R}^{out}$, if $I(V;Y|U,S,\tilde{S})-I(V;Z|U,S,\tilde{S})<0$,
$R_{e}=0$.}
\end{theorem}

\begin{IEEEproof}
\textcolor[rgb]{1.00,0.00,0.00}{The outer bound $\mathcal{R}^{out}$ is achieved by the
following key steps:}
\begin{itemize}

\item \textcolor[rgb]{1.00,0.00,0.00}{First, note that the auxiliary random variable $U_{i}$ in \cite{CK} is defined as $(Y^{i-1},Z_{i+1}^{N})$.
In this paper, in order to introduce the delayed feedback state $S_{i-d}$ into the definition of $U_{i}$, we define $U_{i}\triangleq (Y^{i-1},Z_{i+1}^{N},S^{N})$.
Here note that $S_{i-d}$ is included in the $S^{N}$.}

\item \textcolor[rgb]{1.00,0.00,0.00}{Using Fano's inequality, the transmission rate $R$ and the equivocation rate $R_{e}$ can be upper bounded by $\frac{1}{N}I(W;Y^{N}|S^{N})$ and
$\frac{1}{N}(I(W;Y^{N}|S^{N})-I(W;Z^{N}|S^{N}))$, respectively.}

\item \textcolor[rgb]{1.00,0.00,0.00}{Then, using chain rule and
the following Csisz$\acute{a}$r's equalities
\begin{eqnarray}\label{wow.1}
&&\sum_{i=1}^{N}I(Y_{i};Z_{i+1}^{N}|Y^{i-1},S^{N})=\sum_{i=1}^{N}I(Z_{i};Y^{i-1}|Z_{i+1}^{N},S^{N})
\end{eqnarray}
and
\begin{eqnarray}\label{wow.2}
&&\sum_{i=1}^{N}I(Y_{i};Z_{i+1}^{N}|Y^{i-1},S^{N},W)=\sum_{i=1}^{N}I(Z_{i};Y^{i-1}|Z_{i+1}^{N},S^{N},W),
\end{eqnarray}
to eliminate some identities of the bound on the equivocation rate $R_{e}$, the outer bound $\mathcal{R}^{out}$ is obtained.}

\end{itemize}

The details of the proof are in
Appendix \ref{appen2}.
\end{IEEEproof}

\begin{remark}\label{R1}
There are some notes on Theorem \ref{T3} and Theorem \ref{T3.1}, see the followings.
\begin{itemize}
\item Here note that the inner bound $\mathcal{R}^{in}$ is almost the same as the outer bound $\mathcal{R}^{out}$, except the definitions of
the joint probability $P_{UVS\tilde{S}XYZ}(u,v,s,\tilde{s},x,y,z)$ in $\mathcal{R}^{in}$ and $\mathcal{R}^{out}$. To be specific, in $\mathcal{R}^{in}$,
the definition of $P_{UVS\tilde{S}XYZ}(u,v,s,\tilde{s},x,y,z)$ implies the Markov chains
$S\rightarrow (\tilde{S},U,V)\rightarrow X$, $S\rightarrow (\tilde{S},U)\rightarrow V$ and $S\rightarrow \tilde{S}\rightarrow U$, but these chains
are not guaranteed in $\mathcal{R}^{out}$.

\item If the eavesdropper's received symbol $Z^{N}$ is a degraded version of $Y^{N}$, i.e., the Markov chain
$(X^{N},S^{N})\rightarrow Y^{N}\rightarrow Z^{N}$ holds, the outer bound $\mathcal{R}^{out}$ meets with the inner bound $\mathcal{R}^{in}$, and they
reduce to the following region $\mathcal{R}^{*}$, where
\begin{eqnarray}\label{giveup1}
&&\mathcal{R}^{*}=\{(R, R_{e}): R_{e}\leq R,\nonumber\\
&&R\leq I(X;Y|S,\tilde{S}),\nonumber\\
&&R_{e}\leq I(X;Y|S,\tilde{S})-I(X;Z|S,\tilde{S})\},
\end{eqnarray}
and the joint probability $P_{S\tilde{S}XYZ}(s\tilde{s}xyz)$ satisfies
\begin{eqnarray}\label{appengiveup3}
&&P_{S\tilde{S}XYZ}(s\tilde{s}xyz)=P_{Z|Y}(z|y)P_{Y|X,S}(y|x,s)K^{d}(\tilde{s},s)P_{X|\tilde{S}}(x|\tilde{s})P_{\tilde{S}}(\tilde{s}).
\end{eqnarray}
\begin{IEEEproof}
See Appendix \ref{appengiveup1}.
\end{IEEEproof}

\item A rate $R$ is called
achievable with weak secrecy if, for any $\epsilon>0$,
there exists a channel
encoder-decoder $(N, \Delta, P_{e})$ such that
\begin{eqnarray}\label{e202.sp11}
&&\frac{\log M}{N}\geq R-\epsilon, \,\,\,\, \Delta\geq R-\epsilon, \,\,\,\,P_{e}\leq \epsilon.
\end{eqnarray}
The secrecy capacity
is the maximum achievable rate with weak secrecy, and it can be directly obtained by substituting $R_{e}=R$ into the corresponding
capacity-equivocation region and maximizing $R$. Thus, for the degraded case of the model of Figure \ref{f2} with only delayed state feedback, the secrecy capacity
$\mathcal{C}^{*}_{s}$ is given by
\begin{eqnarray}\label{appengiveup2}
&&\mathcal{C}^{*}_{s}=\max_{P_{X|\tilde{S}}(x|\tilde{s})}(I(X;Y|S,\tilde{S})-I(X;Z|S,\tilde{S})).
\end{eqnarray}
Here $\mathcal{C}^{*}_{s}$ is obtained by substituting $R_{e}=R$ into (\ref{giveup1}) and maximizing $R$.

\end{itemize}
\end{remark}

\emph{\textbf{Main results on $\mathcal{R}^{f}$:}}
\begin{theorem}\label{T1}
An inner bound $\mathcal{R}^{fi}$ on the capacity-equivocation region $\mathcal{R}^{f}$ is given by
\begin{eqnarray*}
&&\mathcal{R}^{fi}=\{(R, R_{e}): 0\leq R_{e}\leq R,\\
&&R\leq I(V;Y|S,\tilde{S}),\\
&&R_{e}\leq [I(V;Y|U,S,\tilde{S})-I(V;Z|U,S,\tilde{S})]^{+}+H(Y|V,Z,S,\tilde{S})\},
\end{eqnarray*}
where $[x]^{+}=x$ if $x>0$, $[x]^{+}=0$ if $x\leq 0$, the joint probability mass function $P_{UVS\tilde{S}XYZ}(u,v,s,\tilde{s},x,y,z)$
satisfies
\begin{eqnarray}\label{dota3}
&&P_{UVS\tilde{S}XYZ}(u,v,s,\tilde{s},x,y,z)\nonumber\\
&&=P_{YZ|XS}(y,z|x,s)P_{X|UV\tilde{S}}(x|u,v,\tilde{s})P_{V|U\tilde{S}}(v|u,\tilde{s})\cdot \nonumber\\
&&P_{U|\tilde{S}}(u|\tilde{s})K^{d}(\tilde{s},s)P_{\tilde{S}}(\tilde{s}),
\end{eqnarray}
and $U$ may be assumed to be a (deterministic) function of $V$.
\end{theorem}

\begin{IEEEproof}

\textcolor[rgb]{1.00,0.00,0.00}{The output feedback inner bound $\mathcal{R}^{fi}$ is constructed according to the inner bound $\mathcal{R}^{in}$ in Theorem \ref{T3}, and it is
achieved by the following key steps:}
\begin{itemize}
\item \textcolor[rgb]{1.00,0.00,0.00}{Similar to the construction of the bound $\mathcal{R}^{in}$, we split $W$ into $W_{c}$ and $W_{p}$, and define
$W_{c}=(W_{c,1},...,W_{c,k})$ and $W_{p}=(W_{p,1},...,W_{p,k})$. Furthermore,
for $1\leq \tilde{s}\leq k$, define $W_{p,\tilde{s}}=(W_{p,\tilde{s},1},W_{p,\tilde{s},2})$. The index $\tilde{s}$ is 
the specific value of the delayed state $S_{i-d}$, which is represented by $\tilde{S}$.}

\item \textcolor[rgb]{1.00,0.00,0.00}{The component message $W_{c,\tilde{s}}$ ($1\leq \tilde{s}\leq k$) is encoded as $U^{N_{\tilde{s}}}$
($N_{\tilde{s}}$ is the codeword length for $W_{c,\tilde{s}}$ and $W_{p,\tilde{s}}$). The component message pair $(W_{c,\tilde{s}}, W_{p,\tilde{s}})$ 
and a secret key generated
by the delayed output feedback are encoded as $V^{N_{\tilde{s}}}$. To be specific, the delayed output feedback is used to generate 
a secret key $K^{*}$ which is shared
between the receiver and the transmitter, and this key is used to encrypt $W_{p,\tilde{s},2}$ (part of the $W_{p,\tilde{s}}$), 
i.e., $W_{p,\tilde{s},2}$ is encrypted as
$W_{p,\tilde{s},2}\oplus K^{*}$. Then, the indexes of $V^{N_{\tilde{s}}}$ is chosen as follows. The first and second indexes are chosen 
from $W_{c,\tilde{s}}$ and $W_{p,\tilde{s},1}$, respectively.
The third index is randomly chosen from a bin with index $W_{p,\tilde{s},2}\oplus K^{*}$.}

\item \textcolor[rgb]{1.00,0.00,0.00}{Comparing the above code construction of $\mathcal{R}^{fi}$ with that of $\mathcal{R}^{in}$, 
we see that the encoding and decoding schemes of these two bounds
are almost the same, except that the bin index of $V^{N_{\tilde{s}}}$ is encrypted by $K^{*}$. Thus, we can conclude that
for the sub-messages $W_{c,\tilde{s}}$ and $W_{p,\tilde{s}}$, the bound $\mathcal{R}^{fi}_{\tilde{s}}$ is almost the same as $\mathcal{R}^{in}_{\tilde{s}}$, 
except that
the equivocation rate $R_{e,\tilde{s}}$ of $\mathcal{R}^{fi}_{\tilde{s}}$ is bounded by the sum of two part, see the followings.}
\begin{itemize}
\item \textcolor[rgb]{1.00,0.00,0.00}{The first part is the upper bound on $R_{e,\tilde{s}}$ of
$\mathcal{R}^{in}_{\tilde{s}}$. Here note that in $\mathcal{R}^{in}_{\tilde{s}}$, the bounds $R_{e,\tilde{s}}\geq 0$ and 
$R_{e,\tilde{s}}\leq I(V;Y|U,S,\tilde{S}=\tilde{s})-I(V;Z|U,S,\tilde{S}=\tilde{s})$ 
imply that $R_{e,\tilde{s}}$ is upper bounded by $[I(V;Y|U,S,\tilde{S}=\tilde{s})-I(V;Z|U,S,\tilde{S}=\tilde{s})]^{+}$.}

\item \textcolor[rgb]{1.00,0.00,0.00}{The second part is the upper bound on the rate of the secret key $K^{*}$. Using the balanced coloring lemma introduced by
Ahlswede and Cai \cite{AC}, we conclude that the rate of the secret key $K^{*}$ is bounded by $H(Y|V,Z,S,\tilde{S}=\tilde{s})$.} 
\end{itemize}
\textcolor[rgb]{1.00,0.00,0.00}{Thus, the $R_{e,\tilde{s}}$ of $\mathcal{R}^{fi}_{\tilde{s}}$ is upper bounded by $[I(V;Y|U,S,\tilde{S}=\tilde{s})-I(V;Z|U,S,\tilde{S}=\tilde{s})]^{+}
+H(Y|V,Z,S,\tilde{S}=\tilde{s})$.
Finally, using Fourier-Motzkin elimination to eliminate $R_{c,\tilde{s}}$ and $R_{p,\tilde{s}}$
from $\mathcal{R}^{fi}_{\tilde{s}}$, and
multiplexing all the sub-messages, the region $\mathcal{R}^{fi}$ is obtained.}

\end{itemize}
The details of the proof are in Appendix \ref{appen2.x}.

\end{IEEEproof}

\begin{theorem}\label{T1.1}
An outer bound $\mathcal{R}^{fo}$ on the capacity-equivocation region $\mathcal{R}^{f}$ is given by
\begin{eqnarray*}
&&\mathcal{R}^{fo}=\{(R, R_{e}): 0\leq R_{e}\leq R,\\
&&R\leq I(V;Y|S,\tilde{S}),\\
&&R_{e}\leq H(Y|Z,U,S,\tilde{S})\},
\end{eqnarray*}
where the joint probability mass function $P_{UVS\tilde{S}XYZ}(u,v,s,\tilde{s},x,y,z)$
satisfies
\begin{eqnarray}\label{dota4}
&&P_{UVS\tilde{S}XYZ}(u,v,s,\tilde{s},x,y,z)=P_{YZ|XS}(y,z|x,s)P_{XVUS\tilde{S}}(x,v,u,s,\tilde{s})\textcolor[rgb]{1.00,0.00,0.00}{,}
\end{eqnarray}
and $U$ may be assumed to be a (deterministic) function of $V$.
\end{theorem}

\begin{IEEEproof}
\textcolor[rgb]{1.00,0.00,0.00}{The derivation of $\mathcal{R}^{fo}$ is almost the same as that of $\mathcal{R}^{out}$, except the bound on $R_{e}$, and it is achieved by
the following two steps. First,
by using Fano's inequality, the equivocation rate $R_{e}$ can be upper bounded by $\frac{1}{N}H(Y^{N}|Z^{N},S^{N})$.
Then, using chain rule and the auxiliary random variables defined in the proof of Theorem \ref{T3.1}, the outer bound $\mathcal{R}^{fo}$ is obtained.
The details of the proof are in
Appendix \ref{appen2.xx}.}

\end{IEEEproof}

\begin{remark}\label{R1.x}
There are some notes on Theorem \ref{T1} and Theorem \ref{T1.1}, see the followings.
\begin{itemize}

\item \textcolor[rgb]{1.00,0.00,0.00}{Since the delayed receiver's channel output feedback is not known by the eavesdropper, it can be 
used to generate a secret key shared only between
the receiver and the transmitter. Comparing $\mathcal{R}^{fi}$ with $\mathcal{R}^{in}$, it is easy to see that
this secret key helps to enhance the achievable rate-equivocation region
of the FSM-WC with only delayed state feedback. Here note that the delayed state is also shared by the receiver and the transmitter, but it is  
known by the eavesdropper, and thus we can not use it to generate a secret key.}

\item If the eavesdropper's received symbol $Z^{N}$ is a degraded version of $Y^{N}$, i.e., the Markov chain
$(X^{N},S^{N})\rightarrow Y^{N}\rightarrow Z^{N}$ holds, the outer bound $\mathcal{R}^{fo}$ meets with the inner bound $\mathcal{R}^{fi}$, and they
reduce to the following region $\mathcal{R}^{f*}$, where
\begin{eqnarray}\label{giveup1.rmb}
&&\mathcal{R}^{f*}=\{(R, R_{e}): R_{e}\leq R,\nonumber\\
&&R\leq I(X;Y|S,\tilde{S}),\nonumber\\
&&R_{e}\leq H(Y|Z,S,\tilde{S})\},
\end{eqnarray}
and the joint probability $P_{S\tilde{S}XYZ}(s\tilde{s}xyz)$ satisfies
\begin{eqnarray}\label{appengiveup3.xxrmb}
&&P_{S\tilde{S}XYZ}(s\tilde{s}xyz)=P_{Z|Y}(z|y)P_{Y|X,S}(y|x,s)K^{d}(\tilde{s},s)P_{X|\tilde{S}}(x|\tilde{s})P_{\tilde{S}}(\tilde{s}).
\end{eqnarray}
\begin{IEEEproof}
See Appendix \ref{appengiveup1.xx}.
\end{IEEEproof}

\item For the degraded case of the model of Figure \ref{f2} with delayed state and receiver's channel output feedback, the secrecy capacity
$\mathcal{C}^{*f}_{s}$ can be directly obtained from the above $\mathcal{R}^{f*}$, and it is given by
\begin{eqnarray}\label{appengiveup2.xx}
&&\mathcal{C}^{*f}_{s}=\max_{P_{X|\tilde{S}}(x|\tilde{s})}\min\{I(X;Y|S,\tilde{S}), H(Y|Z,S,\tilde{S})\}.
\end{eqnarray}
Note that (\ref{appengiveup2.xx}) can also be re-written as
\begin{eqnarray}\label{ppd1}
&&\mathcal{C}^{*f}_{s}=\max_{P_{X|\tilde{S}}(x|\tilde{s})}\min\{I(X;Y|S,\tilde{S}), I(X;Y|S,\tilde{S})-I(X;Z|S,\tilde{S})+H(Y|X,Z,S,\tilde{S})\},
\end{eqnarray}
and this is because
\begin{eqnarray}\label{ppd2}
&&I(X;Y|S,\tilde{S})-I(X;Z|S,\tilde{S})+H(Y|X,Z,S,\tilde{S})=-H(X|S,\tilde{S},Y)+H(X|S,\tilde{S},Z)+H(Y|X,Z,S,\tilde{S})\nonumber\\
&&\stackrel{(1)}=-H(X|S,\tilde{S},Y,Z)+H(X|S,\tilde{S},Z)+H(Y|X,Z,S,\tilde{S})\nonumber\\
&&=I(X;Y|S,\tilde{S},Z)+H(Y|X,Z,S,\tilde{S})\nonumber\\
&&=H(Y|S,\tilde{S},Z),
\end{eqnarray}
where (1) is from the Markov chain $X\rightarrow (S,\tilde{S},Y)\rightarrow Z$. Comparing (\ref{ppd1}) with (\ref{appengiveup2}),
it is easy to see that the delayed receiver's channel output feedback helps to enhance the secrecy capacity of the degraded FSM-WC
with only delayed state feedback.

\end{itemize}
\end{remark}

\section{\textcolor[rgb]{1.00,0.00,0.00}{Secrecy Capacities for Two Special Cases of the Model of Figure \ref{f2}}}\label{secIII}

\subsection{Secrecy Capacity for the Degraded Gaussian Case of the model of Figure \ref{f2} with or without Delayed Receiver's Channel Output Feedback}\label{sub31}

In this subsection, we compute the secrecy capacities for the degraded Gaussian case of Figure \ref{f2} with or without delayed receiver's channel
output feedback, and
investigate how this delayed feedback and channel memory affect the secrecy capacities.
At the $i$-th time ($1\leq i\leq N$), the inputs and outputs of the channel satisfy
\begin{equation}\label{e301}
Y_{i}=X_{i}+N_{S_{i}}, \,\, Z_{i}=Y_{i}+N_{w,i}.
\end{equation}
Here note that $N_{S_{i}}$
is Gaussian distributed with zero mean, and the variance depends on the $i$-th time state $S_{i}=s_{i}$ (denoted by $\sigma^{2}_{s_{i}}$).
The random variable $N_{w,i}$ ($1\leq i\leq N$) is also Gaussian distributed with zero mean and constant variance $\sigma^{2}_{w}$
($N_{w,i}\sim \mathcal{N}(0, \sigma^{2}_{w})$ for all $i\in \{1,2,...,N\}$).
At time $i$, the receiver has access to the state $S_{i}$ and the output $Y_{i}$. The state $S_{i}$ is fed back to
the transmitter through a noiseless feedback channel with a delay time $d$.
The state undergoes a Markov process with steady probability distribution $\pi(s)$ and $1$-step transition
probability matrix $K$. The power constraint of the transmitter is given by
\begin{equation}\label{e304}
\sum_{\tilde{s}}\pi(\tilde{s})E_{P_{X|\tilde{S}}(x|\tilde{s})}[X^{2}|\tilde{s}]\leq \mathcal{P}_{0}.
\end{equation}

\emph{\textbf{Secrecy capacity for the degraded Gaussian case of the model of Figure \ref{f2} with only delayed state feedback:}}
\begin{theorem}\label{Ts1}
\textcolor[rgb]{1.00,0.00,0.00}{For the degraded Gaussian case of the model of Figure \ref{f2} with only delayed state feedback, the secrecy capacity $C_{s}^{(g)}$ is given by
\begin{equation}\label{e306}
C_{s}^{(g)}=\max_{\mathcal{P}(\tilde{s}): \sum_{\tilde{s}}\pi(\tilde{s})\mathcal{P}(\tilde{s})\leq \mathcal{P}_{0}}
\sum_{\tilde{s}}\sum_{s}\pi(\tilde{s})K^{d}(\tilde{s},s)(\frac{1}{2}\log (1+\frac{\mathcal{P}(\tilde{s})}{\sigma^{2}_{s}})
-\frac{1}{2}\log (1+\frac{\mathcal{P}(\tilde{s})}{\sigma^{2}_{s}+\sigma^{2}_{w}})),
\end{equation}
where $\mathcal{P}(\tilde{s})$ is the transmitter's power for the state $\tilde{s}$, and
$\sigma^{2}_{s}$ is the variance of the noise $N_{S}$ given the state $S=s$. Here note that the definition of $\mathcal{P}(\tilde{s})$ is the same as that of the
finite state additive Gaussian noise channel \cite{vis}.}

\end{theorem}

\begin{IEEEproof}

\emph{(Converse part:)} Using (\ref{appengiveup2}), the secrecy capacity $C_{s}^{(g)}$ can be re-written by
\begin{equation}\label{e303}
C_{s}^{(g)}=\max_{P_{X|\tilde{S}}(x|\tilde{s})}\sum_{\tilde{s}}\pi(\tilde{s})\sum_{s}
K^{d}(\tilde{s},s)(I(X;Y|S=s,\tilde{S}=\tilde{s})-I(X;Z|S=s,\tilde{S}=\tilde{s})).
\end{equation}
Letting $\mathcal{P}(\tilde{s})$ be the transmitter's power for the state $\tilde{s}$ satisfying (\ref{e304}),
and $\sigma^{2}_{s}$ be the variance of the noise $N_{S}$ given the state $S=s$,
then we have
\begin{eqnarray}\label{e305}
&&I(X;Y|S=s,\tilde{S}=\tilde{s})-I(X;Z|S=s,\tilde{S}=\tilde{s})\nonumber\\
&&=h(Y|S=s,\tilde{S}=\tilde{s})-h(Y|X,S=s,\tilde{S}=\tilde{s})-h(Z|S=s,\tilde{S}=\tilde{s})+h(Z|X,S=s,\tilde{S}=\tilde{s})\nonumber\\
&&=h(X_{\tilde{s}}+N_{s})-h(N_{s})-h(X_{\tilde{s}}+N_{s}+N_{w})+h(N_{s}+N_{w})\nonumber\\
&&\stackrel{(a)}\leq h(X_{\tilde{s}}+N_{s})-h(N_{s})-\frac{1}{2}\log(2^{2h(X_{\tilde{s}}+N_{s})}+2^{2h(N_{w})})+h(N_{s}+N_{w})\nonumber\\
&&\stackrel{(b)}\leq \frac{1}{2}\log (1+\frac{\mathcal{P}(\tilde{s})}{\sigma^{2}_{s}})
-\frac{1}{2}\log (1+\frac{\mathcal{P}(\tilde{s})}{\sigma^{2}_{s}+\sigma^{2}_{w}}),
\end{eqnarray}
where (a) is from the entropy power inequality,
(b) is from $h(X_{\tilde{s}}+N_{s})-\frac{1}{2}\log(2^{2h(X_{\tilde{s}}+N_{s})}+2^{2h(N_{w})})$ is increasing while $h(X_{\tilde{s}}+N_{s})$
is increasing, and the fact that for a given variance, the largest entropy is achieved if the random variable is Gaussian distributed.
Furthermore, the ``='' in (a) is achieved if $X_{\tilde{s}}\sim \mathcal{N}(0, \mathcal{P}(\tilde{s}))$
and $X_{\tilde{s}}$ is independent of $N_{s}$.
Applying (\ref{e305}) to (\ref{e303}), the converse part of \textcolor[rgb]{1.00,0.00,0.00}{Theorem \ref{Ts1}} is proved.

\emph{(Direct part:)}
Letting $X_{\tilde{s}}$ be the random variable $X$ given the delayed state $\tilde{s}$, and
substituting
$X_{\tilde{s}}\sim \mathcal{N}(0, \mathcal{P}(\tilde{s}))$ and (\ref{e301}) into (\ref{e303}),
the achievability proof of \textcolor[rgb]{1.00,0.00,0.00}{Theorem \ref{Ts1}} is along the lines of that of (\ref{appengiveup2}) 
(see Appendix \ref{appengiveup1}), and thus we omit the proof here.

The proof of \textcolor[rgb]{1.00,0.00,0.00}{Theorem \ref{Ts1}} is completed.

\end{IEEEproof}

\emph{\textbf{Secrecy capacity for the degraded Gaussian case of the model of Figure \ref{f2} with delayed state and receiver's channel output feedback:}}

\begin{theorem}\label{Ts2}
\textcolor[rgb]{1.00,0.00,0.00}{For the degraded Gaussian case of the model of Figure \ref{f2} with delayed state and receiver's channel output feedback, the secrecy capacity $C_{s}^{(gf)}$ is given by
\begin{equation}\label{efm1}
C_{s}^{(gf)}=\max_{\mathcal{P}(\tilde{s}): \sum_{\tilde{s}}\pi(\tilde{s})\mathcal{P}(\tilde{s})\leq \mathcal{P}_{0}}
\sum_{\tilde{s}}\sum_{s}\pi(\tilde{s})K^{d}(\tilde{s},s)\min\{\frac{1}{2}\log(1+\frac{\mathcal{P}(\tilde{s})}{\sigma^{2}_{s}}),
\frac{1}{2}\log\frac{2\pi e\sigma^{2}_{w}(\mathcal{P}(\tilde{s})+\sigma^{2}_{s})}{\mathcal{P}(\tilde{s})+\sigma^{2}_{s}+\sigma^{2}_{w}}\}.
\end{equation}}

\end{theorem}

\begin{IEEEproof}
Defining $\mathcal{P}(\tilde{s})$ as the transmitter's power for the state $\tilde{s}$,
the secrecy capacity
$\mathcal{C}_{s}^{*f}$ in (\ref{appengiveup2.xx}) can be re-written as
\begin{equation}\label{efm2}
\mathcal{C}_{s}^{*f}=\max_{\mathcal{P}(\tilde{s}): \sum_{\tilde{s}}\pi(\tilde{s})\mathcal{P}(\tilde{s})\leq \mathcal{P}_{0}}
\sum_{\tilde{s}}\sum_{s}\pi(\tilde{s})K^{d}(\tilde{s},s)\min\{I(X;Y|S=s,\tilde{S}=\tilde{s}), H(Y|Z,S=s,\tilde{S}=\tilde{s})\}.
\end{equation}

\emph{(Converse part:)}
Defining
$\sigma^{2}_{s}$ as the variance of the noise $N_{S}$ given the state $S=s$, the mutual information $I(X;Y|S=s,\tilde{S}=\tilde{s})$
in (\ref{efm2}) can be further bounded by
\begin{eqnarray}\label{efm2.1}
I(X;Y|S=s,\tilde{S}=\tilde{s})&=&h(Y|S=s,\tilde{S}=\tilde{s})-h(Y|S=s,\tilde{S}=\tilde{s},X)\nonumber\\
&\leq&h(X_{\tilde{s}}+N_{s})-h(Y|S=s,\tilde{S}=\tilde{s},X)\nonumber\\
&=&h(X_{\tilde{s}}+N_{s})-h(N_{s})\nonumber\\
&\stackrel{(a)}\leq&\frac{1}{2}\log(1+\frac{\mathcal{P}(\tilde{s})}{\sigma^{2}_{s}}),
\end{eqnarray}
where (a) is from
the fact that for a given variance, the largest entropy is achieved if the random variable is Gaussian distributed.

Moreover, the differential conditional entropy $h(Y|Z,S=s,\tilde{S}=\tilde{s})$ can be further bounded by
\begin{eqnarray}\label{efm2.2}
h(Y|Z,S=s,\tilde{S}=\tilde{s})&=&h(Y,Z,S=s,\tilde{S}=\tilde{s})-h(Z,S=s,\tilde{S}=\tilde{s})\nonumber\\
&\stackrel{(b)}=&h(Z|Y)+h(Y,S=s,\tilde{S}=\tilde{s})-h(Z,S=s,\tilde{S}=\tilde{s})\nonumber\\
&=&h(Z|Y)+h(Y|S=s,\tilde{S}=\tilde{s})-h(Z|S=s,\tilde{S}=\tilde{s})\nonumber\\
&\stackrel{(c)}=&h(N_{w})+h(Y|S=s,\tilde{S}=\tilde{s})-h(Y+N_{w}|S=s,\tilde{S}=\tilde{s})\nonumber\\
&\stackrel{(d)}\leq&h(N_{w})+h(Y|S=s,\tilde{S}=\tilde{s})-\frac{1}{2}\log(2^{2h(Y|S=s,\tilde{S}=\tilde{s})}+2^{2h(N_{w})})\nonumber\\
&=&\frac{1}{2}\log(2\pi e\sigma^{2}_{w})+h(Y|S=s,\tilde{S}=\tilde{s})-\frac{1}{2}\log(2^{2h(Y|S=s,\tilde{S}=\tilde{s})}+2\pi e\sigma^{2}_{w})\nonumber\\
&\stackrel{(e)}\leq&\frac{1}{2}\log(2\pi e\sigma^{2}_{w})+\frac{1}{2}\log(2\pi e(\mathcal{P}(\tilde{s})+\sigma^{2}_{s}))
-\frac{1}{2}\log(2\pi e(\mathcal{P}(\tilde{s})+\sigma^{2}_{s}+\sigma^{2}_{w}))\nonumber\\
&=&\frac{1}{2}\log\frac{2\pi e\sigma^{2}_{w}(\mathcal{P}(\tilde{s})+\sigma^{2}_{s})}{\mathcal{P}(\tilde{s})+\sigma^{2}_{s}+\sigma^{2}_{w}},
\end{eqnarray}
where (b) is from the Markov chain $(S,\tilde{S})\rightarrow Y\rightarrow Z$, (c) is from the fact that $Z=Y+N_{w}$,
(d) is from the entropy power inequality, and (e) is from the fact that
$h(Y|S=s,\tilde{S}=\tilde{s})-\frac{1}{2}\log(2^{2h(Y|S=s,\tilde{S}=\tilde{s})}+2\pi e\sigma^{2}_{w})$ is increasing while $h(Y|S=s,\tilde{S}=\tilde{s})$
is increasing, and
\begin{eqnarray}\label{efm2.3}
&&h(Y|S=s,\tilde{S}=\tilde{s})\leq h(X_{\tilde{s}}+N_{s})\leq \frac{1}{2}\log(2\pi e(\mathcal{P}(\tilde{s})+\sigma^{2}_{s})).
\end{eqnarray}
Applying (\ref{efm2.1}) and (\ref{efm2.2}) to (\ref{efm2}), the converse proof of \textcolor[rgb]{1.00,0.00,0.00}{Theorem \ref{Ts2}} is completed.

\emph{(Direct part:)}
Letting $X_{\tilde{s}}$ be the random variable $X$ given the delayed state $\tilde{s}$, and
substituting
$X_{\tilde{s}}\sim \mathcal{N}(0, \mathcal{P}(\tilde{s}))$ and (\ref{e301}) into (\ref{efm2}),
the achievability proof of \textcolor[rgb]{1.00,0.00,0.00}{Theorem \ref{Ts2}} is along the lines of that of Theorem \ref{T1}, and thus we omit the details here.

The proof of \textcolor[rgb]{1.00,0.00,0.00}{Theorem \ref{Ts2}} is completed.

\end{IEEEproof}

\emph{\textbf{Numerical results of \textcolor[rgb]{1.00,0.00,0.00}{$C_{s}^{(g)}$ and $C_{s}^{(gf)}$}}}

\begin{figure}[htb]
\centerline{\includegraphics[scale=0.55]{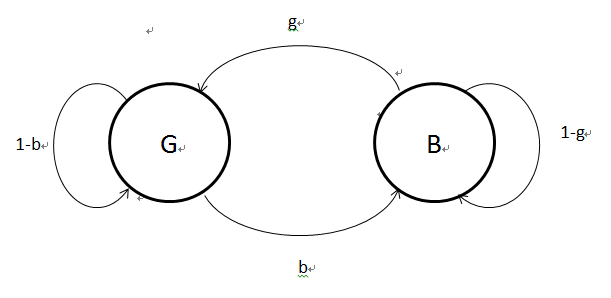}}
\caption{\textcolor[rgb]{1.00,0.00,0.00}{The state process of the two-state case}}
\label{f5}
\end{figure}

In order to gain some intuition on the secrecy capacities \textcolor[rgb]{1.00,0.00,0.00}{$C_{s}^{(g)}$ and $C_{s}^{(gf)}$}, we consider a simple
case that the state alphabet $\mathcal{S}$ is composed of only two elements.
At each time instant, the state of the channel is $G$ (good state) or $B$ (bad state).
For the state $G$, the noise variance of the channel is $\sigma^{2}_{G}$. Analogously, for the state $B$,
the noise variance of the channel is $\sigma^{2}_{B}$.
Here note that
$\sigma^{2}_{B}>\sigma^{2}_{G}$. The state process is shown in Figure \ref{f5}, where
\begin{equation}\label{e307}
P(G|G)=1-b, \,\, P(B|G)=b, \,\, P(B|B)=1-g, \,\, P(G|B)=g.
\end{equation}
The steady state probabilities $\pi(G)$ and $\pi(B)$ are given by
\begin{equation}\label{e308}
\pi(G)=\frac{g}{g+b},\,\, \pi(B)=\frac{b}{g+b}.
\end{equation}

\begin{figure}[htb]
\centerline{\includegraphics[scale=0.45]{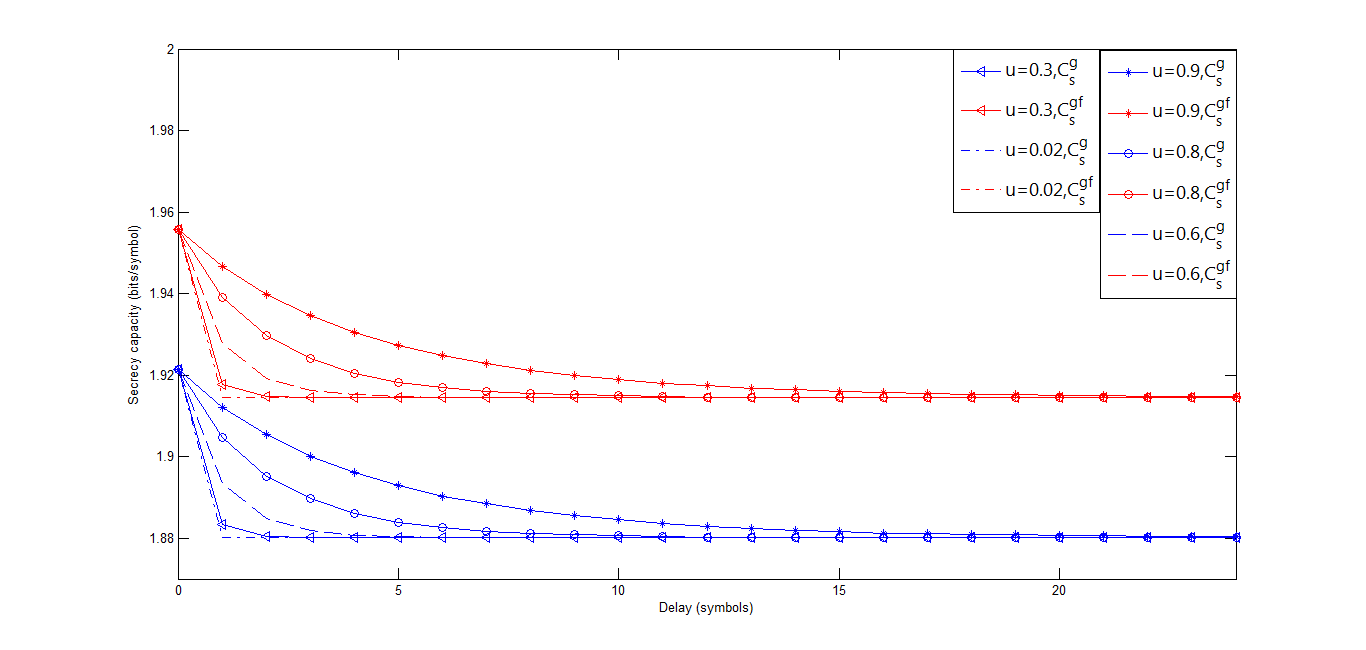}}
\caption{\textcolor[rgb]{1.00,0.00,0.00}{The secrecy
capacities $C_{s}^{(g)}$ and $C_{s}^{(gf)}$ for
$\mathcal{P}_{0}=100$, $\sigma^{2}_{G}=1$, $\sigma^{2}_{B}=100$, $\sigma^{2}_{w}=2000$, $c=1$ and several values of $u$}}
\label{f3}
\end{figure}

\begin{figure}[htb]
\centerline{\includegraphics[scale=0.45]{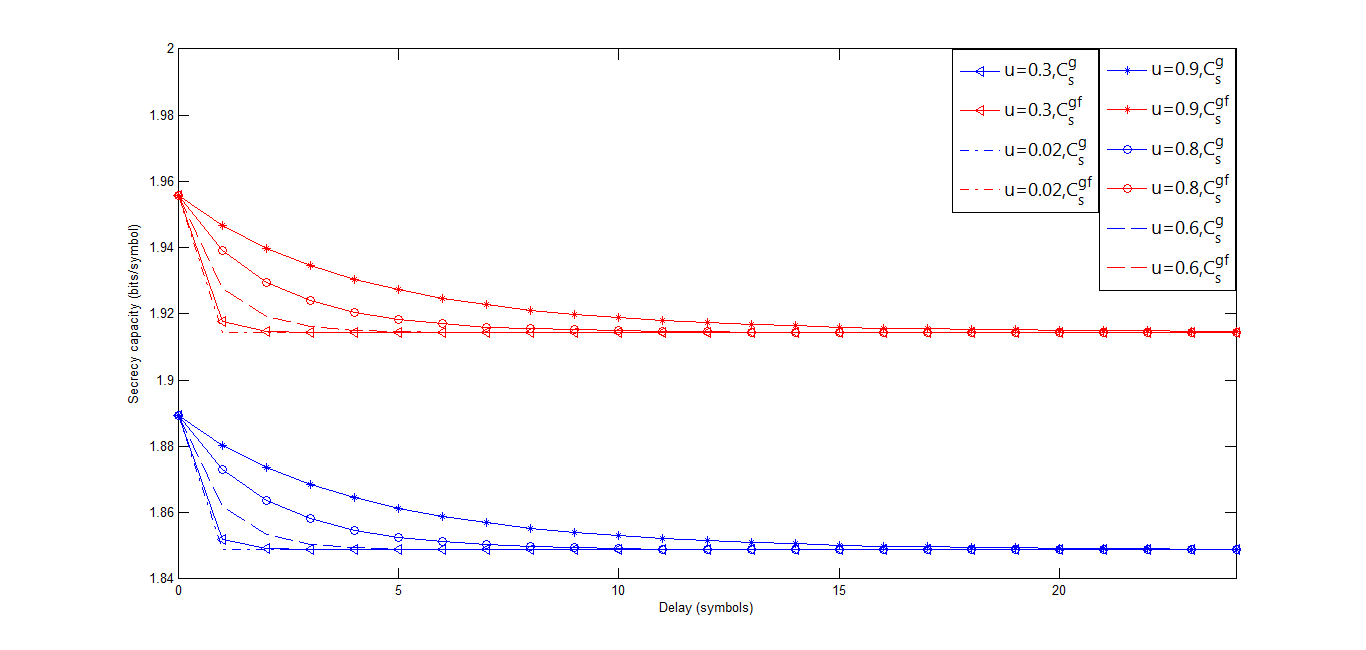}}
\caption{\textcolor[rgb]{1.00,0.00,0.00}{The secrecy
capacities $C_{s}^{(g)}$ and $C_{s}^{(gf)}$ for
$\mathcal{P}_{0}=100$, $\sigma^{2}_{G}=1$, $\sigma^{2}_{B}=100$, $\sigma^{2}_{w}=1000$, $c=1$ and several values of $u$}}
\label{f3.1}
\end{figure}

Define $u=1-g-b$ and $c=g/b$. The parameter $u$ is related to the channel memory,
\footnote{Mushkin and Bar-David \cite{mush} has already shown that the channel memory is increasing while $u$ is increasing.}
and the parameter $c$ controls the steady state distributions (see \ref{e308}).
Fixing $c$ (for example, $c=1$),
we can choose different $u$ and $d$ to investigate the effects of channel memory and feedback delay on the secrecy capacities $C_{s}^{(g)}$ and $C_{s}^{(gf)}$.
Figure \ref{f3} and Figure \ref{f3.1} show the effect of the feedback
delay on the secrecy capacities for $\mathcal{P}_{0}=100$, $\sigma^{2}_{G}=1$, $\sigma^{2}_{B}=100$, $\sigma^{2}_{w}=2000$ ($\sigma^{2}_{w}=1000$) , $c=1$ and
several values of $u$. As we can see in Figure \ref{f3} and Figure \ref{f3.1}, when the channel is changing rapidly
(which implies that the channel memory $u$ is small, for example, $u=0.02$),
the secrecy capacity goes to the infinite asymptote even if $d=1$.
However, when the channel is changing slowly (which implies that the channel memory $u$ is large, for example, $u=0.9$),
a larger feedback delay is tolerable since the secrecy capacity loss compared with feedback without delay ($d=0$) is smaller.
Moreover, it is easy to see that the delayed receiver's channel output feedback enhances the secrecy capacity $C_{s}^{(g)}$ of the degraded Gaussian case of
the FSM-WC with only delayed state feedback. Furthermore, comparing these two figures, we can see that for fixed $\mathcal{P}_{0}$, $\sigma^{2}_{G}$,
$\sigma^{2}_{B}$ and $c$, the gap between $C_{s}^{(g)}$ and $C_{s}^{(gf)}$ is increasing while $\sigma^{2}_{w}$ is decreasing.

\subsection{Secrecy Capacity for the Degraded Gaussian Fading Case of Figure \ref{f2}}\label{sub32}

\begin{figure}[htb]
\centerline{\includegraphics[scale=0.45]{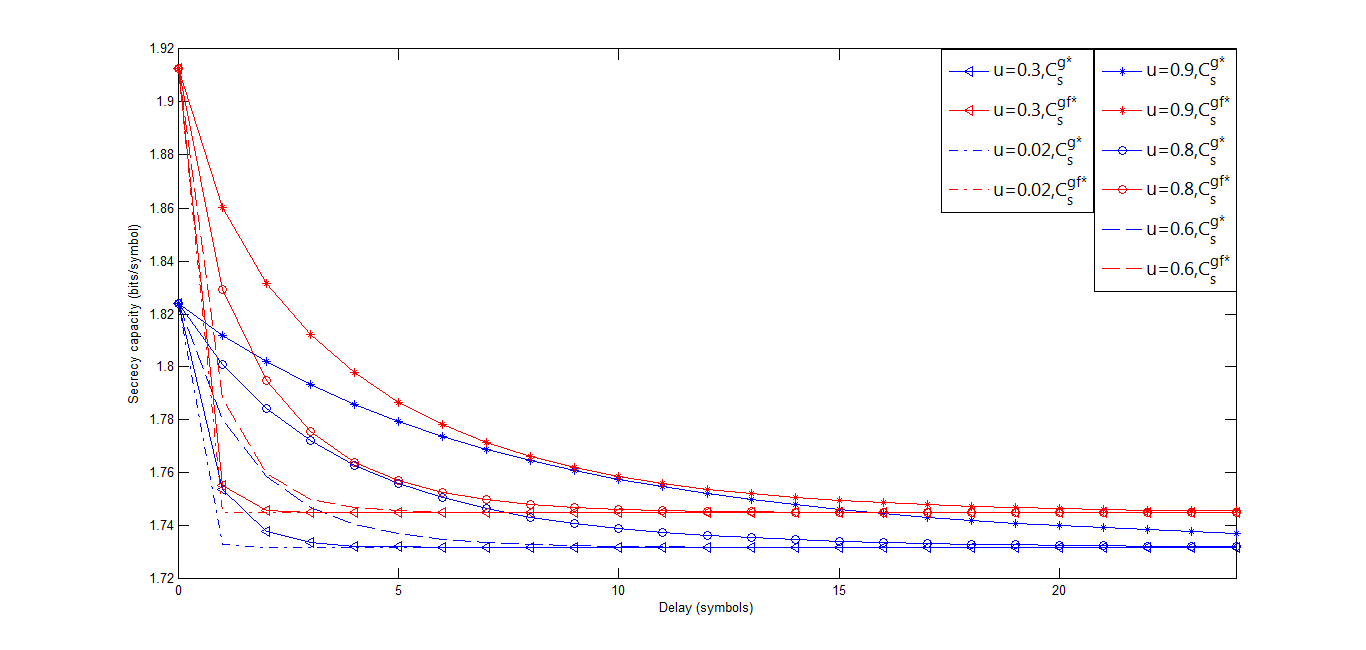}}
\caption{\textcolor[rgb]{1.00,0.00,0.00}{The secrecy
capacities $C_{s}^{(g*)}$ and $C_{s}^{(gf*)}$ for
$\mathcal{P}_{0}=100$, $\sigma^{2}_{G}=1$, $\sigma^{2}_{B}=100$, $\sigma^{2}_{w}=200$, $c=1$, $g(G)=1$, $g(B)=0.5$, 
$l(G)=0.8$, $l(B)=0.2$ and several values of $u$}}
\label{f4ddc1}
\end{figure}

\begin{figure}[htb]
\centerline{\includegraphics[scale=0.45]{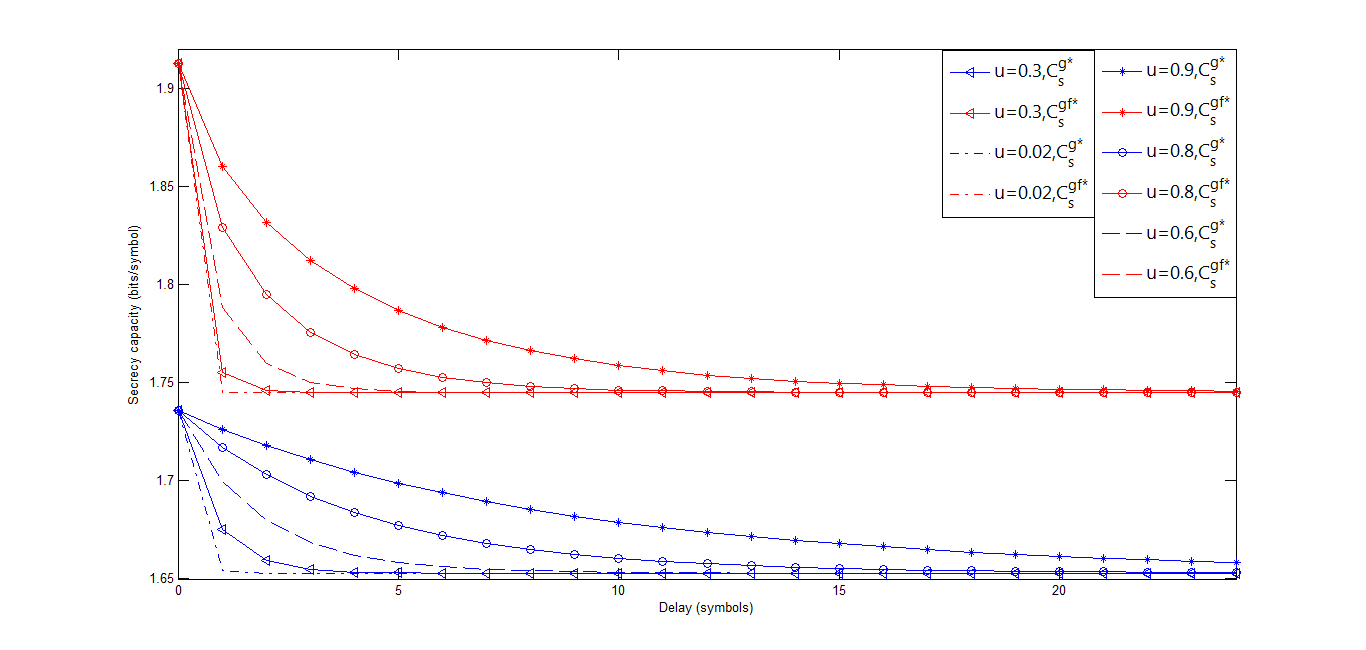}}
\caption{\textcolor[rgb]{1.00,0.00,0.00}{The secrecy
capacities $C_{s}^{(g*)}$ and $C_{s}^{(gf*)}$ for
$\mathcal{P}_{0}=100$, $\sigma^{2}_{G}=1$, $\sigma^{2}_{B}=100$, $\sigma^{2}_{w}=100$, $c=1$, 
$g(G)=1$, $g(B)=0.5$, $l(G)=0.8$, $l(B)=0.2$ and several values of $u$}}
\label{f4ddc2}
\end{figure}

\begin{figure}[htb]
\centerline{\includegraphics[scale=0.45]{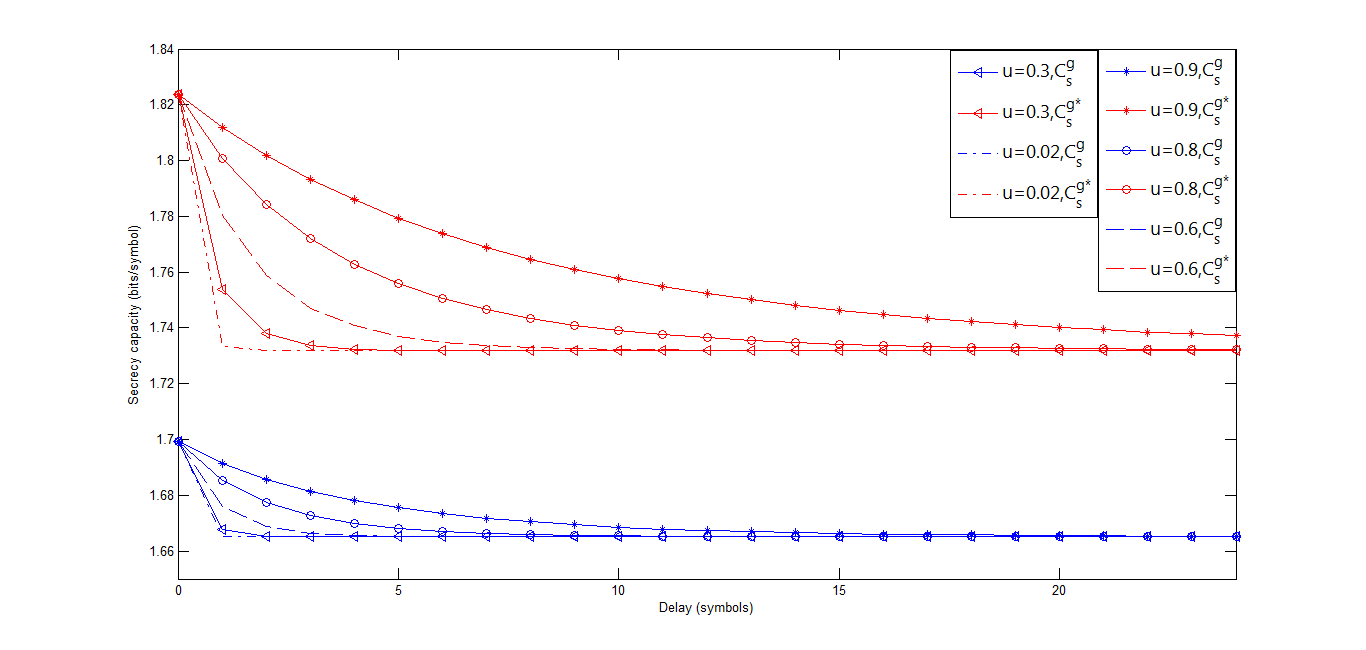}}
\caption{\textcolor[rgb]{1.00,0.00,0.00}{The comparison of the secrecy
capacities $C_{s}^{(g*)}$ and $C_{s}^{(g)}$ for
$\mathcal{P}_{0}=100$, $\sigma^{2}_{G}=1$, $\sigma^{2}_{B}=100$, $\sigma^{2}_{w}=200$, 
$c=1$, $g(G)=1$, $g(B)=0.5$, $l(G)=0.8$, $l(B)=0.2$ and several values of $u$}}
\label{f4ddc3}
\end{figure}

\begin{figure}[htb]
\centerline{\includegraphics[scale=0.45]{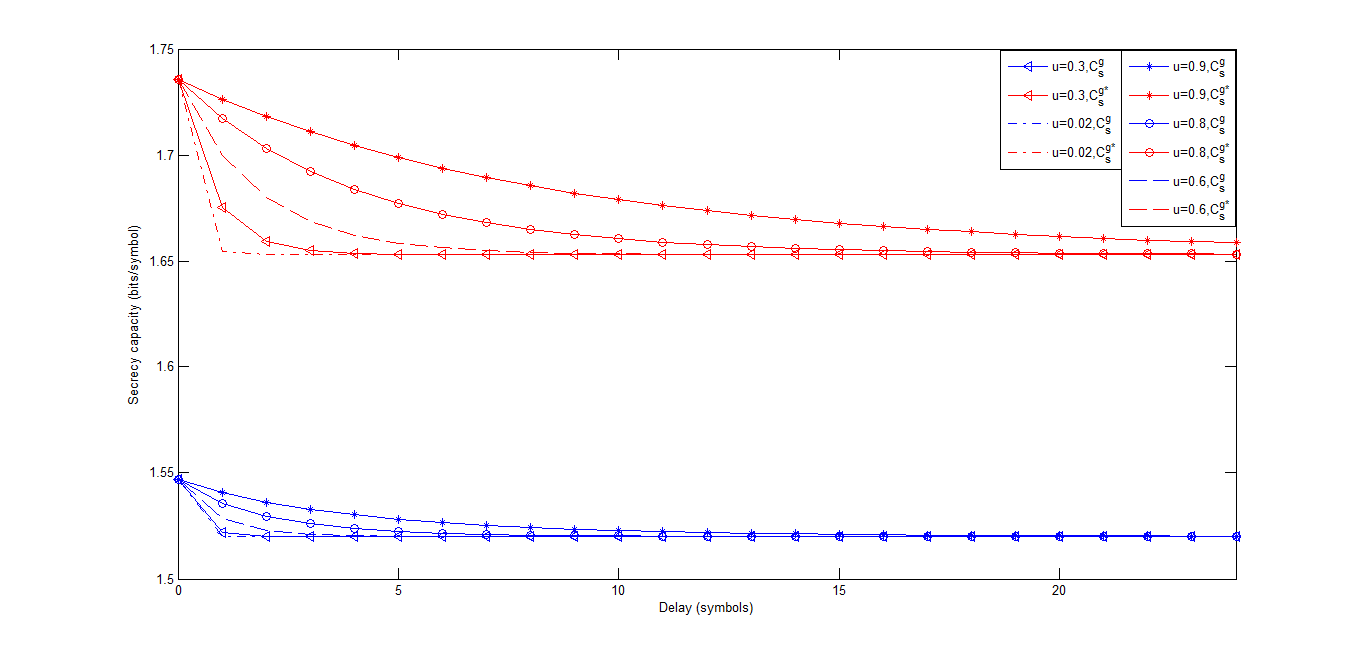}}
\caption{\textcolor[rgb]{1.00,0.00,0.00}{The comparison of the secrecy
capacities $C_{s}^{(g*)}$ and $C_{s}^{(g)}$ for
$\mathcal{P}_{0}=100$, $\sigma^{2}_{G}=1$, $\sigma^{2}_{B}=100$, $\sigma^{2}_{w}=100$, $c=1$, 
$g(G)=1$, $g(B)=0.5$, $l(G)=0.8$, $l(B)=0.2$ and several values of $u$}}
\label{f4ddc4}
\end{figure}

\begin{figure}[htb]
\centerline{\includegraphics[scale=0.45]{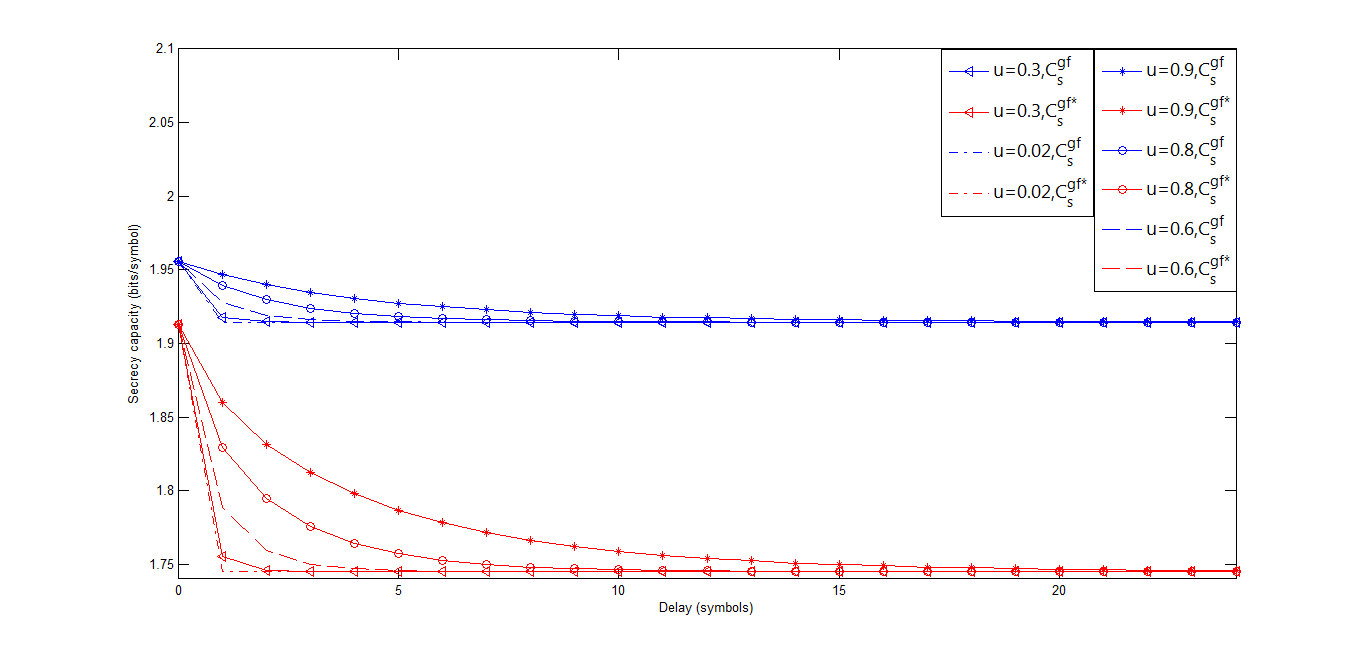}}
\caption{\textcolor[rgb]{1.00,0.00,0.00}{The comparison of the secrecy
capacities $C_{s}^{(gf*)}$ and $C_{s}^{(gf)}$ for
$\mathcal{P}_{0}=100$, $\sigma^{2}_{G}=1$, $\sigma^{2}_{B}=100$, $\sigma^{2}_{w}=200$, 
$c=1$, $g(G)=1$, $g(B)=0.5$, $l(G)=0.8$, $l(B)=0.2$ and several values of $u$}}
\label{f4ddc5}
\end{figure}

\begin{figure}[htb]
\centerline{\includegraphics[scale=0.45]{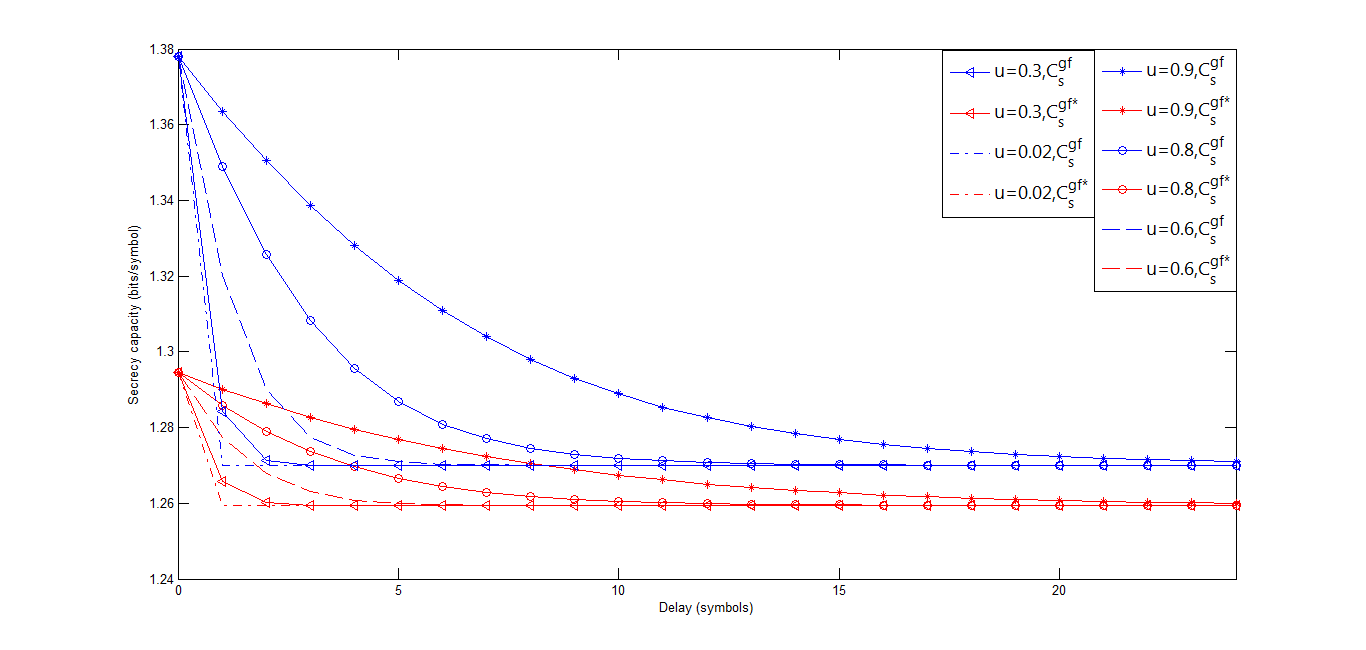}}
\caption{\textcolor[rgb]{1.00,0.00,0.00}{The comparison of the secrecy
capacities $C_{s}^{(gf*)}$ and $C_{s}^{(gf)}$ for
$\mathcal{P}_{0}=100$, $\sigma^{2}_{G}=1$, $\sigma^{2}_{B}=100$, $\sigma^{2}_{w}=1$, 
$c=1$, $g(G)=1$, $g(B)=0.5$, $l(G)=0.8$, $l(B)=0.2$ and several values of $u$}}
\label{f4ddc6}
\end{figure}

In this subsection, we compute the secrecy capacities for the degraded Gaussian fading case of Figure \ref{f2}.
At the $i$-th time ($1\leq i\leq N$), the inputs and the outputs of the channel satisfy
\begin{equation}\label{e310}
Y_{i}=g(s_{i})X_{i}+N_{S_{i}}, \,\, Z_{i}=\textcolor[rgb]{1.00,0.00,0.00}{l(s_{i})}Y_{i}+N_{w,i}.
\end{equation}
\textcolor[rgb]{1.00,0.00,0.00}{Here $g(s_{i})$ and $l(s_{i})$ are the fading processes of the channels for the receiver
and the eavesdropper, respectively, and they are deterministic functions of $s_{i}$.}
The noise $N_{S_{i}}$
is Gaussian distributed with zero mean, and the variance depends on the $i$-th time state $S_{i}$ of the channel.
The random variable $N_{w,i}$ ($1\leq i\leq N$) is also Gaussian distributed with zero mean and constant variance $\sigma^{2}_{w}$
($N_{w,i}\sim \mathcal{N}(0, \sigma^{2}_{w})$ for all $i\in \{1,2,...,N\}$).
Now we apply (\ref{appengiveup2}) to determine
the secrecy capacities of this degraded Gaussian fading model with or without delayed receiver's channel output feedback, see the remainder of this subsection.

\emph{\textbf{Secrecy capacity for the degraded Gaussian fading case of the model of Figure \ref{f2} with only delayed state feedback:}}
\begin{theorem}\label{Ts3}
\textcolor[rgb]{1.00,0.00,0.00}{For the degraded Gaussian fading case of the model of Figure \ref{f2} with only delayed state feedback, 
the secrecy capacity $C_{s}^{(g*)}$ is given by
\begin{equation}\label{e313}
C_{s}^{(g*)}=\max_{\mathcal{P}(\tilde{s}): \sum_{\tilde{s}}\pi(\tilde{s})\mathcal{P}(\tilde{s})\leq \mathcal{P}_{0}}
\frac{1}{2}\sum_{\tilde{s}}\sum_{s}\pi(\tilde{s})K^{d}(\tilde{s},s)(\frac{1}{2}\log (1+\frac{g^{2}(s)\mathcal{P}(\tilde{s})}{\sigma^{2}_{s}})
-\frac{1}{2}\log (1+\frac{g^{2}(s)\textcolor[rgb]{1.00,0.00,0.00}{l^{2}(s)}\mathcal{P}(\tilde{s})}{\textcolor[rgb]{1.00,0.00,0.00}{l^{2}(s)}\sigma^{2}_{s}+\sigma^{2}_{w}})).
\end{equation}}

\end{theorem}

\begin{IEEEproof}

Similar to Subsection \ref{sub31}, let $\mathcal{P}(\tilde{s})$ be the power for the state $\tilde{s}$, and
$\sigma^{2}_{s}$ be the variance of the noise $N_{S}$ given $S=s$, and thus we have
\begin{eqnarray}\label{e312}
&&I(X;Y|S=s,\tilde{S}=\tilde{s})-I(X;Z|S=s,\tilde{S}=\tilde{s})\nonumber\\
&&=h(Y|S=s,\tilde{S}=\tilde{s})-h(Y|X,S=s,\tilde{S}=\tilde{s})-h(Z|S=s,\tilde{S}=\tilde{s})+h(Z|X,S=s,\tilde{S}=\tilde{s})\nonumber\\
&&=h(g(s)X_{\tilde{s}}+N_{s})-h(N_{s})-h(\textcolor[rgb]{1.00,0.00,0.00}{l(s)}(g(s)X_{\tilde{s}}+N_{s})+N_{w})
+h(\textcolor[rgb]{1.00,0.00,0.00}{l(s)}N_{s}+N_{w})\nonumber\\
&&\stackrel{(a)}\leq h(g(s)X_{\tilde{s}}+N_{s})-h(N_{s})-\frac{1}{2}\log(2^{2h(g(s)X_{\tilde{s}}+N_{s})}\textcolor[rgb]{1.00,0.00,0.00}{l^{2}(s)}
+2^{2h(N_{w})})+h(\textcolor[rgb]{1.00,0.00,0.00}{l(s)}N_{s}+N_{w})\nonumber\\
&&\stackrel{(b)}\leq \frac{1}{2}\log (1+\frac{g^{2}(s)\mathcal{P}(\tilde{s})}{\sigma^{2}_{s}})
-\frac{1}{2}\log (1+\frac{g^{2}(s)\textcolor[rgb]{1.00,0.00,0.00}{l^{2}(s)}\mathcal{P}(\tilde{s})}{\textcolor[rgb]{1.00,0.00,0.00}{l^{2}(s)}\sigma^{2}_{s}+\sigma^{2}_{w}}),
\end{eqnarray}
where (a) is from the entropy power inequality and the property that $h(aX)=h(X)+\log a$,
and (b) is from $h(g(s)X_{\tilde{s}}+N_{s})-\frac{1}{2}\log(2^{2h(g(s)X_{\tilde{s}}+N_{s})}\textcolor[rgb]{1.00,0.00,0.00}{l^{2}(s)}+2^{2h(N_{w})})$ is 
increasing while $h(g(s)X_{\tilde{s}}+N_{s})$
is increasing, and
the fact that for a given variance, the largest entropy is achieved if the random variable is Gaussian distributed.
Furthermore, the ``='' in (a) is achieved if $X_{\tilde{s}}\sim \mathcal{N}(0, \mathcal{P}(\tilde{s}))$
and $X_{\tilde{s}}$ is independent of $N_{s}$.
Applying (\ref{e312}) to (\ref{e303}), \textcolor[rgb]{1.00,0.00,0.00}{the converse proof of Theorem \ref{Ts3} is completed.}

Here note that replacing $X_{i}$ by $g(s_{i})X_{i}$, and $Y_{i}$ by $\textcolor[rgb]{1.00,0.00,0.00}{l(s_{i})}Y_{i}$,
the achievability proof of \textcolor[rgb]{1.00,0.00,0.00}{Theorem \ref{Ts3}} is along the lines of that of \textcolor[rgb]{1.00,0.00,0.00}{Theorem \ref{Ts1}}, 
and thus we omit the proof here.

\textcolor[rgb]{1.00,0.00,0.00}{The proof of Theorem \ref{Ts3} is completed.}
\end{IEEEproof}

\emph{\textbf{Secrecy capacity for the degraded Gaussian fading case of the model of Figure \ref{f2} with delayed state and receiver's channel output feedback:}}

\begin{theorem}\label{Ts4}
\textcolor[rgb]{1.00,0.00,0.00}{For the degraded Gaussian fading case of the model of Figure \ref{f2} with delayed state and receiver's channel output feedback,
the secrecy capacity $C_{s}^{(gf*)}$ is given by
\begin{equation}\label{marry1}
C_{s}^{(gf*)}=\max_{\mathcal{P}(\tilde{s}): \sum_{\tilde{s}}\pi(\tilde{s})\mathcal{P}(\tilde{s})\leq \mathcal{P}_{0}}
\sum_{\tilde{s}}\sum_{s}\pi(\tilde{s})K^{d}(\tilde{s},s)\min\{\frac{1}{2}\log(1+\frac{g^{2}(s)\mathcal{P}(\tilde{s})}{\sigma^{2}_{s}}),
\frac{1}{2}\log\frac{2\pi e\sigma^{2}_{w}(g^{2}(s)\mathcal{P}(\tilde{s})+\sigma^{2}_{s})}{g^{2}(s)
\textcolor[rgb]{1.00,0.00,0.00}{l^{2}(s)}\mathcal{P}(\tilde{s})+\textcolor[rgb]{1.00,0.00,0.00}{l^{2}(s)}\sigma^{2}_{s}+\sigma^{2}_{w}}\}.
\end{equation}}

\end{theorem}

\begin{IEEEproof}
Replacing $X_{i}$ by $g(s_{i})X_{i}$, and $Y_{i}$ by $\textcolor[rgb]{1.00,0.00,0.00}{l(s_{i})}Y_{i}$, the proof of \textcolor[rgb]{1.00,0.00,0.00}{Theorem \ref{Ts4}}
is along the lines of that of \textcolor[rgb]{1.00,0.00,0.00}{Theorem \ref{Ts2}}, and thus we omit the proof here.
\end{IEEEproof}

\emph{\textbf{Numerical results of \textcolor[rgb]{1.00,0.00,0.00}{$C_{s}^{(g*)}$ and $C_{s}^{(gf*)}$}}}

We consider a simple two-state
case where the state process is the same as that in Subsection \ref{sub31}, see Figure \ref{f5}.
Define $g(G)=1$, $g(B)=0.5$, \textcolor[rgb]{1.00,0.00,0.00}{$l(G)=0.8$, $l(B)=0.2$}, $u=1-g-b$ and $c=g/b$.
By choosing $c=1$, Figure \ref{f4ddc1} and Figure \ref{f4ddc2}
show the effect of the feedback
delay ($d$) and channel memory ($u$) on the secrecy capacities $C_{s}^{(g*)}$ and $C_{s}^{(gf*)}$
for $\mathcal{P}_{0}=100$, $\sigma^{2}_{G}=1$, $\sigma^{2}_{B}=100$, $\sigma^{2}_{w}=200$ ($\sigma^{2}_{w}=100$) and
several values of $u$.
Similar to the numerical result of Subsection \ref{sub31}, we find that
when the channel is changing rapidly
(which implies that the channel memory $u$ is small, for example, $u=0.02$),
the secrecy capacity goes to the infinite asymptote even if $d=1$.
However, when the channel is changing slowly (which implies that the channel memory $u$ is large, for example, $u=0.9$),
a larger feedback delay is tolerable since the secrecy capacity loss compared with feedback without delay ($d=0$) is smaller.
Moreover, it is easy to see that the delayed
receiver's channel output feedback enhances the secrecy capacity $C_{s}^{(g*)}$ of the degraded Gaussian fading case of
the FSM-WC with only delayed state feedback. Furthermore, comparing these two figures, we can see that for fixed $\mathcal{P}_{0}$, $\sigma^{2}_{G}$,
$\sigma^{2}_{B}$ and $c$, the gap between $C_{s}^{(g*)}$ and $C_{s}^{(gf*)}$ is increasing while $\sigma^{2}_{w}$ is decreasing.

\emph{\textbf{\textcolor[rgb]{1.00,0.00,0.00}{Comparison of the fading and non-fading cases}}}

\textcolor[rgb]{1.00,0.00,0.00}{The comparison of the fading and no-fading cases is shown in the following Figure \ref{f4ddc3} to Figure \ref{f4ddc6}.}
In Figure \ref{f4ddc3} and Figure \ref{f4ddc4}, we see that
$C_{s}^{(g*)}$ dominates $C_{s}^{(g)}$ (which implies that the fading may enhance the security of the
degraded Gaussian model of Figure \ref{f2} with only delayed state feedback), and the gap between $C_{s}^{(g*)}$ and $C_{s}^{(g)}$ is increasing while
$\sigma^{2}_{w}$ is decreasing.

In Figure \ref{f4ddc5} and Figure \ref{f4ddc6}, we see that $C_{s}^{(gf)}$ dominates $C_{s}^{(gf*)}$
(which implies that the fading may weaken the security of the
degraded Gaussian model of Figure \ref{f2} with delayed state and receiver's channel output feedback), and the gap between $C_{s}^{(gf)}$ and $C_{s}^{(gf*)}$ is increasing while
$\sigma^{2}_{w}$ is increasing.

\section{Conclusions\label{secIV}}

In this paper, we provide inner and outer bounds on the capacity-equivocation regions of
the FSM-WC with delayed state feedback, and with or without delayed receiver's channel output feedback.
We find that these bounds meet
if the channel output for the eavesdropper is a degraded version of that for the legitimate receiver.
\textcolor[rgb]{1.00,0.00,0.00}{In the proof of these bounds, we show that the delayed receiver's channel output feedback is used to generate 
a secret key shared between the receiver and the transmitter, and this key}
helps to enhance the rate-equivocation region of the FSM-WC with only delayed state feedback.
The results of this paper are further explained via degraded Gaussian and degraded Gaussian fading examples.
In these examples, we show that when the channel is changing rapidly,
the secrecy capacities go to the infinite asymptote even if the delayed time $d$ is very small,
and when the channel is changing slowly,
a larger feedback delay is tolerable since the secrecy capacity loss compared with feedback without delay ($d=0$) is smaller.
Moreover, comparing these two examples,
we find that the fading may enhance the security of the
degraded Gaussian FSM-WC with only delayed state feedback, and
the fading may weaken the security of the
degraded Gaussian FSM-WC with delayed state and receiver's channel output feedback.

\section*{Acknowledgement}

The authors would like to thank Professor Xuming Fang
for his valuable suggestions on improving this paper.

\renewcommand{\theequation}{A\arabic{equation}}
\appendices\section{Proof of Theorem \ref{T3}\label{appen1}}
\setcounter{equation}{0}

The main idea of the proof of Theorem \ref{T3} is to construct a hybrid encoding-decoding scheme, which
combines the rate splitting technique, Wyner's random binning technique \cite{Wy} with
the classical multiplexing coding for the finite state Markov channel \cite{vis}. The details of the proof are as follows.

\subsection*{A. Definitions}

\begin{itemize}

\item The transmitted message $W$ is split into a common message $W_{c}$ and a private message $W_{p}$, i.e., $W=(W_{c}, W_{p})$.
Here $W_{c}$ and $W_{p}$ are uniformly distributed in the sets  $\{1,2,...,2^{NR_{c}}\}$ and $\{1,2,...,2^{NR_{p}}\}$,
respectively. Since $W$ is uniformly distributed in the set $\{1,2,...,2^{NR}\}$, we have $R=R_{c}+R_{p}$. In the remainder of this section, we first prove that
the region $\mathcal{R}_{1}$
\begin{eqnarray*}
&&\mathcal{R}_{1}=\{(R, R_{e}): 0\leq R=R_{c}+R_{p},\\
&&0\leq R_{c}\leq \min\{I(U;Y|S,\tilde{S}), I(U;Z|S,\tilde{S})\},\\
&&0\leq R_{p}\leq I(V;Y|U,S,\tilde{S}),\\
&&0\leq R_{e}\leq R_{p},\\
&&R_{e}\leq I(V;Y|U,S,\tilde{S})-I(V;Z|U,S,\tilde{S})\}
\end{eqnarray*}
is achievable.
Then, using Fourier-Motzkin elimination (see e.g., \cite{lall}) to eliminate $R_{c}$ and $R_{p}$ from $\mathcal{R}_{1}$, it is easy to see that the region
$\mathcal{R}$ is achievable.

\item Without loss of generality, we assume that the state takes values in $\mathcal{S}=\{1,2,...,k\}$ and that the steady state probability
$\pi(l)>0$ for all $l\in \mathcal{S}$. Let $N_{\tilde{s}}$ ($1\leq \tilde{s}\leq k$) be the number satisfying
\begin{eqnarray}\label{c3.q2}
&&N_{\tilde{s}}=N(\pi(\tilde{s})-\epsilon^{'}),
\end{eqnarray}
where $0\leq\epsilon^{'}<\min\{\pi(\tilde{s}); \tilde{s}\in \{1,2,...,k\}\}$ and $\epsilon^{'}\rightarrow 0$ as $N\rightarrow \infty$.
Denote the transmission rates $R_{c}$ and $R_{p}$ for a given $\tilde{s}$ by $R_{c}(\tilde{s})$ and $R_{p}(\tilde{s})$ ($1\leq \tilde{s}\leq k$), respectively,
and they satisfy
\begin{eqnarray}\label{c1.q1.rmb1}
&&\sum_{\tilde{s}=1}^{k}\pi(\tilde{s})R_{c}(\tilde{s})=R_{c},
\end{eqnarray}
and
\begin{eqnarray}\label{c1.q1}
&&\sum_{\tilde{s}=1}^{k}\pi(\tilde{s})R_{p}(\tilde{s})=R_{p}.
\end{eqnarray}

\item Divide the common message $W_{c}$ into $k$ sub-messages $W_{c,1}$,...,$W_{c,k}$, and each sub-message $W_{c,\textcolor[rgb]{1.00,0.00,0.00}{\tilde{s}}}$ 
($1\leq \textcolor[rgb]{1.00,0.00,0.00}{\tilde{s}}\leq k$) takes values
in the set $\mathcal{W}_{c,\textcolor[rgb]{1.00,0.00,0.00}{\tilde{s}}}=\{1,2,...,
2^{N_{\textcolor[rgb]{1.00,0.00,0.00}{\tilde{s}}}R_{c}(\textcolor[rgb]{1.00,0.00,0.00}{\tilde{s}})}\}$. Since the actual transmission rate $R^{*}_{c}$ of the common message $W_{c}$ is denoted by
\begin{eqnarray}\label{c1.q1.1.rmb1}
R^{*}_{c}&=&\frac{H(W_{c})}{N}=\frac{\sum_{\textcolor[rgb]{1.00,0.00,0.00}{\tilde{s}}=1}^{k}H(W_{c,\textcolor[rgb]{1.00,0.00,0.00}{\tilde{s}}})}{N}
=\frac{\sum_{\textcolor[rgb]{1.00,0.00,0.00}{\tilde{s}}=1}^{k}N_{\textcolor[rgb]{1.00,0.00,0.00}{\tilde{s}}}R_{c}(\textcolor[rgb]{1.00,0.00,0.00}{\tilde{s}})}{N}\nonumber\\
&\stackrel{(a)}=&\frac{\sum_{\textcolor[rgb]{1.00,0.00,0.00}{\tilde{s}}=1}^{k}N(\pi(\textcolor[rgb]{1.00,0.00,0.00}{\tilde{s}})
-\epsilon^{'})R_{c}(\textcolor[rgb]{1.00,0.00,0.00}{\tilde{s}})}{N}\nonumber\\
&=&\sum_{\textcolor[rgb]{1.00,0.00,0.00}{\tilde{s}}=1}^{k}(\pi(\textcolor[rgb]{1.00,0.00,0.00}{\tilde{s}})-\epsilon^{'})R_{c}(\textcolor[rgb]{1.00,0.00,0.00}{\tilde{s}})\nonumber\\
&=&\sum_{\textcolor[rgb]{1.00,0.00,0.00}{\tilde{s}}=1}^{k}\pi(\textcolor[rgb]{1.00,0.00,0.00}{\tilde{s}})R_{c}(\textcolor[rgb]{1.00,0.00,0.00}{\tilde{s}})
-\epsilon^{'}\sum_{\textcolor[rgb]{1.00,0.00,0.00}{\tilde{s}}=1}^{k}R_{c}(\textcolor[rgb]{1.00,0.00,0.00}{\tilde{s}}),
\end{eqnarray}
where (a) is from (\ref{c3.q2}). From (\ref{c1.q1.rmb1}) and (\ref{c1.q1.1.rmb1}), it is easy to see that $R^{*}_{c}$ tends to be $R_{c}$
while $\epsilon^{'}\rightarrow 0$.

\item Divide the private message $W_{p}$ into $k$ sub-messages $W_{p,1}$,...,$W_{p,k}$, and each sub-message 
$W_{p,\textcolor[rgb]{1.00,0.00,0.00}{\tilde{s}}}$ ($1\leq \textcolor[rgb]{1.00,0.00,0.00}{\tilde{s}}\leq k$) takes values
in the set $\mathcal{W}_{p,\textcolor[rgb]{1.00,0.00,0.00}{\tilde{s}}}
=\{1,2,...,2^{N_{\textcolor[rgb]{1.00,0.00,0.00}{\tilde{s}}}R_{p}(\textcolor[rgb]{1.00,0.00,0.00}{\tilde{s}})}\}$. 
Similar to (\ref{c1.q1.1.rmb1}), the actual transmission rate $R^{*}_{p}$ of the private message $W_{p}$
tends to be $R_{p}$ while $\epsilon^{'}\rightarrow 0$.
\end{itemize}

\subsection*{B. Construction of the code-books}

Fix the joint probability mass function
$P_{UVS\tilde{S}XYZ}(u,v,s,\tilde{s},x,y,z)$ satisfying (\ref{dota1}).

\begin{itemize}

\item \textbf{Construction of $U^{N}$}: Construct $k$ code-books $\mathcal{U}^{\tilde{s}}$ of $U^{N}$ for all $\tilde{s}\in \mathcal{S}$.
In each code-book $\mathcal{U}^{\tilde{s}}$, randomly generate $2^{N_{\tilde{s}}R_{c}(\tilde{s})}$ i.i.d. sequences
$u^{N_{\tilde{s}}}$ according to the probability mass function $P_{U|\tilde{S}}(u|\tilde{s})$, and index these sequences as $u^{N_{\tilde{s}}}(i)$, where
$1\leq i\leq 2^{N_{\tilde{s}}R_{c}(\tilde{s})}$.

\item \textbf{Construction of $V^{N}$}: Construct $k$ code-books $\mathcal{V}^{\tilde{s}}$ of $V^{N}$ for all $\tilde{s}\in \mathcal{S}$.
In each code-book $\mathcal{V}^{\tilde{s}}$, randomly generate $2^{N_{\tilde{s}}(I(V;Y|U,S,\tilde{S}=\tilde{s})+R_{c}(\tilde{s}))}$ i.i.d. sequences
$v^{N_{\tilde{s}}}$ according to the probability mass function $P_{V|U,\tilde{S}}(v|u,\tilde{s})$. Index these sequences of the code-book $\mathcal{V}^{\tilde{s}}$
as $v^{N_{\tilde{s}}}(i_{\tilde{s}},a_{\tilde{s}},b_{\tilde{s}})$, where $1\leq i_{\tilde{s}}\leq 2^{N_{\tilde{s}}R_{c}(\tilde{s})}$,
$a_{\tilde{s}}\in \mathcal{A}_{\tilde{s}}=\{1,2,...,A_{\tilde{s}}\}$,
$b_{\tilde{s}}\in \mathcal{B}_{\tilde{s}}=\{1,2,...,B_{\tilde{s}}\}$,
\begin{eqnarray}\label{miyue1}
A_{\tilde{s}}=2^{N_{\tilde{s}}(I(V;Y|U,S,\tilde{S}=\tilde{s})-I(V;Z|U,S,\tilde{S}=\tilde{s}))},
\end{eqnarray}
and
\begin{eqnarray}\label{miyue2}
B_{\tilde{s}}=2^{N_{\tilde{s}}I(V;Z|U,S,\tilde{S}=\tilde{s})}.
\end{eqnarray}

\item \textbf{Construction of $X^{N}$}: For each $\tilde{s}$, the sequence $x^{N_{\tilde{s}}}$ is i.i.d. generated according to a new discrete memoryless
channel (DMC) with transition probability $P_{X|U,V,\tilde{S}}(x|u,v,\tilde{s})$. The inputs of this new DMC are $u^{N_{\tilde{s}}}$ and $v^{N_{\tilde{s}}}$,
while the output is $x^{N_{\tilde{s}}}$.
\end{itemize}

\subsection*{C. Encoding scheme}

For a fixed length $N$, let $L_{\tilde{s}}$ be the number of times during the $N$ symbols for which
the delayed feedback state at the transmitter is $\tilde{S}=\tilde{s}$. Every time that the corresponding delayed state is $\tilde{S}=\tilde{s}$,
the transmitter chooses the next symbols of $u^{N}$ and $v^{N}$ from the component code-books $\mathcal{U}^{\tilde{s}}$ and $\mathcal{V}^{\tilde{s}}$, respectively.
Since $L_{\tilde{s}}$ is not necessarily equivalent to
$N_{\tilde{s}}$, an error is declared if $L_{\tilde{s}}< N_{\tilde{s}}$, and the codes are filled with zero if $L_{\tilde{s}}> N_{\tilde{s}}$. Therefore,
we can send a total of $2^{\sum_{i=1}^{k}N_{i}(R_{c}(i)+R_{p}(i))}$  messages. Since the state process is stationary and ergodic
$\lim_{N\rightarrow \infty}\frac{L_{\tilde{s}}}{N}=Pr\{\tilde{S}=\tilde{s}\}$ in probability. Thus, we have
\begin{eqnarray}\label{miyue0}
&&Pr\{L_{\tilde{s}}< N_{\tilde{s}}\}\rightarrow 0, \,\, \mbox{as}\,\, N\rightarrow \infty.
\end{eqnarray}

For each $\tilde{s}\in \mathcal{S}$, define
$\mathcal{W}_{p,\tilde{s}}=\mathcal{A}_{\tilde{s}}\times \mathcal{J}_{\tilde{s}}$, where $\mathcal{J}_{\tilde{s}}=\{1,2,...,J_{\tilde{s}}\}$ and
$J_{\tilde{s}}=2^{N_{\tilde{s}}(R_{p}(\tilde{s})-I(V;Y|U,S,\tilde{S}=\tilde{s})+I(V;Z|U,S,\tilde{S}=\tilde{s}))}$. Furthermore, we define
the mapping $g_{\tilde{s}}: \mathcal{B}_{\tilde{s}}\rightarrow \mathcal{J}_{\tilde{s}}$, and partition $\mathcal{B}_{\tilde{s}}$ into $J_{\tilde{s}}$ subsets
with nearly equal size. Here
the ``nearly equal size'' means
\begin{eqnarray}\label{miyue3}
&&\|g_{\tilde{s}}^{-1}(j_{1})\|\leq 2\|g_{\tilde{s}}^{-1}(j_{2})\|, \,\, \forall j_{1}, j_{2}\in \mathcal{J}_{\tilde{s}}.
\end{eqnarray}
The transmitted codewords $u^{N}$ and $v^{N}$ are obtained by multiplexing
the different component codewords. Specifically, first,
suppose that a message $w=(w_{c},w_{p})=(w_{c,1},...,w_{c,k},w_{p,1},...,w_{p,k})$ is
transmitted, and here we denote
$w_{p,\tilde{s}}$ ($1\leq \tilde{s}\leq k$) by $(a_{\tilde{s}},j_{\tilde{s}})$, where $a_{\tilde{s}}\in \mathcal{A}_{\tilde{s}}$ and
$j_{\tilde{s}}\in \mathcal{J}_{\tilde{s}}$.
Second, in each component code-book $\mathcal{U}^{\tilde{s}}$ ($1\leq \tilde{s}\leq k$),
the transmitter chooses $u^{N_{\tilde{s}}}(w_{c,\tilde{s}})$ as the $\tilde{s}$-th component codeword of the transmitted $u^{N}$.
Third, in each component code-book $\mathcal{V}^{\tilde{s}}$ ($1\leq \tilde{s}\leq k$),
the transmitter chooses $v^{N_{\tilde{s}}}(i^{*}_{\tilde{s}},a^{*}_{\tilde{s}},b^{*}_{\tilde{s}})$
as the $\tilde{s}$-th component codeword of the transmitted $v^{N}$,
where $i^{*}_{\tilde{s}}=w_{c,\tilde{s}}$, $a^{*}_{\tilde{s}}=a_{\tilde{s}}$, and $b^{*}_{\tilde{s}}$ is randomly chosen from
the sub-set $j_{\tilde{s}}$ of $\mathcal{B}_{\tilde{s}}$.

\subsection*{D. Decoding scheme}

\begin{itemize}

\item (\textbf{Decoding scheme for the receiver}:)

\begin{itemize}

\item (\textbf{Decoding the common message $w_{c}$:}) The delayed feedback state $\tilde{S}$ at the transmitter, which is used to multiplex
the component codewords, is also available at the receiver. Thus once the receiver receives $y^{N}$ and the state sequence $s^{N}$,
he first demultiplexes them into outputs corresponding to the
component code-books and separately decodes each component codeword. To be specific, in each code-book $\mathcal{U}^{\tilde{s}}$, the receiver has
$(y^{N_{\tilde{s}}}, s^{N_{\tilde{s}}})$ and tries to search a unique $u^{N_{\tilde{s}}}$ such that $(u^{N_{\tilde{s}}}, y^{N_{\tilde{s}}}, s^{N_{\tilde{s}}})$
are strongly jointly typical sequences \cite{coverx}, i.e.,
\begin{eqnarray}\label{miyue4.1}
&&(u^{N_{\tilde{s}}}, y^{N_{\tilde{s}}}, s^{N_{\tilde{s}}})\in T^{N_{\tilde{s}}}_{U,S,Y|\tilde{S}}(\epsilon).
\end{eqnarray}
If there exists such a unique $u^{N_{\tilde{s}}}$, put out the corresponding index $\hat{w}_{c,\tilde{s}}$.
Otherwise, i.e., if no such sequence exists or multiple sequences have different message indices,
declare a decoding error. If for all $1\leq \tilde{s}\leq k$, there exist unique sequences $u^{N_{\tilde{s}}}$ such that (\ref{miyue4.1}) is satisfied,
the receiver
declares that $\hat{w}_{c}=(\hat{w}_{c,1},\hat{w}_{c,2},...,\hat{w}_{c,k})$ is sent.
Based on the AEP, the error probability $Pr\{\hat{w}_{c,\tilde{s}}\neq w_{c,\tilde{s}}\}$ ($1\leq \tilde{s}\leq k$)
goes to $0$ if
\begin{eqnarray}\label{miyue4}
&&R_{c}(\tilde{s})\leq I(U;Y|S,\tilde{S}=\tilde{s}).
\end{eqnarray}

\item (\textbf{Decoding the private message $w_{p}$:}) After decoding $u^{N_{\tilde{s}}}(\hat{w}_{c,\tilde{s}})$
and $\hat{w}_{c,\tilde{s}}$ for all $1\leq \tilde{s}\leq k$,
in each component code-book $\mathcal{V}^{\tilde{s}}$,
the receiver tries to find a unique sequence $v^{N_{\tilde{s}}}$ such that
\begin{eqnarray}\label{miyue4.2}
&&(v^{N_{\tilde{s}}}, u^{N_{\tilde{s}}}, y^{N_{\tilde{s}}}, s^{N_{\tilde{s}}})\in T^{N_{\tilde{s}}}_{U,V,S,Y|\tilde{S}}(\epsilon).
\end{eqnarray}
If there exists such a unique $v^{N_{\tilde{s}}}$, put out the corresponding indexes $\hat{i}_{\tilde{s}}$, $\hat{a}_{\tilde{s}}$ and $\hat{b}_{\tilde{s}}$.
Otherwise, i.e., if no such sequence exists or multiple sequences have different message indices,
declare a decoding error. After the receiver obtains the index $\hat{b}_{\tilde{s}}$, he also knows $\hat{j}_{\tilde{s}}$ since it is
the index of the sub-set which $\hat{b}_{\tilde{s}}$ belongs to. Thus, for $1\leq \tilde{s}\leq k$, the receiver has an estimation
$\hat{w}_{p,\tilde{s}}$ of the private message
$w_{p,\tilde{s}}$ by letting $\hat{w}_{p,\tilde{s}}=(\hat{a}_{\tilde{s}},\hat{j}_{\tilde{s}})$.
If for all $1\leq \tilde{s}\leq k$, there exist unique sequences $v^{N_{\tilde{s}}}$ such that (\ref{miyue4.2}) is satisfied, the receiver
declares that $\hat{w}_{p}=(\hat{w}_{p,1},\hat{w}_{p,2},...,\hat{w}_{p,k})$ is sent.
Based on the AEP, the error probability $Pr\{\hat{w}_{p,\tilde{s}}\neq w_{p,\tilde{s}}\}$ ($1\leq \tilde{s}\leq k$)
goes to $0$ if
\begin{eqnarray}\label{miyue5}
&&R_{p}(\tilde{s})\leq I(V;Y|U,S,\tilde{S}=\tilde{s}).
\end{eqnarray}

\end{itemize}

\item (\textbf{Decoding scheme for the eavesdropper}:)

\begin{itemize}

\item (\textbf{Decoding the common message $w_{c}$:}) The delayed feedback state $\tilde{S}$ at the transmitter,
is also available at the eavesdropper. Thus once the eavesdropper receives $z^{N}$ and the state sequence $s^{N}$,
he first demultiplexes them into outputs corresponding to the
component code-books and separately decodes each component codeword. To be specific, in each code-book $\mathcal{U}^{\tilde{s}}$, the eavesdropper has
$(z^{N_{\tilde{s}}}, s^{N_{\tilde{s}}})$ and tries to search a unique $u^{N_{\tilde{s}}}$ such that $(u^{N_{\tilde{s}}}, z^{N_{\tilde{s}}}, s^{N_{\tilde{s}}})$
are strongly jointly typical sequences \cite{coverx}, i.e.,
\begin{eqnarray}\label{miyue6.1}
&&(u^{N_{\tilde{s}}}, z^{N_{\tilde{s}}}, s^{N_{\tilde{s}}})\in T^{N_{\tilde{s}}}_{U,S,Z|\tilde{S}}(\epsilon).
\end{eqnarray}
If there exists such a unique $u^{N_{\tilde{s}}}$, put out the corresponding index $\check{w}_{c,\tilde{s}}$.
Otherwise, i.e., if no such sequence exists or multiple sequences have different message indices,
declare a decoding error. If for all $1\leq \tilde{s}\leq k$, there exist unique sequences $u^{N_{\tilde{s}}}$ such that (\ref{miyue6.1}) is satisfied,
the eavesdropper
declares that $\check{w}_{c}=(\check{w}_{c,1},\check{w}_{c,2},...,\check{w}_{c,k})$ is sent.
Based on the AEP, the error probability $Pr\{\check{w}_{c,\tilde{s}}\neq w_{c,\tilde{s}}\}$ ($1\leq \tilde{s}\leq k$)
goes to $0$ if
\begin{eqnarray}\label{miyue7}
&&R_{c}(\tilde{s})\leq I(U;Z|S,\tilde{S}=\tilde{s}).
\end{eqnarray}

\item (\textbf{Given $w_{c}$ and $w_{p}$, decoding $v^{N}$:}) In each component code-book $\mathcal{V}^{\tilde{s}}$
($1\leq \tilde{s}\leq k$), given $\tilde{S}=\tilde{s}$, $s^{N_{\tilde{s}}}$,
$u^{N_{\tilde{s}}}(w_{c,\tilde{s}})$ and $w_{p,\tilde{s}}=(a_{\tilde{s}},j_{\tilde{s}})$, the eavesdropper
tries to find a unique $\check{b}_{\tilde{s}}$ such that
\begin{eqnarray}\label{miyue8}
&&(v^{N_{\tilde{s}}}(w_{c,\tilde{s}},a_{\tilde{s}},\check{b}_{\tilde{s}}), u^{N_{\tilde{s}}}(w_{c,\tilde{s}}), z^{N_{\tilde{s}}}, s^{N_{\tilde{s}}})\in
T^{N_{\tilde{s}}}_{U,V,S,Z|\tilde{S}}(\epsilon).
\end{eqnarray}
Since the index $b^{*}_{\tilde{s}}$ of the transmitted $v^{N_{\tilde{s}}}$ is randomly chosen from
the sub-set $j_{\tilde{s}}$ of $\mathcal{B}_{\tilde{s}}$ and there are $2^{N_{\tilde{s}}(I(V;Y|U,S,\tilde{S}=\tilde{s})-R_{p}(\tilde{s}))}$ sequences
of $v^{N_{\tilde{s}}}$ in the sub-set $j_{\tilde{s}}$,
based on the AEP, the error probability
$Pr\{\check{b}_{\tilde{s}}\neq b^{*}_{\tilde{s}}\}$ ($1\leq \tilde{s}\leq k$)
goes to $0$ if
\begin{eqnarray}\label{miyue9}
&&I(V;Y|U,S,\tilde{S}=\tilde{s})-R_{p}(\tilde{s})\leq I(V;Z|U,S,\tilde{S}=\tilde{s}).
\end{eqnarray}

\end{itemize}

\end{itemize}

Combining (\ref{c1.q1.rmb1}) with (\ref{miyue4}) and (\ref{miyue7}), we have
\begin{eqnarray}\label{miyue10}
R_{c}&=&\sum_{\tilde{s}=1}^{k}\pi(\tilde{s})R_{c}(\tilde{s})\nonumber\\
&\leq&\sum_{\tilde{s}=1}^{k}\pi(\tilde{s})\min\{I(U;Y|S,\tilde{S}=\tilde{s}),I(U;Z|S,\tilde{S}=\tilde{s})\}\nonumber\\
&=&\min\{I(U;Y|S,\tilde{S}),I(U;Z|S,\tilde{S})\},
\end{eqnarray}
and combining (\ref{c1.q1}) with (\ref{miyue5}), we have
\begin{eqnarray}\label{miyue11}
R_{p}&=&\sum_{\tilde{s}=1}^{k}\pi(\tilde{s})R_{p}(\tilde{s})\nonumber\\
&\leq&\sum_{\tilde{s}=1}^{k}\pi(\tilde{s})I(V;Y|U,S,\tilde{S}=\tilde{s})\nonumber\\
&=&I(V;Y|U,S,\tilde{S}).
\end{eqnarray}
It remains to show that $R_{e}\leq I(V;Y|U,S,\tilde{S})-I(V;Z|U,S,\tilde{S})$ and $R_{e}\leq R_{p}$, see
the followings.

\subsection*{E. Equivocation analysis:}

Since the eavesdropper also knows the state $S^{N}$ and the delayed time $d$,
the equivocation $\Delta$ is bounded by
\begin{eqnarray}\label{eapp5}
\Delta&=&\frac{1}{N}H(W|Z^{N},S^{N})=\frac{1}{N}H(W_{c},W_{p}|Z^{N},S^{N})\nonumber\\
&\geq&\frac{1}{N}H(W_{p}|Z^{N},S^{N},W_{c})\geq\frac{1}{N}H(W_{p}|Z^{N},S^{N},W_{c},U^{N})\nonumber\\
&\stackrel{(a)}=&\frac{1}{N}H(W_{p}|Z^{N},S^{N},U^{N})=\frac{1}{N}H(W_{p,1},W_{p,2},...,W_{p,k}|Z^{N},S^{N},U^{N})\nonumber\\
&=&\frac{1}{N}\sum_{\tilde{s}=1}^{k}H(W_{p,\tilde{s}}|Z^{N},S^{N},U^{N},W_{p,1},...,W_{p,\tilde{s}-1})\nonumber\\
&\geq&\frac{1}{N}\sum_{\tilde{s}=1}^{k}H(W_{p,\tilde{s}}|Z^{N},S^{N},U^{N},W_{p,1},...,W_{p,\tilde{s}-1},\tilde{S}=\tilde{s})\nonumber\\
&\stackrel{(b)}=&\frac{1}{N}\sum_{\tilde{s}=1}^{k}H(W_{p,\tilde{s}}|Z^{N_{\tilde{s}}},S^{N_{\tilde{s}}},U^{N_{\tilde{s}}},\tilde{S}=\tilde{s})\nonumber\\
&=&\frac{1}{N}\sum_{\tilde{s}=1}^{k}(H(W_{p,\tilde{s}},Z^{N_{\tilde{s}}},S^{N_{\tilde{s}}},U^{N_{\tilde{s}}},\tilde{S}=\tilde{s})
-H(Z^{N_{\tilde{s}}},S^{N_{\tilde{s}}},U^{N_{\tilde{s}}},\tilde{S}=\tilde{s}))\nonumber\\
&=&\frac{1}{N}\sum_{\tilde{s}=1}^{k}(H(W_{p,\tilde{s}},Z^{N_{\tilde{s}}},S^{N_{\tilde{s}}},U^{N_{\tilde{s}}},V^{N_{\tilde{s}}},\tilde{S}=\tilde{s})
-H(V^{N_{\tilde{s}}}|W_{p,\tilde{s}},Z^{N_{\tilde{s}}},S^{N_{\tilde{s}}},U^{N_{\tilde{s}}},\tilde{S}=\tilde{s})\nonumber\\
&&-H(Z^{N_{\tilde{s}}},S^{N_{\tilde{s}}},U^{N_{\tilde{s}}},\tilde{S}=\tilde{s}))\nonumber\\
&\stackrel{(c)}=&\frac{1}{N}\sum_{\tilde{s}=1}^{k}(H(Z^{N_{\tilde{s}}}|S^{N_{\tilde{s}}},U^{N_{\tilde{s}}},V^{N_{\tilde{s}}},\tilde{S}=\tilde{s})
+H(S^{N_{\tilde{s}}},U^{N_{\tilde{s}}},V^{N_{\tilde{s}}},\tilde{S}=\tilde{s})\nonumber\\
&&-H(Z^{N_{\tilde{s}}}|S^{N_{\tilde{s}}},U^{N_{\tilde{s}}},\tilde{S}=\tilde{s})
-H(S^{N_{\tilde{s}}},U^{N_{\tilde{s}}},\tilde{S}=\tilde{s})
-H(V^{N_{\tilde{s}}}|W_{p,\tilde{s}},Z^{N_{\tilde{s}}},S^{N_{\tilde{s}}},U^{N_{\tilde{s}}},\tilde{S}=\tilde{s}))\nonumber\\
&\stackrel{(d)}\geq&\frac{1}{N}\sum_{\tilde{s}=1}^{k}(N_{\tilde{s}}H(Z|S,U,V,\tilde{S}=\tilde{s})-H(Z^{N_{\tilde{s}}}|S^{N_{\tilde{s}}},U^{N_{\tilde{s}}},\tilde{S}=\tilde{s})
+H(S^{N_{\tilde{s}}},U^{N_{\tilde{s}}},V^{N_{\tilde{s}}},\tilde{S}=\tilde{s})-H(S^{N_{\tilde{s}}},U^{N_{\tilde{s}}},\tilde{S}=\tilde{s})\nonumber\\
&&-H(V^{N_{\tilde{s}}}|W_{p,\tilde{s}},Z^{N_{\tilde{s}}},S^{N_{\tilde{s}}},U^{N_{\tilde{s}}},\tilde{S}=\tilde{s}))\nonumber\\
&\geq&\frac{1}{N}\sum_{\tilde{s}=1}^{k}(N_{\tilde{s}}H(Z|S,U,V,\tilde{S}=\tilde{s})-N_{\tilde{s}}H(Z|S,U,\tilde{S}=\tilde{s})
+H(S^{N_{\tilde{s}}},U^{N_{\tilde{s}}},V^{N_{\tilde{s}}},\tilde{S}=\tilde{s})-H(S^{N_{\tilde{s}}},U^{N_{\tilde{s}}},\tilde{S}=\tilde{s})\nonumber\\
&&-H(V^{N_{\tilde{s}}}|W_{p,\tilde{s}},Z^{N_{\tilde{s}}},S^{N_{\tilde{s}}},U^{N_{\tilde{s}}},\tilde{S}=\tilde{s}))\nonumber\\
&=&\frac{1}{N}\sum_{\tilde{s}=1}^{k}(N_{\tilde{s}}H(Z|S,U,V,\tilde{S}=\tilde{s})-N_{\tilde{s}}H(Z|S,U,\tilde{S}=\tilde{s})\nonumber\\
&&+H(V^{N_{\tilde{s}}}|S^{N_{\tilde{s}}},U^{N_{\tilde{s}}},\tilde{S}=\tilde{s})
-H(V^{N_{\tilde{s}}}|W_{p,\tilde{s}},Z^{N_{\tilde{s}}},S^{N_{\tilde{s}}},U^{N_{\tilde{s}}},\tilde{S}=\tilde{s}))\nonumber\\
&\stackrel{(e)}\geq&\frac{1}{N}\sum_{\tilde{s}=1}^{k}(N_{\tilde{s}}H(Z|S,U,V,\tilde{S})-N_{\tilde{s}}H(Z|S,U,\tilde{S}=\tilde{s})\nonumber\\
&&+N_{\tilde{s}}I(V;Y|U,S,\tilde{S}=\tilde{s})-1
-H(V^{N_{\tilde{s}}}|W_{p,\tilde{s}},Z^{N_{\tilde{s}}},S^{N_{\tilde{s}}},U^{N_{\tilde{s}}},\tilde{S}=\tilde{s}))\nonumber\\
&\stackrel{(f)}\geq&\frac{1}{N}\sum_{\tilde{s}=1}^{k}(N_{\tilde{s}}H(Z|S,U,V,\tilde{S}=\tilde{s})-N_{\tilde{s}}H(Z|S,U,\tilde{S}=\tilde{s})
+N_{\tilde{s}}I(V;Y|U,S,\tilde{S})-1-N_{\tilde{s}}\epsilon_{1})\nonumber\\
&=&\sum_{\tilde{s}=1}^{k}\frac{N_{\tilde{s}}}{N}(I(V;Y|U,S,\tilde{S}=\tilde{s})-I(V;Z|U,S,\tilde{S}=\tilde{s})-\frac{1}{N_{\tilde{s}}}-\epsilon_{1})\nonumber\\
&\stackrel{(g)}=&\sum_{\tilde{s}=1}^{k}(\pi(\tilde{s})-\epsilon^{'})
(I(V;Y|U,S,\tilde{S}=\tilde{s})-I(V;Z|U,S,\tilde{S}=\tilde{s})-\frac{1}{N_{\tilde{s}}}-\epsilon_{1})\nonumber\\
&=&I(V;Y|U,S,\tilde{S})-I(V;Z|U,S,\tilde{S})-\sum_{\tilde{s}=1}^{k}(\pi(\tilde{s})-\epsilon^{'})(\frac{1}{N_{\tilde{s}}}+\epsilon_{1})\nonumber\\
&&-\epsilon^{'}\sum_{\tilde{s}=1}^{k}(I(V;Y|U,S,\tilde{S})-I(V;Z|U,S,\tilde{S})),
\end{eqnarray}
where (a) is from the fact that $H(W_{c}|U^{N})=0$, (b) is from the the Markov chain
$(Z^{N_{\tilde{1}}},...,Z^{N_{\tilde{s}-1}},Z^{N_{\tilde{s}+1}},...,Z^{N_{k}},\\U^{N_{\tilde{1}}},...,U^{N_{\tilde{s}-1}},U^{N_{\tilde{s}+1}},...,U^{N_{k}}
,S^{N_{\tilde{1}}},...,S^{N_{\tilde{s}-1}},S^{N_{\tilde{s}+1}},...,S^{N_{k}})
\rightarrow (Z^{N_{\tilde{s}}},S^{N_{\tilde{s}}},U^{N_{\tilde{s}}},\tilde{S}=\tilde{s})\rightarrow W_{p,\tilde{s}}$,
which implies that
given the $\tilde{s}$-th component of the sequences $Z^{N}$, $U^{N}$ and $S^{N}$, $W_{p,\tilde{s}}$ is independent of the other parts of
$Z^{N}$, $U^{N}$ and $S^{N}$, (c) is from the fact that $H(W_{p,\tilde{s}}|V^{N_{\tilde{s}}})=0$, (d) is from the fact that
the channel is a DMC with transition probability $P_{Y,Z|X,S}(y,z|x,s)$, and for each $\tilde{s}$,
$X^{N_{\tilde{s}}}$ is i.i.d. generated according to a new DMC with transition probability $P_{X|U,V,\tilde{S}}(x|u,v,\tilde{s})$, thus we have
$H(Z^{N_{\tilde{s}}}|S^{N_{\tilde{s}}},U^{N_{\tilde{s}}},V^{N_{\tilde{s}}},\tilde{S}=\tilde{s})=N_{\tilde{s}}H(Z|S,U,V,\tilde{S}=\tilde{s})$,
(e) is from the fact that for given $\tilde{s}$, $u^{N_{\tilde{s}}}$ and $s^{N_{\tilde{s}}}$, $V^{N_{\tilde{s}}}$ has $A_{\tilde{s}}\cdot B_{\tilde{s}}$
possible values, and the encoding mapping function $g_{\tilde{s}}$ partitions $\mathcal{B}_{\tilde{s}}$ into $j_{\tilde{s}}$ subsets
with ``nearly equal size'' (see (\ref{miyue3})), using a similar lemma in \cite{CK}, we have
\begin{eqnarray}\label{db1}
&&\frac{1}{N_{\tilde{s}}}H(V^{N_{\tilde{s}}}|S^{N_{\tilde{s}}},U^{N_{\tilde{s}}},\tilde{S}=\tilde{s})\geq \frac{1}{N_{\tilde{s}}}\log A_{\tilde{s}}
+\frac{1}{N_{\tilde{s}}}\log B_{\tilde{s}}-\frac{1}{N_{\tilde{s}}},
\end{eqnarray}
(f) is from the fact that given $\tilde{S}=\tilde{s}$, $s^{N_{\tilde{s}}}$,
$u^{N_{\tilde{s}}}(w_{c,\tilde{s}})$ and $w_{p,\tilde{s}}=(a_{\tilde{s}},j_{\tilde{s}})$, the eavesdropper's decoding error probability of $v^{N_{\tilde{s}}}$
tends to zero if (\ref{miyue9}) is satisfied, and thus, by using Fano's inequality, we have
\begin{eqnarray}\label{db2}
&&\frac{1}{N_{\tilde{s}}}H(V^{N_{\tilde{s}}}|W_{p,\tilde{s}},Z^{N_{\tilde{s}}},S^{N_{\tilde{s}}},U^{N_{\tilde{s}}},\tilde{S}=\tilde{s})\leq \epsilon_{1},
\end{eqnarray}
where $\epsilon_{1}\rightarrow 0$ as $N_{\tilde{s}}\rightarrow \infty$, and (g) is from (\ref{c3.q2}).

From (\ref{eapp5}), we have
\begin{eqnarray}\label{db3}
&&\Delta\geq I(V;Y|U,S,\tilde{S})-I(V;Z|U,S,\tilde{S})-\epsilon_{2},
\end{eqnarray}
where $\epsilon_{2}$ is small for sufficiently large $N$. By the definition of $R_{e}$, we can conclude that
$R_{e}\leq I(V;Y|U,S,\tilde{S})-I(V;Z|U,S,\tilde{S})$.

In addition, we know that (\ref{db2}) holds if (\ref{miyue9}) is satisfied, and this implies that
\begin{eqnarray}\label{db4}
R_{p}&=&\sum_{\tilde{s}=1}^{k}\pi(\tilde{s})R_{p}(\tilde{s})\nonumber\\
&\geq&\sum_{\tilde{s}=1}^{k}\pi(\tilde{s})(I(V;Y|U,S,\tilde{S}=\tilde{s})-I(V;Z|U,S,\tilde{S}=\tilde{s}))\nonumber\\
&=&I(V;Y|U,S,\tilde{S})-I(V;Z|U,S,\tilde{S})\geq R_{e}.
\end{eqnarray}
Thus, $R_{e}\leq I(V;Y|U,S,\tilde{S})-I(V;Z|U,S,\tilde{S})$ and $R_{e}\leq R_{p}$ are proved, and the
achievability proof of the region $\mathcal{R}_{1}$ is completed.
Finally, using Fourier-Motzkin elimination (see e.g., \cite{lall}) to eliminate $R_{c}$ and $R_{p}$ from $\mathcal{R}_{1}$,
the proof of Theorem \ref{T3} is completed.

\section{Proof of Theorem \ref{T3.1}\label{appen2}}

In this section, we will prove Theorem \ref{T3.1}: all the achievable $(R, R_{e})$ pairs are contained in
the set $\mathcal{R}^{out}$. Since $R_{e}\leq R$ is obvious, we only need to prove the
inequalities $R\leq I(V;Y|S,\tilde{S})$ and $R_{e}\leq I(V;Y|U,S,\tilde{S})-I(V;Z|U,S,\tilde{S})$
of Theorem \ref{T3.1} in the remainder of this section.

First, define the following auxiliary random variables,
\begin{eqnarray}\label{jmds1}
&&U\triangleq (Y^{J-1}, Z_{J+1}^{N}, S^{N}, J), V\triangleq (U, W), S\triangleq S_{J}, \tilde{S}\triangleq S_{J-d}, Y\triangleq Y_{J}, Z\triangleq Z_{J},
\end{eqnarray}
where $J$ is a random variable uniformly distributed over $\{1, 2, ,...,N\}$, and it is independent of
$Y^{N}$, $Z^{N}$, $W$ and $S^{N}$.

\textbf{Proof of $R\leq I(V;Y|S,\tilde{S})$:} Note that
\begin{eqnarray}\label{jmds2}
R-\epsilon&\stackrel{(a)}\leq&\frac{1}{N}H(W)\nonumber\\
&\stackrel{(b)}=&\frac{1}{N}H(W|S^{N})\nonumber\\
&=&\frac{1}{N}(I(W;Y^{N}|S^{N})+H(W|Y^{N},S^{N}))\nonumber\\
&\stackrel{(c)}\leq&\frac{1}{N}(I(W;Y^{N}|S^{N})+\delta(P_{e}))\nonumber\\
&=&\frac{1}{N}\sum_{i=1}^{N}(H(Y_{i}|Y^{i-1},S^{N})-H(Y_{i}|Y^{i-1},S^{N},W))+\frac{\delta(P_{e})}{N}\nonumber\\
&\leq&\frac{1}{N}\sum_{i=1}^{N}(H(Y_{i}|S_{i},S_{i-d})-H(Y_{i}|Y^{i-1},Z_{i+1}^{N},S^{N},W))+\frac{\delta(P_{e})}{N}\nonumber\\
&\stackrel{(d)}=&\frac{1}{N}\sum_{i=1}^{N}(H(Y_{i}|S_{i},S_{i-d})-H(Y_{i}|Y^{i-1},Z_{i+1}^{N},S^{N},W,S_{i},S_{i-d}))+\frac{\delta(P_{e})}{N}\nonumber\\
&\stackrel{(e)}=&\frac{1}{N}\sum_{i=1}^{N}(H(Y_{i}|S_{i},S_{i-d},J=i)-H(Y_{i}|Y^{i-1},Z_{i+1}^{N},S^{N},W,S_{i},S_{i-d},J=i))+\frac{\delta(P_{e})}{N}\nonumber\\
&\stackrel{(f)}=&H(Y_{J}|S_{J},S_{J-d},J)-H(Y_{J}|S_{J},S_{J-d},W,Y^{J-1},Z_{J+1}^{N},S^{N},J)+\frac{\delta(P_{e})}{N}\nonumber\\
&\leq&H(Y_{J}|S_{J},S_{J-d})-H(Y_{J}|S_{J},S_{J-d},W,Y^{J-1},Z_{J+1}^{N},S^{N},J)+\frac{\delta(P_{e})}{N}\nonumber\\
&\stackrel{(g)}=&H(Y|S,\tilde{S})-H(Y|S,\tilde{S},V)+\frac{\delta(P_{e})}{N}\nonumber\\
&\stackrel{(h)}\leq&I(V;Y|S,\tilde{S})+\frac{\delta(\epsilon)}{N},
\end{eqnarray}
where (a) is from (\ref{e211}), (b) is from the fact that $W$ is independent of $S^{N}$, (c) is from the Fano's inequality,
(d) is from the fact that $S_{i}$ and $S_{i-d}$ (here $S_{i-d}=const$ when $i\leq d$) are included in $S^{N}$, and thus there exists a Markov chain
$(S_{i},S_{i-d})\rightarrow (Y^{i-1},Z_{i+1}^{N},S^{N},W)\rightarrow Y_{i}$, (e) is from
the fact that $J$ is a random variable (uniformly distributed
over $\{1,2,...,N\}$), and it is independent of $Y^{N}$, $Z^{N}$, $W$ and $S^{N}$, (f) is from $J$ is
uniformly distributed over $\{1,2,...,N\}$, (g) is from the definitions in (\ref{jmds1}), and (h)
is from $\delta(P_{e})$ is increasing while $P_{e}$ is increasing, and $P_{e}\leq \epsilon$.
Then, letting $\epsilon\rightarrow 0$, we have $R\leq I(V;Y|S,\tilde{S})$.

\textbf{Proof of $R_{e}\leq I(V;Y|U,S,\tilde{S})-I(V;Z|U,S,\tilde{S})$:}
By using (\ref{e210}) and (\ref{e211}), we have
\begin{eqnarray}\label{jmds3}
R_{e}-\epsilon&\stackrel{(1)}\leq&\frac{1}{N}H(W|Z^{N},S^{N})\nonumber\\
&=&\frac{1}{N}(H(W|S^{N})-I(W;Z^{N}|S^{N}))\nonumber\\
&=&\frac{1}{N}(H(W|S^{N})-H(W|S^{N},Y^{N})+H(W|S^{N},Y^{N})-I(W;Z^{N}|S^{N}))\nonumber\\
&\stackrel{(2)}\leq&\frac{1}{N}(I(W;Y^{N}|S^{N})-I(W;Z^{N}|S^{N})+\delta(P_{e}))\nonumber\\
&=&\frac{1}{N}\sum_{i=1}^{N}(I(W;Y_{i}|Y^{i-1},S^{N})-I(W;Z_{i}|Z_{i+1}^{N},S^{N}))+\frac{\delta(P_{e})}{N},
\end{eqnarray}
where (1) from (\ref{e211}), and (2) is from the Fano's inequality.

The character $I(W;Y_{i}|Y^{i-1},S^{N})$ in (\ref{jmds3}) can be processed as
\begin{eqnarray}\label{jmds4}
&&I(W;Y_{i}|Y^{i-1},S^{N})=H(Y_{i}|Y^{i-1},S^{N})-H(Y_{i}|Y^{i-1},S^{N},W)\nonumber\\
&&=H(Y_{i}|Y^{i-1},S^{N})-H(Y_{i}|Y^{i-1},S^{N},W)-H(Y_{i}|Y^{i-1},Z_{i+1}^{N},S^{N})+H(Y_{i}|Y^{i-1},Z_{i+1}^{N},S^{N})\nonumber\\
&&+H(Y_{i}|Y^{i-1},Z_{i+1}^{N},S^{N},W)-H(Y_{i}|Y^{i-1},Z_{i+1}^{N},S^{N},W)\nonumber\\
&&=I(Y_{i};Z_{i+1}^{N}|Y^{i-1},S^{N})-I(Y_{i};Z_{i+1}^{N}|Y^{i-1},S^{N},W)+I(W;Y_{i}|Y^{i-1},Z_{i+1}^{N},S^{N}),
\end{eqnarray}
and the character $I(W;Z_{i}|Z_{i+1}^{N},S^{N})$ in (\ref{jmds3}) can be processed as
\begin{eqnarray}\label{jmds5}
&&I(W;Z_{i}|Z_{i+1}^{N},S^{N})=H(Z_{i}|Z_{i+1}^{N},S^{N})-H(Z_{i}|Z_{i+1}^{N},S^{N},W)\nonumber\\
&&=H(Z_{i}|Z_{i+1}^{N},S^{N})-H(Z_{i}|Z_{i+1}^{N},S^{N},W)-H(Z_{i}|Y^{i-1},Z_{i+1}^{N},S^{N})+H(Z_{i}|Y^{i-1},Z_{i+1}^{N},S^{N})\nonumber\\
&&+H(Z_{i}|Y^{i-1},Z_{i+1}^{N},S^{N},W)-H(Z_{i}|Y^{i-1},Z_{i+1}^{N},S^{N},W)\nonumber\\
&&=I(Z_{i};Y^{i-1}|Z_{i+1}^{N},S^{N})-I(Z_{i};Y^{i-1}|Z_{i+1}^{N},S^{N},W)+I(W;Z_{i}|Y^{i-1},Z_{i+1}^{N},S^{N}).
\end{eqnarray}
Substituting (\ref{jmds4}) and (\ref{jmds5}) into (\ref{jmds3}), and using the properties
\begin{eqnarray}\label{jmdsx.1}
&&\sum_{i=1}^{N}I(Y_{i};Z_{i+1}^{N}|Y^{i-1},S^{N})=\sum_{i=1}^{N}I(Z_{i};Y^{i-1}|Z_{i+1}^{N},S^{N})
\end{eqnarray}
and
\begin{eqnarray}\label{jmdsx.2}
&&\sum_{i=1}^{N}I(Y_{i};Z_{i+1}^{N}|Y^{i-1},S^{N},W)=\sum_{i=1}^{N}I(Z_{i};Y^{i-1}|Z_{i+1}^{N},S^{N},W),
\end{eqnarray}
we have
\begin{eqnarray}\label{jmds6}
R_{e}-\epsilon&\stackrel{(a)}\leq&\frac{1}{N}\sum_{i=1}^{N}(I(W;Y_{i}|Y^{i-1},Z_{i+1}^{N},S^{N})-I(W;Z_{i}|Y^{i-1},Z_{i+1}^{N},S^{N}))+\frac{\delta(P_{e})}{N}\nonumber\\
&\stackrel{(b)}=&\frac{1}{N}\sum_{i=1}^{N}(I(W;Y_{i}|Y^{i-1},Z_{i+1}^{N},S^{N},S_{i-d},S_{i})-I(W;Z_{i}|Y^{i-1},Z_{i+1}^{N},S^{N},S_{i-d},S_{i}))+\frac{\delta(P_{e})}{N}\nonumber\\
&\stackrel{(c)}=&\frac{1}{N}\sum_{i=1}^{N}(I(W;Y_{i}|Y^{i-1},Z_{i+1}^{N},S^{N},S_{i-d},S_{i},J=i)-I(W;Z_{i}|Y^{i-1},Z_{i+1}^{N},S^{N},S_{i-d},S_{i},J=i))+\frac{\delta(P_{e})}{N}\nonumber\\
&\stackrel{(d)}=&I(W;Y_{J}|Y^{J-1},Z_{J+1}^{N},S^{N},S_{J-d},S_{J},J)-I(W;Z_{J}|Y^{J-1},Z_{J+1}^{N},S^{N},S_{J-d},S_{J},J)+\frac{\delta(P_{e})}{N}\nonumber\\
&\stackrel{(e)}=&I(V;Y|U,\tilde{S},S)-I(V;Z|U,\tilde{S},S)+\frac{\delta(P_{e})}{N}\nonumber\\
&\stackrel{(f)}\leq&I(V;Y|U,\tilde{S},S)-I(V;Z|U,\tilde{S},S)+\frac{\delta(\epsilon)}{N},
\end{eqnarray}
where (a) is from (\ref{jmdsx.1}) and (\ref{jmdsx.2})
(b) is from the fact that $S_{i}$ and $S_{i-d}$ (here $S_{i-d}=const$ when $i\leq d$) are included in $S^{N}$, (c) is from
the fact that $J$ is a random variable (uniformly distributed
over $\{1,2,...,N\}$), and it is independent of $Y^{N}$, $Z^{N}$, $W$ and $S^{N}$, (d) is from $J$ is
uniformly distributed over $\{1,2,...,N\}$, (e) is from the definitions in (\ref{jmds1}), and (f)
is from $\delta(P_{e})$ is increasing while $P_{e}$ is increasing, and $P_{e}\leq \epsilon$.
Letting $\epsilon\rightarrow 0$, we have $R_{e}\leq I(V;Y|U,S,\tilde{S})-I(V;Z|U,S,\tilde{S})$. Now it remains to prove the equalities
(\ref{jmdsx.1}) and (\ref{jmdsx.2}), see the followings.

\begin{IEEEproof}

Using the chain rule, the left parts of (\ref{jmdsx.1}) and (\ref{jmdsx.2}) can be re-written as
\begin{eqnarray}\label{jmdsx.bv1}
&&\sum_{i=1}^{N}I(Y_{i};Z_{i+1}^{N}|Y^{i-1},S^{N})=\sum_{i=1}^{N}\sum_{j=i+1}^{N}I(Y_{i};Z_{j}|Y^{i-1},S^{N},Z_{j+1}^{N}),
\end{eqnarray}
and
\begin{eqnarray}\label{jmdsx.bv2}
&&\sum_{i=1}^{N}I(Y_{i};Z_{i+1}^{N}|Y^{i-1},S^{N},W)=\sum_{i=1}^{N}\sum_{j=i+1}^{N}I(Y_{i};Z_{j}|Y^{i-1},S^{N},Z_{j+1}^{N},W).
\end{eqnarray}
The right parts of (\ref{jmdsx.1}) and (\ref{jmdsx.2}) can be re-written as
\begin{eqnarray}\label{jmdsx.bv3}
&&\sum_{i=1}^{N}I(Z_{i};Y^{i-1}|Z_{i+1}^{N},S^{N})=\sum_{i=1}^{N}\sum_{j=1}^{i-1}I(Y_{j};Z_{i}|Y^{j-1},S^{N},Z_{i+1}^{N})\nonumber\\
&&=\sum_{j=1}^{N}\sum_{i=1}^{j-1}I(Y_{i};Z_{j}|Y^{i-1},S^{N},Z_{j+1}^{N})\nonumber\\
&&=\sum_{j=i+1}^{N}\sum_{i=1}^{N}I(Y_{i};Z_{j}|Y^{i-1},S^{N},Z_{j+1}^{N}),
\end{eqnarray}
and
\begin{eqnarray}\label{jmdsx.bv4}
&&\sum_{i=1}^{N}I(Z_{i};Y^{i-1}|Z_{i+1}^{N},S^{N},W)=\sum_{i=1}^{N}\sum_{j=1}^{i-1}I(Y_{j};Z_{i}|Y^{j-1},S^{N},Z_{i+1}^{N},W)\nonumber\\
&&=\sum_{j=1}^{N}\sum_{i=1}^{j-1}I(Y_{i};Z_{j}|Y^{i-1},S^{N},Z_{j+1}^{N},W)\nonumber\\
&&=\sum_{j=i+1}^{N}\sum_{i=1}^{N}I(Y_{i};Z_{j}|Y^{i-1},S^{N},Z_{j+1}^{N},W).
\end{eqnarray}
By checking (\ref{jmdsx.bv1})-(\ref{jmdsx.bv4}), it is easy to see that (\ref{jmdsx.1}) and (\ref{jmdsx.2}) hold, and the proof is completed.

\end{IEEEproof}

The proof of Theorem \ref{T3.1} is completed.

\section{Proof of (\ref{giveup1})\label{appengiveup1}}

Replacing $V^{N}$ by $X^{N}$, and letting $W_{c}$, $U^{N}$ be constants, the achievability of (\ref{giveup1}) is along the lines of the direct proof of Theorem \ref{T3}
(see Appendix \ref{appen1}), and thus we only need to show the converse proof of (\ref{giveup1}). Since $R_{e}\leq R$ is obvious,
it remains to show that $R\leq I(X;Y|S,\tilde{S})$ and $R_{e}\leq I(X;Y|S,\tilde{S})-I(X;Z|S,\tilde{S})$, see the followings.

Note that
\begin{eqnarray}\label{jmds2}
R-\epsilon&\leq&\frac{1}{N}H(W)\leq\frac{1}{N}(I(W;Y^{N}|S^{N})+\delta(P_{e}))\nonumber\\
&\stackrel{(a)}\leq&\frac{1}{N}(I(X^{N};Y^{N}|S^{N})+\delta(P_{e}))\nonumber\\
&=&\frac{1}{N}\sum_{i=1}^{N}(H(Y_{i}|Y^{i-1},S^{N})-H(Y_{i}|Y^{i-1},S^{N},X^{N}))+\frac{\delta(P_{e})}{N}\nonumber\\
&\leq&\frac{1}{N}\sum_{i=1}^{N}(H(Y_{i}|S_{i},S_{i-d})-H(Y_{i}|Y^{i-1},S^{N},X^{N}))+\frac{\delta(P_{e})}{N}\nonumber\\
&\stackrel{(b)}=&\frac{1}{N}\sum_{i=1}^{N}(H(Y_{i}|S_{i},S_{i-d})-H(Y_{i}|S_{i},X_{i}))+\frac{\delta(P_{e})}{N}\nonumber\\
&\stackrel{(c)}=&\frac{1}{N}\sum_{i=1}^{N}(H(Y_{i}|S_{i},S_{i-d})-H(Y_{i}|S_{i},X_{i},S_{i-d}))+\frac{\delta(P_{e})}{N}\nonumber\\
&\stackrel{(d)}=&H(Y_{J}|S_{J},S_{J-d},J)-H(Y_{J}|S_{J},S_{J-d},X_{J},J)+\frac{\delta(P_{e})}{N}\nonumber\\
&\stackrel{(e)}\leq&H(Y_{J}|S_{J},S_{J-d})-H(Y_{J}|S_{J},S_{J-d},X_{J})+\frac{\delta(P_{e})}{N}\nonumber\\
&\stackrel{(f)}\leq&I(X;Y|S,\tilde{S})+\frac{\delta(\epsilon)}{N},
\end{eqnarray}
where (a) is from $H(W|X^{N})=0$, (b) is from the Markov chain
$(Y^{i-1},S^{i-1},S_{i+1}^{N},X^{i-1},X_{i+1}^{N})\rightarrow (S_{i},X_{i})\rightarrow Y_{i}$, (c) is from the Markov chain
$S_{i-d}\rightarrow (S_{i},X_{i})\rightarrow Y_{i}$,
(d) is from the fact that $J$ is a random variable (uniformly distributed
over $\{1,2,...,N\}$), and it is independent of $Y^{N}$, $Z^{N}$, $W$ and $S^{N}$, (e) is from the Markov chains
$(J,S_{J-d})\rightarrow (S_{J},X_{J})\rightarrow Y_{J}$ and $S_{J-d}\rightarrow (S_{J},X_{J})\rightarrow Y_{J}$,
and (f) is from the definitions in (\ref{jmds1}), $X\triangleq X_{J}$ and the fact that $\delta(P_{e})\leq \delta(\epsilon)$.
Then, letting $\epsilon\rightarrow 0$, we have $R\leq I(X;Y|S,\tilde{S})$.

Similarly, note that
\begin{eqnarray}\label{b1}
R_{e}-\epsilon&\stackrel{(1)}\leq& \frac{H(W|Z^{N},S^{N})}{N}\nonumber\\
&=&\frac{1}{N}(H(W|Z^{N},S^{N})-H(W|Z^{N},S^{N},Y^{N})+H(W|Z^{N},S^{N},Y^{N}))\nonumber\\
&\stackrel{(2)}\leq&\frac{1}{N}(I(W;Y^{N}|Z^{N},S^{N})+\delta(P_{e}))\nonumber\\
&\leq&\frac{1}{N}(H(Y^{N}|Z^{N},S^{N})-H(Y^{N}|Z^{N},S^{N},W,X^{N})+\delta(P_{e}))\nonumber\\
&\stackrel{(3)}=&\frac{1}{N}(H(Y^{N}|Z^{N},S^{N})-H(Y^{N}|Z^{N},S^{N},X^{N})+\delta(P_{e}))\nonumber\\
&=&\frac{1}{N}(I(X^{N};Y^{N}|Z^{N},S^{N})+\delta(P_{e}))\nonumber\\
&\stackrel{(4)}=&\frac{1}{N}(H(X^{N}|Z^{N},S^{N})-H(X^{N}|Y^{N},S^{N})+H(X^{N}|S^{N})-H(X^{N}|S^{N})+\delta(P_{e}))\nonumber\\
&=&\frac{1}{N}(I(X^{N};Y^{N}|S^{N})-I(X^{N};Z^{N}|S^{N})+\delta(P_{e}))\nonumber\\,
&\stackrel{(5)}\leq&\frac{1}{N}(I(X^{N};Y^{N}|S^{N})-I(X^{N};Z^{N}|S^{N})+\delta(\epsilon)),
\end{eqnarray}
where (1) is from (\ref{e211}), (2) is from Fano's inequality, (3) is from the fact that $H(W|X^{N})=0$,
(4) is from the Markov chain $X^{N}\rightarrow (Y^{N},S^{N})\rightarrow Z^{N}$, and (5) is from the fact that $P_{e}\leq \epsilon$ and
$\delta(P_{e})$ is increasing while $P_{e}$ is increasing.

The character $I(X^{N};Y^{N}|S^{N})-I(X^{N};Z^{N}|S^{N})$ in (\ref{b1}) can be further bounded by
\begin{eqnarray}\label{b2}
&&\frac{1}{N}I(X^{N};Y^{N}|S^{N})-I(X^{N};Z^{N}|S^{N})\nonumber\\
&&\stackrel{(a)}=\frac{1}{N}\sum_{i=1}^{N}(H(Y_{i}|Y^{i-1},S^{N})-H(Y_{i}|X_{i},S_{i})-H(Z_{i}|Z^{i-1},S^{N})+H(Z_{i}|X_{i},S_{i}))\nonumber\\
&&\stackrel{(b)}=\frac{1}{N}\sum_{i=1}^{N}(H(Y_{i}|Y^{i-1},S^{N},Z^{i-1})-H(Y_{i}|X_{i},S_{i})-H(Z_{i}|Z^{i-1},S^{N})+H(Z_{i}|X_{i},S_{i}))\nonumber\\
&&\stackrel{(c)}\leq\frac{1}{N}\sum_{i=1}^{N}(H(Y_{i}|S_{i},S_{i-d},S^{N},Z^{i-1})-H(Y_{i}|X_{i},S_{i},S_{i-d})
-H(Z_{i}|Z^{i-1},S_{i},S_{i-d},S^{N})+H(Z_{i}|X_{i},S_{i},S_{i-d}))\nonumber\\
&&\stackrel{(d)}\leq\frac{1}{N}\sum_{i=1}^{N}(H(Y_{i}|S_{i},S_{i-d})-H(Y_{i}|X_{i},S_{i},S_{i-d})-H(Z_{i}|S_{i},S_{i-d})+H(Z_{i}|X_{i},S_{i},S_{i-d}))\nonumber\\
&&=\frac{1}{N}\sum_{i=1}^{N}(I(X_{i};Y_{i}|S_{i},S_{i-d})-I(X_{i};Z_{i}|S_{i},S_{i-d}))\nonumber\\
&&\stackrel{(e)}=I(X_{J};Y_{J}|S_{J},S_{J-d},J)-I(X_{J};Z_{J}|S_{J},S_{J-d},J)\nonumber\\
&&\stackrel{(f)}\leq I(X_{J};Y_{J}|S_{J},S_{J-d})-I(X_{J};Z_{J}|S_{J},S_{J-d})\nonumber\\
&&\stackrel{(g)}=I(X;Y|S,\tilde{S})-I(X;Z|S,\tilde{S}),
\end{eqnarray}
where (a) is from the Markov chains $(Y^{i-1},S^{i-1},S_{i+1}^{N},X^{i-1},X_{i+1}^{N})\rightarrow (S_{i},X_{i})\rightarrow Y_{i}$
and $(Z^{i-1},S^{i-1},S_{i+1}^{N},X^{i-1},\\X_{i+1}^{N})\rightarrow (S_{i},X_{i})\rightarrow Z_{i}$,
(b) is from the Markov chain $Y_{i}\rightarrow (Y^{i-1},S^{N})\rightarrow Z^{i-1}$,
(c) is from the Markov chains $S_{i-d}\rightarrow (X_{i},S_{i})\rightarrow Y_{i}$ and $S_{i-d}\rightarrow (X_{i},S_{i})\rightarrow Z_{i}$,
and the fact that $S_{i}$ and $S_{i-d}$ are a part of $S^{N}$ (here note that $S_{i-d}=const$ if $i\leq d$), (d) is from
\begin{eqnarray}\label{b4}
&&H(Y_{i}|S_{i},S_{i-d},S^{N},Z^{i-1})-H(Z_{i}|Z^{i-1},S_{i},S_{i-d},S^{N})\leq H(Y_{i}|S_{i},S_{i-d})-H(Z_{i}|S_{i},S_{i-d}),
\end{eqnarray}
(e) is from the fact that $J$ is a random variable (uniformly distributed
over $\{1,2,...,N\}$), and it is independent of $Y^{N}$, $Z^{N}$, $W$ and $S^{N}$, (f) is from the Markov chains
$(J,S_{J-d})\rightarrow (S_{J},X_{J})\rightarrow Y_{J}$, $S_{J-d}\rightarrow (S_{J},X_{J})\rightarrow Y_{J}$,
$(J,S_{J-d})\rightarrow (S_{J},X_{J})\rightarrow Z_{J}$, $S_{J-d}\rightarrow (S_{J},X_{J})\rightarrow Z_{J}$ and the fact that
\begin{eqnarray}\label{b4.1}
&&H(Y_{J}|S_{J},S_{J-d},J)-H(Z_{J}|S_{J},S_{J-d},J)\leq H(Y_{J}|S_{J},S_{J-d})-H(Z_{J}|S_{J},S_{J-d}),
\end{eqnarray}
and (g) is from the definitions in (\ref{jmds1}) and $X\triangleq X_{J}$.
Here note that the proof of (\ref{b4.1}) is analogous to that of (\ref{b4}), and thus we only need to prove the above (\ref{b4}), see the followings.

\textbf{Proof of (\ref{b4}):}
\begin{IEEEproof}
Note that (\ref{b4}) is equivalent to
\begin{eqnarray}\label{b6}
&&I(Z_{i};Z^{i-1},S^{N}|S_{i},S_{i-d})\leq I(Y_{i};S^{N},Z^{i-1}|S_{i},S_{i-d}).
\end{eqnarray}
Since
\begin{eqnarray}\label{b8}
I(Z_{i};Z^{i-1},S^{N}|S_{i},S_{i-d})&=&H(Z^{i-1},S^{N}|S_{i},S_{i-d})-H(Z^{i-1},S^{N}|S_{i},S_{i-d},Z_{i})\nonumber\\
&\leq&H(Z^{i-1},S^{N}|S_{i},S_{i-d})-H(Z^{i-1},S^{N}|S_{i},S_{i-d},Z_{i},Y_{i})\nonumber\\
&\stackrel{(1)}=&H(Z^{i-1},S^{N}|S_{i},S_{i-d})-H(Z^{i-1},S^{N}|S_{i},S_{i-d},Y_{i})\nonumber\\
&=&I(Y_{i};S^{N},Z^{i-1}|S_{i},S_{i-d}),
\end{eqnarray}
where (1)
is from the Markov chain $(Z^{i-1},S^{N})\rightarrow (S_{i},S_{i-d},Y_{i})\rightarrow Z_{i}$.
Then it is easy to see that (\ref{b6}) is proved, and thus the proof of (\ref{b4}) is completed.
\end{IEEEproof}

Substituting (\ref{b2}) into (\ref{b1}), and letting $\epsilon\rightarrow 0$,
$R_{e}\leq I(X;Y|S,\tilde{S})-I(X;Z|S,\tilde{S})$ is proved.
The converse and entire proof of (\ref{giveup1}) is completed.

\section{Proof of Theorem \ref{T1}\label{appen2.x}}

Rate splitting, block Markov coding, multiplexing random binning, and the idea of
using the delayed receiver's channel output feedback as a secret key \cite{AC}
are combined to show the achievability of $\mathcal{R}^{fi}$ in Theorem \ref{T1}.
The outline of the proof is as follows. Notations and definitions are given in Subsection \ref{sub-x1}, the construction of the code-books
are shown in Subsection \ref{sub-x2}, the encoding and decoding schemes are respectively introduced in Subsection \ref{sub-x3} and Subsection \ref{sub-x4},
and the equivocation analysis is shown in Subsection \ref{sub-x5}.

\subsection{Definitions}\label{sub-x1}

\begin{itemize}

\item The state takes values in $\mathcal{S}=\{1,2,...,k\}$ and the steady state probability
$\pi(l)>0$ for all $l\in \mathcal{S}$. Let $N_{\tilde{s}}$ ($1\leq \tilde{s}\leq k$) be the number satisfying
\begin{eqnarray}\label{swjtu.1}
&&N_{\tilde{s}}=N(\pi(\tilde{s})-\epsilon^{'}),
\end{eqnarray}
where $0\leq\epsilon^{'}<\min\{\pi(\tilde{s}); \tilde{s}\in \{1,2,...,k\}\}$ and $\epsilon^{'}\rightarrow 0$ as $N\rightarrow \infty$.

\item The message $W=(W_{1},...,W_{n})$ is transmitted through $n$ blocks, and similar to
the definitions in Appendix \ref{appen1}, the uniformly distributed message $W$ is divided into
a common message $W_{c}$ and a private message $W_{p}$ ($W=(W_{c},W_{p})$), and
$W$, $W_{c}$ and $W_{p}$ take values in the sets $\{1,2,...,2^{nNR}\}$, $\{1,2,...,2^{nNR_{c}}\}$ and $\{1,2,...,2^{nNR_{p}}\}$, respectively. Here
$R=R_{c}+R_{p}$. In the remainder of this section, we first prove
\begin{eqnarray}\label{final-1}
&&\mathcal{R}^{fi\diamond}=\{(R_{c},R_{p}, R_{e}): 0\leq R_{e}\leq R_{p},\nonumber\\
&&R_{c}\leq \min\{I(U;Y|S,\tilde{S}),I(U;Z|S,\tilde{S})\},\nonumber\\
&&R_{p}\leq I(V;Y|U,S,\tilde{S}),\nonumber\\
&&R_{e}\leq [I(V;Y|U,S,\tilde{S})-I(V;Z|U,S,\tilde{S})]^{+}+H(Y|V,Z,S,\tilde{S})\},
\end{eqnarray}
is achievable. Then, using Fourier-Motzkin elimination to eliminate $R_{c}$ and $R_{p}$ from $\mathcal{R}^{fi\diamond}$,
$\mathcal{R}^{fi}$ is directly obtained.

\item In order to prove $\mathcal{R}^{fi\diamond}$ is achievable, it is sufficient to show the following two cases are achievable.
\begin{itemize}

\item (Case 1:) for the case that $I(V;Y|U,S,\tilde{S})\geq I(V;Z|U,S,\tilde{S})$, we only need to show that
$(R_{c}=\min\{I(U;Y|S,\tilde{S}),I(U;Z|S,\tilde{S})\}, R_{p}=I(V;Y|U,S,\tilde{S}),R_{e}=I(V;Y|U,S,\tilde{S})-I(V;Z|U,S,\tilde{S})+R_{f})$
is achievable, where
\begin{eqnarray}\label{final-2.1}
&&R_{f}=\min\{H(Y|V,Z,S,\tilde{S}), I(V;Z|U,S,\tilde{S})\}.
\end{eqnarray}

\item (Case 2:) for the case that $I(V;Y|U,S,\tilde{S})<I(V;Z|U,S,\tilde{S})$, we only need to show that
$(R_{c}=\min\{I(U;Y|S,\tilde{S}),I(U;Z|S,\tilde{S})\}, R_{p}=I(V;Y|U,S,\tilde{S}),R_{e}=R^{*}_{f})$ is achievable, where
\begin{eqnarray}\label{final-2.2}
&&R^{*}_{f}=\min\{H(Y|V,Z,S,\tilde{S}), I(V;Y|U,S,\tilde{S})\}.
\end{eqnarray}
\end{itemize}

\item Define
\begin{eqnarray}\label{final-2.x}
&&R_{p,1}=[I(V;Y|U,S,\tilde{S})-I(V;Z|U,S,\tilde{S})]^{+},
\end{eqnarray}
and
\begin{eqnarray}\label{final-2.xx}
&&R_{p}=R_{p,1}+R_{p,2}.
\end{eqnarray}

\item In block $i$ ($1\leq i\leq n$), the message $W_{i}$ is divided into
$k$ sub-messages, i.e., $W_{i}=(W_{i,1},...,W_{i,k})$, where $W_{i,\textcolor[rgb]{1.00,0.00,0.00}{\tilde{s}}}
=(W_{i,\textcolor[rgb]{1.00,0.00,0.00}{\tilde{s}},c},W_{i,\textcolor[rgb]{1.00,0.00,0.00}{\tilde{s}},p,1},W_{i,\textcolor[rgb]{1.00,0.00,0.00}{\tilde{s}},p,2})$ 
($1\leq \textcolor[rgb]{1.00,0.00,0.00}{\tilde{s}}\leq k$),
$W_{i,\textcolor[rgb]{1.00,0.00,0.00}{\tilde{s}},c}$, $W_{i,\textcolor[rgb]{1.00,0.00,0.00}{\tilde{s}},p,1}$ 
and $W_{i,\textcolor[rgb]{1.00,0.00,0.00}{\tilde{s}},p,2}$ take values in the sets $\{1,2,...,2^{N_{\textcolor[rgb]{1.00,0.00,0.00}{\tilde{s}}}
R_{c}(\textcolor[rgb]{1.00,0.00,0.00}{\tilde{s}})}\}$, $\{1,2,...,2^{N_{\textcolor[rgb]{1.00,0.00,0.00}{\tilde{s}}}R_{p,1}(\textcolor[rgb]{1.00,0.00,0.00}{\tilde{s}})}\}$
and $\{1,2,...,2^{N_{\textcolor[rgb]{1.00,0.00,0.00}{\tilde{s}}}R_{p,2}(\textcolor[rgb]{1.00,0.00,0.00}{\tilde{s}})}\}$, respectively, 
and $N_{\textcolor[rgb]{1.00,0.00,0.00}{\tilde{s}}}$ satisfies (\ref{swjtu.1}). Here
\begin{eqnarray}\label{final-2}
&&R_{c}(\textcolor[rgb]{1.00,0.00,0.00}{\tilde{s}})=
\min\{I(U;Y|S,\tilde{S}=\textcolor[rgb]{1.00,0.00,0.00}{\tilde{s}}),I(U;Z|S,\tilde{S}=\textcolor[rgb]{1.00,0.00,0.00}{\tilde{s}})\},
\end{eqnarray}
\begin{eqnarray}\label{final-3}
&&R_{p,1}(\textcolor[rgb]{1.00,0.00,0.00}{\tilde{s}})=[I(V;Y|U,S,\tilde{S}=\textcolor[rgb]{1.00,0.00,0.00}{\tilde{s}})
-I(V;Z|U,S,\tilde{S}=\textcolor[rgb]{1.00,0.00,0.00}{\tilde{s}})]^{+},
\end{eqnarray}
\begin{eqnarray}\label{final-4}
R_{p,2}(\textcolor[rgb]{1.00,0.00,0.00}{\tilde{s}})&=&R_{p}(\textcolor[rgb]{1.00,0.00,0.00}{\tilde{s}})-R_{p,1}(\textcolor[rgb]{1.00,0.00,0.00}{\tilde{s}})\nonumber\\
&=&I(V;Y|U,S,\tilde{S}=\textcolor[rgb]{1.00,0.00,0.00}{\tilde{s}})-[I(V;Y|U,S,\tilde{S}=\textcolor[rgb]{1.00,0.00,0.00}{\tilde{s}})
-I(V;Z|U,S,\tilde{S}=\textcolor[rgb]{1.00,0.00,0.00}{\tilde{s}})]^{+}\nonumber\\
&=&\min\{I(V;Y|U,S,\tilde{S}=\textcolor[rgb]{1.00,0.00,0.00}{\tilde{s}}), I(V;Z|U,S,\tilde{S}=\textcolor[rgb]{1.00,0.00,0.00}{\tilde{s}})\}.
\end{eqnarray}
Note that $R_{c}(\textcolor[rgb]{1.00,0.00,0.00}{\tilde{s}})$, $R_{p,1}(\textcolor[rgb]{1.00,0.00,0.00}{\tilde{s}})$ 
and $R_{p,2}(\textcolor[rgb]{1.00,0.00,0.00}{\tilde{s}})$ are the transmission rates $R_{c}$, $R_{p,1}$ and $R_{p,2}$ for a given $\tilde{s}$,
respectively. Furthermore, it is easy to see that
\begin{eqnarray}\label{wing.1}
&&\sum_{\tilde{s}=1}^{k}\pi(\tilde{s})R_{c}(\tilde{s})=R_{c},\,\, \sum_{\tilde{s}=1}^{k}\pi(\tilde{s})R_{p,1}(\tilde{s})=R_{p,1},\,\,
\sum_{\tilde{s}=1}^{k}\pi(\tilde{s})R_{p,2}(\tilde{s})=R_{p,2}.
\end{eqnarray}
From the above definitions, it is easy to see that $W_{c}=(W_{1,1,c},...,W_{1,k,c},W_{2,1,c},...,W_{2,k,c},...,W_{n,1,c},...,W_{n,k,c})$
and $W_{p}=(W_{p,1},W_{p,2})$, where
$W_{p,1}=(W_{1,1,p,1},...,W_{1,k,p,1},W_{2,1,p,1},...,W_{2,k,p,1},...,W_{n,1,p,1},...,W_{n,k,p,1})$ and
$W_{p,2}=(W_{1,1,p,2},...,W_{1,k,p,2},W_{2,1,p,2},...,W_{2,k,p,2},...,W_{n,1,p,2},...,W_{n,k,p,2})$.

\item
The transmission rate $R^{*}_{c}$ of the common message $W_{c}$ is denoted by
\begin{eqnarray}\label{swjtu.2}
R^{*}_{c}&=&\frac{H(W_{c})}{nN}=\frac{\sum_{i=1}^{n}\sum_{\textcolor[rgb]{1.00,0.00,0.00}{\tilde{s}}=1}^{k}
H(W_{i,\textcolor[rgb]{1.00,0.00,0.00}{\tilde{s}},c})}{nN}=\frac{\sum_{i=1}^{n}\sum_{\textcolor[rgb]{1.00,0.00,0.00}{\tilde{s}}=1}^{k}
N_{\textcolor[rgb]{1.00,0.00,0.00}{\tilde{s}}}R_{c}(\textcolor[rgb]{1.00,0.00,0.00}{\tilde{s}})}{nN}\nonumber\\
&\stackrel{(a)}=&\frac{\sum_{i=1}^{n}\sum_{\textcolor[rgb]{1.00,0.00,0.00}{\tilde{s}}=1}^{k}N(\pi(\textcolor[rgb]{1.00,0.00,0.00}{\tilde{s}})
-\epsilon^{'})R_{c}(\textcolor[rgb]{1.00,0.00,0.00}{\tilde{s}})}{nN}\nonumber\\
&=&\sum_{\textcolor[rgb]{1.00,0.00,0.00}{\tilde{s}}=1}^{k}(\pi(\textcolor[rgb]{1.00,0.00,0.00}{\tilde{s}})-\epsilon^{'})R_{c}(\textcolor[rgb]{1.00,0.00,0.00}{\tilde{s}})\nonumber\\
&=&\sum_{\textcolor[rgb]{1.00,0.00,0.00}{\tilde{s}}=1}^{k}\pi(\textcolor[rgb]{1.00,0.00,0.00}{\tilde{s}})
R_{c}(\textcolor[rgb]{1.00,0.00,0.00}{\tilde{s}})-\epsilon^{'}\sum_{\textcolor[rgb]{1.00,0.00,0.00}{\tilde{s}}=1}^{k}R_{c}(\textcolor[rgb]{1.00,0.00,0.00}{\tilde{s}}),
\end{eqnarray}
where (a) is from (\ref{swjtu.1}). From (\ref{final-2}) and (\ref{swjtu.2}), it is easy to see that $R^{*}_{c}$ tends to be $R_{c}$
while $\epsilon^{'}\rightarrow 0$.
Similarly, the transmission rate $R^{*}_{p}$ of the private message $W_{p}$ tends to be $R_{p}$ while $\epsilon^{'}\rightarrow 0$.

\item Let $\widetilde{U}_{i}$ ($1\leq i\leq n$) be the random vector with length $N$ for block $i$ and $U^{n}=(\widetilde{U}_{1},...,\widetilde{U}_{n})$.
Similarly, $S^{n}=(\widetilde{S}_{1},...,\widetilde{S}_{n})$, $V^{n}=(\widetilde{V}_{1},...,\widetilde{V}_{n})$, $X^{n}=(\widetilde{X}_{1},...,\widetilde{X}_{n})$,
$Y^{n}=(\widetilde{Y}_{1},...,\widetilde{Y}_{n})$ and $Z^{n}=(\widetilde{Z}_{1},...,\widetilde{Z}_{n})$.
The specific values of the above random vectors are denoted by lower case letters.

\end{itemize}

\subsection{Construction of the code-books}\label{sub-x2}

Fix the joint probability mass function
$P_{UVS\tilde{S}XYZ}(u,v,s,\tilde{s},x,y,z)$ satisfying (\ref{dota3}).

\begin{itemize}

\item \textbf{Construction of $U^{N}$}: Construct $k$ code-books $\mathcal{U}^{\tilde{s}}$ of $U^{N}$ for all $\tilde{s}\in \mathcal{S}$.
In each code-book $\mathcal{U}^{\tilde{s}}$, randomly generate $2^{N_{\tilde{s}}R_{c}(\tilde{s})}$ i.i.d. sequences
$u^{N_{\tilde{s}}}$ according to the probability mass function $P_{U|\tilde{S}}(u|\tilde{s})$, and index these sequences as $u^{N_{\tilde{s}}}(i)$, where
$1\leq i\leq 2^{N_{\tilde{s}}R_{c}(\tilde{s})}$.

\item \textbf{Construction of $V^{N}$}: Construct $k$ code-books $\mathcal{V}^{\tilde{s}}$ of $V^{N}$ for all $\tilde{s}\in \mathcal{S}$.
In each code-book $\mathcal{V}^{\tilde{s}}$, randomly generate $2^{N_{\tilde{s}}(R_{p}(\tilde{s})+R_{c}(\tilde{s}))}$ i.i.d. sequences
$v^{N_{\tilde{s}}}$ according to the probability mass function $P_{V|U,\tilde{S}}(v|u,\tilde{s})$. Index these sequences of the code-book $\mathcal{V}^{\tilde{s}}$
as $v^{N_{\tilde{s}}}(i_{\tilde{s}},a_{\tilde{s}},b_{\tilde{s}})$, where $1\leq i_{\tilde{s}}\leq 2^{N_{\tilde{s}}R_{c}(\tilde{s})}$,
$a_{\tilde{s}}\in \mathcal{A}_{\tilde{s}}=\{1,2,...,A_{\tilde{s}}\}$,
$b_{\tilde{s}}\in \mathcal{B}_{\tilde{s}}=\{1,2,...,B_{\tilde{s}}\}$,
\begin{eqnarray}\label{swjtu.3}
&&A_{\tilde{s}}=2^{N_{\tilde{s}}[I(V;Y|U,S,\tilde{S}=\tilde{s})-I(V;Z|U,S,\tilde{S}=\tilde{s})]^{+}},
\end{eqnarray}
and
\begin{eqnarray}\label{swjtu.4}
&&B_{\tilde{s}}=2^{N_{\tilde{s}}I(V;Z|U,S,\tilde{S}=\tilde{s})}.
\end{eqnarray}
From (\ref{final-4}) and (\ref{swjtu.4}), it is easy to see that $2^{N_{\tilde{s}}R_{p,2}(\tilde{s})}\leq B_{\tilde{s}}$. Thus we partition
$\mathcal{B}_{\tilde{s}}$ into $2^{N_{\tilde{s}}R_{p,2}(\tilde{s})}$ bins,
and each bin has $2^{N_{\tilde{s}}(I(V;Z|U,S,\tilde{S}=\tilde{s})-R_{p,2}(\tilde{s}))}$ elements.

\item \textbf{Construction of $X^{N}$}: For each $\tilde{s}$, the sequence $x^{N_{\tilde{s}}}$ is i.i.d. generated according to a new discrete memoryless
channel (DMC) with transition probability $P_{X|U,V,\tilde{S}}(x|u,v,\tilde{s})$. The inputs of this new DMC are $u^{N_{\tilde{s}}}$ and $v^{N_{\tilde{s}}}$,
while the output is $x^{N_{\tilde{s}}}$.
\end{itemize}

\subsection{Encoding scheme}\label{sub-x3}

The codeword in each block has length $N$.
Let $L_{\tilde{s}}$ be the number of times during the $N$ symbols for which
the delayed feedback state at the transmitter is $\tilde{S}=\tilde{s}$. Every time that the corresponding delayed state is $\tilde{S}=\tilde{s}$,
the transmitter chooses the next symbols of $u^{N}$ and $v^{N}$ from the component code-books $\mathcal{U}^{\tilde{s}}$ and $\mathcal{V}^{\tilde{s}}$, respectively.
Since $L_{\tilde{s}}$ is not necessarily equivalent to
$N_{\tilde{s}}$, an error is declared if $L_{\tilde{s}}< N_{\tilde{s}}$, and the codes are filled with zero if $L_{\tilde{s}}> N_{\tilde{s}}$. Since the state process is stationary and ergodic
$\lim_{N\rightarrow \infty}\frac{L_{\tilde{s}}}{N}=Pr\{\tilde{S}=\tilde{s}\}$ in probability. Thus, we have
\begin{eqnarray}\label{dota.m1}
&&Pr\{L_{\tilde{s}}< N_{\tilde{s}}\}\rightarrow 0, \,\, \mbox{as}\,\, N\rightarrow \infty.
\end{eqnarray}
For the $i$-th block ($1\leq i\leq n$), the transmitted message is
$w_{i}=(w_{i,1,c},w_{i,1,p,1},w_{i,1,p,2},...,w_{i,k,c},w_{i,k,p,1},w_{i,k,p,2})$.
The encoding scheme is considered into two steps. First,
for block $1\leq i\leq 2d$, the encoding scheme is as follows.
\begin{itemize}
\item (Choosing $\widetilde{u}_{i}$:) In each component code-book $\mathcal{U}^{\tilde{s}}$ ($1\leq \tilde{s}\leq k$),
the transmitter chooses $\widetilde{u}_{i}^{N_{\tilde{s}}}(w_{i,\tilde{s},c})$ as the $\tilde{s}$-th component codeword of the transmitted $\widetilde{u}_{i}$.
The transmitted codeword $\widetilde{u}_{i}$ is obtained by multiplexing
the different component codewords.

\item (Choosing $\widetilde{v}_{i}$:) In each component code-book $\mathcal{V}^{\tilde{s}}$ ($1\leq \tilde{s}\leq k$),
the transmitter chooses $\widetilde{v}_{i}^{N_{\tilde{s}}}(i^{*}_{\tilde{s}},a^{*}_{\tilde{s}},b^{*}_{\tilde{s}})$
as the $\tilde{s}$-th component codeword of the transmitted $\widetilde{v}_{i}$,
where $i^{*}_{\tilde{s}}=w_{i,\tilde{s},c}$, $a^{*}_{\tilde{s}}=w_{i,\tilde{s},p,1}$, and $b^{*}_{\tilde{s}}$ is randomly chosen from
the bin $w_{i,\tilde{s},p,2}$ of $\mathcal{B}_{\tilde{s}}$. The transmitted codeword $\widetilde{v}_{i}$ is obtained by multiplexing
the different component codewords.
\end{itemize}
Second, for block $2d+1\leq i\leq n$,
the encoding scheme is as follows.
\begin{itemize}
\item The choosing of $\widetilde{u}_{i}$ for block $2d+1\leq i\leq n$ is the same as that in block $1\leq i\leq 2d$.

\item (Generation of the key:)
In block $2d+1\leq i\leq n$, the transmitter has already known $\widetilde{s}_{i-2d}$, and
it is used to multiplex
the component codewords $\widetilde{u}_{i-d}$, $\widetilde{v}_{i-d}$ and vectors $\widetilde{s}_{i-d}$,
$\widetilde{x}_{i-d}$ $\widetilde{y}_{i-d}$ and $\widetilde{z}_{i-d}$.
Once the transmitter receives the delayed feedback $\widetilde{y}_{i-d}$ and $\widetilde{s}_{i-d}$,
he first demultiplexes them into $\widetilde{y}^{N_{1}}_{i-d}$, $\widetilde{y}^{N_{2}}_{i-d}$,...,
$\widetilde{y}^{N_{k}}_{i-d}$
and $\widetilde{s}^{N_{1}}_{i-d}$, $\widetilde{s}^{N_{2}}_{i-d}$,...,$\widetilde{s}^{N_{k}}_{i-d}$.
Then, when the transmitter receives $\widetilde{y}^{N_{j}}_{i-d}$ ($1\leq j\leq k$),
he gives up if $\widetilde{y}^{N_{j}}_{i-d}\notin T^{N_{j}}_{Y|V,S,\tilde{S}}
(\widetilde{v}^{N_{j}}_{i-d},\widetilde{s}^{N_{j}}_{i-d},\tilde{s}=j)$.
It is easy to see that for $\tilde{s}=j$, the probability for giving up at the $i-d$-th block tends to $0$ as $N\rightarrow\infty$
(here $N_{j}=N(\pi(j)-\epsilon^{'})$).
In the case $\widetilde{y}^{N_{j}}_{i-d}\in T^{N_{j}}_{Y|V,S,\tilde{S}}(\widetilde{v}^{N_{j}}_{i-d},
\widetilde{s}^{N_{j}}_{i-d},\tilde{s}=j)$,
generate a mapping
\begin{eqnarray}\label{dead1.1}
&&g_{i,j}: \widetilde{y}^{N_{j}}_{i-d}
\rightarrow \{1,2,...,2^{N_{j}R_{f}(j)}\}
\end{eqnarray}
for case 1, and
\begin{eqnarray}\label{dead2.1}
&&g_{i,j}: \widetilde{y}^{N_{j}}_{i-d}
\rightarrow \{1,2,...,2^{N_{j}R^{*}_{f}(j)}\}
\end{eqnarray}
for case 2. Here note that
\begin{eqnarray}\label{dead1}
&&R_{f}(j)=\min\{H(Y|V,Z,S,\tilde{S}=j), I(V;Z|U,S,\tilde{S}=j)\},
\end{eqnarray}
\begin{eqnarray}\label{dead2}
&&R^{*}_{f}(j)=\min\{H(Y|V,Z,S,\tilde{S}=j), I(V;Y|U,S,\tilde{S}=j)\}.
\end{eqnarray}
Define a random variable $K_{i,j}^{*}=g_{i,j}(\widetilde{Y}^{N_{j}}_{i-d})$ ($2d+1\leq i\leq n$),
which is uniformly
distributed over $\{1,2,...,\\ 2^{N_{j}R_{f}(j)}\}$ or $\{1,2,...,2^{N_{j}R^{*}_{f}(j)}\}$, and $K_{i,j}^{*}$ is independent of
$\widetilde{U}_{i}$, $\widetilde{V}_{i}$, $\widetilde{S}_{i}$,
$\widetilde{X}_{i}$ $\widetilde{Y}_{i}$, $\widetilde{Z}_{i}$ and $W_{i}$. Here note that $K_{i,j}^{*}$ is used as
a secret key shared by the transmitter and the receiver, and $k_{i,j}^{*}$ is a specific value of $K_{i,j}^{*}$.
Reveal the mapping $g_{i,j}$ to the transmitter, receiver and the eavesdropper.

\item (Choosing $\widetilde{v}_{i}$:) From (\ref{final-4}), (\ref{dead1}) and (\ref{dead2}), it is easy to see that
$R_{p,2}(j)\geq R_{f}(j)$ for case 1, and $R_{p,2}(j)\geq R^{*}_{f}(j)$ for case 2. Thus, for block $2d+1\leq i\leq n$ and $\tilde{s}=j$ ($1\leq j\leq k$),
divide the component message $w_{i,j,p,2}$ into $w^{*}_{i,j,p,2}$ and $w^{**}_{i,j,p,2}$, i.e., $w_{i,j,p,2}=(w^{*}_{i,j,p,2},w^{**}_{i,j,p,2})$,
where $w^{*}_{i,j,p,2}\in\{1,2,...,2^{N_{j}R_{f}(j)}\}$, $w^{**}_{i,j,p,2}\in\{1,2,...,2^{N_{j}(R_{p,2}(j)-R_{f}(j))}\}$ for case 1, and
$w^{*}_{i,j,p,2}\in\{1,2,...,2^{N_{j}R^{*}_{f}(j)}\}$, $w^{**}_{i,j,p,2}\in\{1,2,...,2^{N_{j}(R_{p,2}(j)-R^{*}_{f}(j))}\}$ for case 2.
For both cases, in each component code-book $\mathcal{V}^{\tilde{s}}$ ($1\leq \tilde{s}\leq k$),
the transmitter chooses $\widetilde{v}_{i}^{N_{\tilde{s}}}(i^{*}_{\tilde{s}},a^{*}_{\tilde{s}},b^{*}_{\tilde{s}})$
as the $\tilde{s}$-th component codeword of the transmitted $\widetilde{v}_{i}$,
where $i^{*}_{\tilde{s}}=w_{i,\tilde{s},c}$, $a^{*}_{\tilde{s}}=w_{i,\tilde{s},p,1}$, and $b^{*}_{\tilde{s}}$ is randomly chosen from
the bin $(w^{*}_{i,j,p,2}\oplus k_{i,j}^{*},w^{**}_{i,j,p,2})$ of $\mathcal{B}_{\tilde{s}}$, where $\oplus$ is the modulo addition over
$\{1,2,...,2^{N_{j}R_{f}(j)}\}$ for case 1 and $\{1,2,...,2^{N_{j}R^{*}_{f}(j)}\}$ for case 2.
Here note that since $K_{i,j}^{*}$ and $W^{*}_{i,j,p,2}$ are independent and uniformly distributed over the same alphabet,
$K_{i,j}^{*}\oplus W^{*}_{i,j,p,2}$ is also independent of $K_{i,j}^{*}$ and $W^{*}_{i,j,p,2}$, and it is also uniformly distributed over
the same alphabet as that of $K_{i,j}^{*}$ and $W^{*}_{i,j,p,2}$.
The transmitted codeword $\widetilde{v}_{i}$ is obtained by multiplexing
the different component codewords.
\end{itemize}

\subsection{Decoding scheme}\label{sub-x4}

\begin{itemize}

\item (\textbf{Decoding scheme for the receiver}:)

\begin{itemize}

\item (\textbf{Decoding the common message $w_{i,c}$ for block $1\leq i\leq n$:}) The delayed feedback state $\tilde{S}$ at the transmitter, which is used to multiplex
the component codewords, is also available at the receiver. For block $1\leq i\leq n$, once the receiver receives $\widetilde{y}_{i}$
and the state sequence $\widetilde{s}_{i}$,
he first demultiplexes them into outputs corresponding to the
component code-books and separately decodes each component codeword. To be specific, in each code-book $\mathcal{U}^{\tilde{s}}$, the receiver has
$(\widetilde{y}_{i}^{N_{\tilde{s}}}, \widetilde{s}_{i}^{N_{\tilde{s}}})$ and tries to search a unique $\widetilde{u}_{i}^{N_{\tilde{s}}}$
such that
\begin{eqnarray}\label{dead3.1}
&&(\widetilde{u}_{i}^{N_{\tilde{s}}}, \widetilde{y}_{i}^{N_{\tilde{s}}}, \widetilde{s}_{i}^{N_{\tilde{s}}})\in
T^{N_{\tilde{s}}}_{UYS|\tilde{S}}(\epsilon).
\end{eqnarray}
If there exists such a unique $\widetilde{u}_{i}^{N_{\tilde{s}}}$, put out the corresponding index $\hat{w}_{i,\tilde{s},c}$.
Otherwise, i.e., if no such sequence exists or multiple sequences have different message indices,
declare a decoding error. If
for all $1\leq \tilde{s}\leq k$, there exist unique sequences $\widetilde{u}_{i}^{N_{\tilde{s}}}$ satisfying (\ref{dead3.1}),
the receiver
declares that $\hat{w}_{i,c}=(\hat{w}_{i,1,c},\hat{w}_{i,2,c},...,\hat{w}_{i,k,c})$ is sent in block $i$.
Based on the AEP and (\ref{final-2}), it is easy to see that the error probability $Pr\{\hat{w}_{i,\tilde{s},c}\neq w_{i,\tilde{s},c}\}$ ($1\leq \tilde{s}\leq k$)
goes to $0$.

\item (\textbf{Decoding the private message $w_{i,p}$ for block $1\leq i\leq 2d$:})
After decoding $\widetilde{u}_{i}^{N_{\tilde{s}}}$
for all $1\leq \tilde{s}\leq k$,
in each component code-book $\mathcal{V}^{\tilde{s}}$,
the receiver tries to find a unique sequence $\widetilde{v}_{i}^{N_{\tilde{s}}}$ such that
\begin{eqnarray}\label{dead3.2}
&&(\widetilde{v}_{i}^{N_{\tilde{s}}}, \widetilde{u}_{i}^{N_{\tilde{s}}}, \widetilde{y}_{i}^{N_{\tilde{s}}}, \widetilde{s}_{i}^{N_{\tilde{s}}})\in
T^{N_{\tilde{s}}}_{VUYS|\tilde{S}}(\epsilon).
\end{eqnarray}
If there exists such a unique $\widetilde{v}_{i}^{N_{\tilde{s}}}$, put out the corresponding indexes
$\hat{i}^{*}_{\tilde{s}}$, $\hat{a}^{*}_{\tilde{s}}$ and $\hat{b}^{*}_{\tilde{s}}$.
Otherwise, i.e., if no such sequence exists or multiple sequences have different message indices,
declare a decoding error. For block $1\leq i\leq 2d$, after the receiver obtains the index $\hat{b}^{*}_{\tilde{s}}$, he also knows $\hat{w}_{i,\tilde{s},p,2}$
since it is
the index of the bin which $\hat{b}^{*}_{\tilde{s}}$ belongs to. Thus, for $1\leq \tilde{s}\leq k$, the receiver has an estimation
$\hat{w}_{i,\tilde{s},p}$ of the private message
$w_{i,\tilde{s},p}$ by letting $\hat{w}_{i,\tilde{s},p}=(\hat{a}^{*}_{\tilde{s}},\hat{w}_{i,\tilde{s},p,2})$.
If for all $1\leq \tilde{s}\leq k$, there exist unique sequences $\widetilde{v}_{i}^{N_{\tilde{s}}}$ such that (\ref{dead3.2}) is satisfied, the receiver
declares that $\hat{w}_{i,p}=(\hat{w}_{i,1,p},\hat{w}_{i,2,p},...,\hat{w}_{i,k,p})$ is sent for block $i$.
Based on the AEP and $R_{p}(\tilde{s})=I(V;Y|U,S,\tilde{S}=\tilde{s})$, it is easy to see that
the error probability $Pr\{\hat{w}_{i,\tilde{s},p}\neq w_{i,\tilde{s},p}\}$ ($1\leq \tilde{s}\leq k$)
goes to $0$.

\item (\textbf{Decoding the private message $w_{i,p}$ for block $2d+1\leq i\leq n$:})
For block $2d+1\leq i\leq n$ and $1\leq \tilde{s}\leq k$,
after decoding $\widetilde{u}_{i}^{N_{\tilde{s}}}$, first,
the receiver tries to find a unique sequence $\widetilde{v}_{i}^{N_{\tilde{s}}}$ satisfying (\ref{dead3.2}).
If there exists such a unique $\widetilde{v}_{i}^{N_{\tilde{s}}}$, put out the corresponding indexes
$\hat{i}^{*}_{\tilde{s}}$, $\hat{a}^{*}_{\tilde{s}}$ and $\hat{b}^{*}_{\tilde{s}}$.
Otherwise, i.e., if no such sequence exists or multiple sequences have different message indices,
declare a decoding error. After the receiver obtains the index $\hat{b}^{*}_{\tilde{s}}$, he also knows
$(\hat{w}^{*}_{i,\tilde{s},p,2}\oplus k_{i,\tilde{s}}^{*},\hat{w}^{**}_{i,\tilde{s},p,2})$
since it is
the index of the bin which $\hat{b}^{*}_{\tilde{s}}$ belongs to. Then, note that the receiver knows the secret key $k_{i,\tilde{s}}^{*}$,
and thus he can directly obtain
$\hat{w}_{i,\tilde{s},p,2}=(\hat{w}^{*}_{i,\tilde{s},p,2},\hat{w}^{**}_{i,\tilde{s},p,2})$ from
$(\hat{w}^{*}_{i,\tilde{s},p,2}\oplus k_{i,\tilde{s}}^{*},\hat{w}^{**}_{i,\tilde{s},p,2})$ and the key $k_{i,\tilde{s}}^{*}$.
Thus for $1\leq \tilde{s}\leq k$, the receiver has an estimation
$\hat{w}_{i,\tilde{s},p}$ of the private message
$w_{i,\tilde{s},p}$ by letting $\hat{w}_{i,\tilde{s},p}=(\hat{a}^{*}_{\tilde{s}},\hat{w}_{i,\tilde{s},p,2})$.
If for all $1\leq \tilde{s}\leq k$, there exist unique sequences $\widetilde{v}_{i}^{N_{\tilde{s}}}$ such that (\ref{dead3.2}) is satisfied, the receiver
declares that $\hat{w}_{i,p}=(\hat{w}_{i,1,p},\hat{w}_{i,2,p},...,\hat{w}_{i,k,p})$ is sent for block $2d+1\leq i\leq n$.
Based on the AEP and $R_{p}(\tilde{s})=I(V;Y|U,S,\tilde{S}=\tilde{s})$, it is easy to see that
the error probability $Pr\{\hat{w}_{i,\tilde{s},p}\neq w_{i,\tilde{s},p}\}$ ($1\leq \tilde{s}\leq k$)
goes to $0$.
\end{itemize}

\item (\textbf{Decoding scheme for the eavesdropper}:)

\begin{itemize}

\item (\textbf{Decoding the common message $w_{i,c}$ for block $1\leq i\leq n$:}) The delayed feedback state $\tilde{S}$ at the transmitter, which is used to multiplex
the component codewords, is also available at the eavesdropper. For block $1\leq i\leq n$, once the eavesdropper receives $\widetilde{z}_{i}$
and the state sequence $\widetilde{s}_{i}$,
he first demultiplexes them into outputs corresponding to the
component code-books and separately decodes each component codeword. To be specific, in each code-book $\mathcal{U}^{\tilde{s}}$, the eavesdropper has
$(\widetilde{z}_{i}^{N_{\tilde{s}}}, \widetilde{s}_{i}^{N_{\tilde{s}}})$ and tries to search a unique $\widetilde{u}_{i}^{N_{\tilde{s}}}$
such that
\begin{eqnarray}\label{dead4.1}
&&(\widetilde{u}_{i}^{N_{\tilde{s}}}, \widetilde{z}_{i}^{N_{\tilde{s}}}, \widetilde{s}_{i}^{N_{\tilde{s}}})\in
T^{N_{\tilde{s}}}_{UZS|\tilde{S}}(\epsilon).
\end{eqnarray}
If there exists such a unique $\widetilde{u}_{i}^{N_{\tilde{s}}}$, put out the corresponding index $\check{w}_{i,\tilde{s},c}$.
Otherwise, i.e., if no such sequence exists or multiple sequences have different message indices,
declare a decoding error. If
for all $1\leq \tilde{s}\leq k$, there exist unique sequences $\widetilde{u}_{i}^{N_{\tilde{s}}}$ satisfying (\ref{dead4.1}),
the receiver
declares that $\check{w}_{i,c}=(\check{w}_{i,1,c},\check{w}_{i,2,c},...,\check{w}_{i,k,c})$ is sent in block $i$.
Based on the AEP and (\ref{final-2}), it is easy to see that the error probability $Pr\{\check{w}_{i,\tilde{s},c}\neq w_{i,\tilde{s},c}\}$ ($1\leq \tilde{s}\leq k$)
goes to $0$.

\item (\textbf{For block $1\leq i\leq n$, given $\widetilde{z}_{i}$, $\widetilde{u}_{i}$, $\widetilde{s}_{i}$
and $w_{i,p,1}$, decoding $\widetilde{v}_{i}$:}) In each component code-book $\mathcal{V}^{\tilde{s}}$
($1\leq \tilde{s}\leq k$), given $\widetilde{s}_{i}^{N_{\tilde{s}}}$,
$\widetilde{u}_{i}^{N_{\tilde{s}}}(w_{i,\tilde{s},c})$, $\widetilde{z}_{i}^{N_{\tilde{s}}}$ and $w_{i,\tilde{s},p,1}$, the eavesdropper
tries to find a unique $\check{b}^{*}_{\tilde{s}}$ such that
\begin{eqnarray}\label{dead4.2}
&&(\widetilde{v}_{i}^{N_{\tilde{s}}}(w_{i,\tilde{s},c},w_{i,\tilde{s},p,1},\check{b}^{*}_{\tilde{s}}),
\widetilde{u}_{i}^{N_{\tilde{s}}}(w_{i,\tilde{s},c}), \widetilde{z}_{i}^{N_{\tilde{s}}}, \widetilde{s}_{i}^{N_{\tilde{s}}})\in
T^{N_{\tilde{s}}}_{UVSZ|\tilde{S}}(\epsilon).
\end{eqnarray}
Since there are $2^{N_{\tilde{s}}I(V;Z|U,S,\tilde{S}=\tilde{s})}$ possible values of $\check{b}^{*}_{\tilde{s}}$ (see (\ref{swjtu.4})),
based on the AEP, the error probability
\begin{eqnarray}\label{dead4.3}
&&Pr\{\check{b}^{*}_{\tilde{s}}\neq b^{*}_{\tilde{s}}\}\rightarrow 0.
\end{eqnarray}

\item (\textbf{For block $2d+1\leq i\leq n$, given
$\widetilde{v}_{i-d}$, $\widetilde{z}_{i-d}$ and $\widetilde{s}_{i-d}$,
the eavesdropper's equivocation about the secret key:}) For block $2d+1\leq i\leq n$ and $\tilde{S}=\tilde{s}$,
even the eavesdropper knows $\widetilde{v}_{i}^{N_{\tilde{s}}}$,
without the secret key $k_{i,\tilde{s}}^{*}$ he still
can not obtain $w_{i,\tilde{s},p,2}$, and this is because $w_{i,\tilde{s},p,2}=(w^{*}_{i,\tilde{s},p,2}\oplus k_{i,\tilde{s}}^{*},w^{**}_{i,\tilde{s},p,2})$.
The eavesdropper can guess $k_{i,\tilde{s}}^{*}$ from $\widetilde{v}_{i-d}^{N_{\tilde{s}}}$, $\widetilde{z}_{i-d}^{N_{\tilde{s}}}$
and $\widetilde{s}_{i-d}^{N_{\tilde{s}}}$, and his equivocation
about the secret key $k_{i,\tilde{s}}^{*}$ can be bounded by the following balanced coloring lemma introduced by
Ahlswede and Cai \cite{AC}.
\begin{lemma}\label{Lx}
\textbf{(Balanced coloring lemma)} Given $\tilde{S}=\tilde{s}$, for
any $\epsilon, \delta>0$, sufficiently large $N_{\tilde{s}}$, all $N_{\tilde{s}}$-type
$P_{VS\tilde{S}Y}(v,s,\tilde{s},y)$ and all
$\widetilde{v}_{i-d}^{N_{\tilde{s}}}, \widetilde{s}_{i-d}^{N_{\tilde{s}}}\in T_{VS|\tilde{S}}^{N_{\tilde{s}}}$ ($2d+1\leq i\leq n$),
there exists a $\gamma$- coloring $c: T_{Y|V,S,\tilde{S}}^{N_{\tilde{s}}}(\widetilde{v}_{i-d}^{N_{\tilde{s}}},
\widetilde{s}_{i-d}^{N_{\tilde{s}}},\tilde{s})\rightarrow \{1,2,..,\gamma\}$ of $T_{Y|V,S,\tilde{S}}^{N_{\tilde{s}}}(\widetilde{v}_{i-d}^{N_{\tilde{s}}},
\widetilde{s}_{i-d}^{N_{\tilde{s}}},\tilde{s})$
such that for all joint $N_{\tilde{s}}$-type $P_{VS\tilde{S}YZ}(v,s,\tilde{s},y,z)$ with marginal distribution $P_{VS\tilde{S}Z}(v,s,\tilde{s},z)$ and
$\frac{|T_{Y|V,S,\tilde{S},Z}^{N_{\tilde{s}}}(\widetilde{v}_{i-d}^{N_{\tilde{s}}},
\widetilde{s}_{i-d}^{N_{\tilde{s}}},\tilde{s},\widetilde{z}_{i-d}^{N_{\tilde{s}}})|}{\gamma}>2^{N_{\tilde{s}}\epsilon}$,
$\widetilde{v}_{i-d}^{N_{\tilde{s}}}, \widetilde{s}_{i-d}^{N_{\tilde{s}}}, \widetilde{z}_{i-d}^{N_{\tilde{s}}}\in T_{VSZ|\tilde{S}}^{N_{\tilde{s}}}$,
\begin{equation}\label{appen7g}
|c^{-1}(k)|\leq \frac{|T_{Y|V,S,\tilde{S},Z}^{N_{\tilde{s}}}(\widetilde{v}_{i-d}^{N_{\tilde{s}}},
\widetilde{s}_{i-d}^{N_{\tilde{s}}},\tilde{s},\widetilde{z}_{i-d}^{N_{\tilde{s}}})|(1+\delta)}{\gamma},
\end{equation}
for $k=1,2,...,\gamma$, where $c^{-1}$ is the inverse image of $c$.
\end{lemma}
\begin{IEEEproof}
See \cite[p. 260]{AC}.
\end{IEEEproof}
Lemma \ref{Lx} shows that given $\tilde{S}=\tilde{s}$, if $\widetilde{v}_{i-d}^{N_{\tilde{s}}}$, $\widetilde{s}_{i-d}^{N_{\tilde{s}}}$,
$\widetilde{y}_{i-d}^{N_{\tilde{s}}}$
and $\widetilde{z}_{i-d}^{N_{\tilde{s}}}$ are jointly typical, for given $\widetilde{v}_{i-d}^{N_{\tilde{s}}}$, $\widetilde{s}_{i-d}^{N_{\tilde{s}}}$
and $\widetilde{z}_{i-d}^{N_{\tilde{s}}}$, the number of
$\widetilde{y}_{i-d}^{N_{\tilde{s}}}\in T_{Y|V,S,\tilde{S},Z}^{N_{\tilde{s}}}(\widetilde{v}_{i-d}^{N_{\tilde{s}}},
\widetilde{s}_{i-d}^{N_{\tilde{s}}},\tilde{s},\widetilde{z}_{i-d}^{N_{\tilde{s}}})$ for a certain color $k$ ($k=1,2,...,\gamma$), which is denoted as
$|c^{-1}(k)|$, is upper bounded by $\frac{|T_{Y|V,S,\tilde{S},Z}^{N_{\tilde{s}}}(\widetilde{v}_{i-d}^{N_{\tilde{s}}},
\widetilde{s}_{i-d}^{N_{\tilde{s}}},\tilde{s},\widetilde{z}_{i-d}^{N_{\tilde{s}}})|(1+\delta)}{\gamma}$.
By using Lemma \ref{Lx}, it is easy to see that the typical set $T_{Y|V,S,\tilde{S},Z}^{N_{\tilde{s}}}(\widetilde{v}_{i-d}^{N_{\tilde{s}}},
\widetilde{s}_{i-d}^{N_{\tilde{s}}},\tilde{s},\widetilde{z}_{i-d}^{N_{\tilde{s}}})$ maps into at least
\begin{eqnarray}\label{appen8g}
&&\frac{|T_{Y|V,S,\tilde{S},Z}^{N_{\tilde{s}}}(\widetilde{v}_{i-d}^{N_{\tilde{s}}},
\widetilde{s}_{i-d}^{N_{\tilde{s}}},\tilde{s},\widetilde{z}_{i-d}^{N_{\tilde{s}}})|}{\frac{|T_{Y|V,S,\tilde{S},Z}^{N_{\tilde{s}}}(\widetilde{v}_{i-d}^{N_{\tilde{s}}},
\widetilde{s}_{i-d}^{N_{\tilde{s}}},\tilde{s},\widetilde{z}_{i-d}^{N_{\tilde{s}}})|(1+\delta)}{\gamma}}=\frac{\gamma}{1+\delta}
\end{eqnarray}
colors. On the other hand, the typical set $T_{Y|V,S,\tilde{S},Z}^{N_{\tilde{s}}}(\widetilde{v}_{i-d}^{N_{\tilde{s}}},
\widetilde{s}_{i-d}^{N_{\tilde{s}}},\tilde{s},\widetilde{z}_{i-d}^{N_{\tilde{s}}})$ maps into at most $\gamma$ colors.
Thus, given $\tilde{S}=\tilde{s}$, $\widetilde{V}_{i-d}^{N_{\tilde{s}}}$, $\widetilde{Z}_{i-d}^{N_{\tilde{s}}}$, $\widetilde{S}_{i-d}^{N_{\tilde{s}}}$,
the eavesdropper's equivocation $H(K_{i,\tilde{s}}^{*}|\widetilde{V}_{i-d}^{N_{\tilde{s}}},
\widetilde{S}_{i-d}^{N_{\tilde{s}}},\widetilde{Z}_{i-d}^{N_{\tilde{s}}})$ about the secret key $K_{i,\tilde{s}}^{*}$
is lower bounded by
\begin{eqnarray}\label{newbee1}
&&H(K_{i,\tilde{s}}^{*}|\widetilde{V}_{i-d}^{N_{\tilde{s}}},\widetilde{S}_{i-d}^{N_{\tilde{s}}},\widetilde{Z}_{i-d}^{N_{\tilde{s}}}
,\tilde{S}=\tilde{s})\geq \log\frac{\gamma}{1+\delta}.
\end{eqnarray}
Here note that in our encoding scheme, $\gamma=2^{N_{\tilde{s}}R_{f}(\tilde{s})}$ for case 1, and $\gamma=2^{N_{\tilde{s}}R^{*}_{f}(\tilde{s})}$ for case 2,
see (\ref{dead1.1}) and (\ref{dead2.1}). Then, it is easy to see that (\ref{newbee1}) can be re-written as follows.
For case 1,
\begin{eqnarray}\label{newbee2}
&&H(K_{i,\tilde{s}}^{*}|\widetilde{V}_{i-d}^{N_{\tilde{s}}},\widetilde{S}_{i-d}^{N_{\tilde{s}}},\widetilde{Z}_{i-d}^{N_{\tilde{s}}}
,\tilde{S}=\tilde{s})\geq N_{\tilde{s}}R_{f}(\tilde{s})-\log(1+\delta),
\end{eqnarray}
and for case 2,
\begin{eqnarray}\label{newbee3}
&&H(K_{i,\tilde{s}}^{*}|\widetilde{V}_{i-d}^{N_{\tilde{s}}},\widetilde{S}_{i-d}^{N_{\tilde{s}}},\widetilde{Z}_{i-d}^{N_{\tilde{s}}},
\tilde{S}=\tilde{s})\geq N_{\tilde{s}}R^{*}_{f}(\tilde{s})-\log(1+\delta).
\end{eqnarray}

\end{itemize}

\end{itemize}

Now it remains to show that $R_{e}=I(V;Y|U,S,\tilde{S})-I(V;Z|U,S,\tilde{S})+R_{f})$ for case 1 and $R_{e}=R_{f}^{*}$ for case 2, see
the followings.

\subsection{Equivocation analysis:}\label{sub-x5}

\subsubsection*{Equivocation analysis for case 1}\label{sub-x5.1}

For all blocks, the equivocation $\Delta$ is bounded by
\begin{eqnarray}\label{deadgame1}
\Delta&=&\frac{1}{nN}H(W|Z^{n},S^{n})=\frac{1}{nN}H(W_{c},W_{p}|Z^{n},S^{n})\nonumber\\
&\geq&\frac{1}{nN}H(W_{p}|Z^{n},S^{n},W_{c})\geq\frac{1}{nN}H(W_{p}|Z^{n},S^{n},W_{c},U^{n})\nonumber\\
&\stackrel{(a)}=&\frac{1}{nN}H(W_{p}|Z^{n},S^{n},U^{n})=\frac{1}{nN}H(W_{1,p},...,W_{n,p}|Z^{n},S^{n},U^{n})\nonumber\\
&=&\frac{1}{nN}\sum_{i=1}^{n}H(W_{i,p}|Z^{n},S^{n},U^{n},W_{1,p},...,W_{i-1,p})\nonumber\\
&=&\frac{1}{nN}(\sum_{i=1}^{2d}H(W_{i,p}|Z^{n},S^{n},U^{n},W_{1,p},...,W_{i-1,p})\nonumber\\
&&+\sum_{i=2d+1}^{n}H(W_{i,p}|Z^{n},S^{n},U^{n},W_{1,p},...,W_{i-1,p}))\nonumber\\
&\stackrel{(b)}=&\frac{1}{nN}(\sum_{i=1}^{2d}H(W_{i,p}|\widetilde{Z}_{i},\widetilde{S}_{i},\widetilde{U}_{i})
+\sum_{i=2d+1}^{n}H(W_{i,p}|\widetilde{Z}_{i},\widetilde{S}_{i},\widetilde{U}_{i},\widetilde{Z}_{i-d},\widetilde{S}_{i-d},\widetilde{U}_{i-d}))\nonumber\\
&\stackrel{(c)}\geq&\frac{1}{nN}\sum_{i=2d+1}^{n}H(W_{i,p}|\widetilde{Z}_{i},\widetilde{S}_{i},
\widetilde{U}_{i},\widetilde{Z}_{i-d},\widetilde{S}_{i-d},\widetilde{U}_{i-d})\nonumber\\
&=&\frac{1}{nN}\sum_{i=2d+1}^{n}\sum_{\tilde{s}=1}^{k}H(W_{i,\tilde{s},p}|W_{i,1,p},...,W_{i,\tilde{s}-1,p},
\widetilde{Z}_{i},\widetilde{S}_{i},\widetilde{U}_{i},\widetilde{Z}_{i-d},\widetilde{S}_{i-d},\widetilde{U}_{i-d})\nonumber\\
&\stackrel{(d)}=&\frac{1}{nN}\sum_{i=2d+1}^{n}\sum_{\tilde{s}=1}^{k}H(W_{i,\tilde{s},p}|\widetilde{Z}^{N_{\tilde{s}}}_{i},
\widetilde{S}^{N_{\tilde{s}}}_{i},\widetilde{U}^{N_{\tilde{s}}}_{i},
\widetilde{Z}^{N_{\tilde{s}}}_{i-d},\widetilde{S}^{N_{\tilde{s}}}_{i-d},\widetilde{U}^{N_{\tilde{s}}}_{i-d})\nonumber\\
&=&\frac{1}{nN}\sum_{i=2d+1}^{n}\sum_{\tilde{s}=1}^{k}H(W_{i,\tilde{s},p,1},W_{i,\tilde{s},p,2}|\widetilde{Z}^{N_{\tilde{s}}}_{i},\widetilde{S}^{N_{\tilde{s}}}_{i},
\widetilde{U}^{N_{\tilde{s}}}_{i},
\widetilde{Z}^{N_{\tilde{s}}}_{i-d},\widetilde{S}^{N_{\tilde{s}}}_{i-d},\widetilde{U}^{N_{\tilde{s}}}_{i-d})\nonumber\\
&\stackrel{(e)}=&\frac{1}{nN}\sum_{i=2d+1}^{n}\sum_{\tilde{s}=1}^{k}(H(W_{i,\tilde{s},p,1}|\widetilde{Z}^{N_{\tilde{s}}}_{i},\widetilde{S}^{N_{\tilde{s}}}_{i},
\widetilde{U}^{N_{\tilde{s}}}_{i})\nonumber\\
&&+H(W_{i,\tilde{s},p,2}|W_{i,\tilde{s},p,1},\widetilde{Z}^{N_{\tilde{s}}}_{i},\widetilde{S}^{N_{\tilde{s}}}_{i},
\widetilde{U}^{N_{\tilde{s}}}_{i},
\widetilde{Z}^{N_{\tilde{s}}}_{i-d},\widetilde{S}^{N_{\tilde{s}}}_{i-d},\widetilde{U}^{N_{\tilde{s}}}_{i-d})),
\end{eqnarray}
where (a) is from the definition $W_{i,p}=(W_{i,1,p},W_{i,2,p},...,W_{i,k,p})$ ($1\leq i\leq n$), (b) is from the Markov chains
$W_{i,p}\rightarrow (\widetilde{Z}_{i},\widetilde{S}_{i},\widetilde{U}_{i})\rightarrow (W_{1,p},...,W_{i-1,p},
\widetilde{Z}^{i-1},\widetilde{Z}_{i+1}^{n},\widetilde{U}^{i-1},\widetilde{U}_{i+1}^{n},\widetilde{S}^{i-1},\widetilde{S}_{i+1}^{n})$ for block $1\leq i\leq 2d$,
and $W_{i,p}\rightarrow (\widetilde{Z}_{i},\widetilde{S}_{i},\widetilde{U}_{i},\widetilde{Z}_{i-d},\widetilde{S}_{i-d},\widetilde{U}_{i-d})
\rightarrow (W_{1,p},...,W_{i-1,p},
\widetilde{Z}^{i-d-1},\widetilde{Z}_{i-d+1}^{i-1},\widetilde{Z}_{i+1}^{n},
\widetilde{U}^{i-d-1},\widetilde{U}_{i-d+1}^{i-1},\widetilde{U}_{i+1}^{n},
\widetilde{S}^{i-d-1},\widetilde{S}_{i-d+1}^{i-1},\widetilde{S}_{i+1}^{n})$ for block $2d+1\leq i\leq n$,
(c) is from the fact that when $n$ and $N$ tend to infinity, $\frac{1}{nN}\sum_{i=1}^{2d}H(W_{i,p}|\widetilde{Z}_{i},\widetilde{S}_{i},\widetilde{U}_{i})$
tends to zero, and thus we can drop it, (d) is from the Markov chain $W_{i,\tilde{s},p}\rightarrow
(\widetilde{Z}^{N_{\tilde{s}}}_{i},
\widetilde{S}^{N_{\tilde{s}}}_{i},\widetilde{U}^{N_{\tilde{s}}}_{i},
\widetilde{Z}^{N_{\tilde{s}}}_{i-d},\widetilde{S}^{N_{\tilde{s}}}_{i-d},\widetilde{U}^{N_{\tilde{s}}}_{i-d})\rightarrow
(W_{i,1,p},...,W_{i,\tilde{s}-1,p},\widetilde{Z}^{N_{1}}_{i},...,\widetilde{Z}^{N_{\tilde{s}-1}}_{i},\widetilde{Z}^{N_{\tilde{s}+1}}_{i},...,
\widetilde{Z}^{N_{k}}_{i},\widetilde{U}^{N_{1}}_{i},...,\widetilde{U}^{N_{\tilde{s}-1}}_{i},\widetilde{U}^{N_{\tilde{s}+1}}_{i},...,
\widetilde{U}^{N_{k}}_{i},\widetilde{S}^{N_{1}}_{i},...,\widetilde{S}^{N_{\tilde{s}-1}}_{i},\widetilde{S}^{N_{\tilde{s}+1}}_{i},...,
\widetilde{S}^{N_{k}}_{i},\\ \widetilde{Z}^{N_{1}}_{i-d},...,\widetilde{Z}^{N_{\tilde{s}-1}}_{i-d},\widetilde{Z}^{N_{\tilde{s}+1}}_{i-d},...,
\widetilde{Z}^{N_{k}}_{i-d},\widetilde{U}^{N_{1}}_{i-d},...,\widetilde{U}^{N_{\tilde{s}-1}}_{i-d},\widetilde{U}^{N_{\tilde{s}+1}}_{i-d},...,
\widetilde{U}^{N_{k}}_{i-d},\widetilde{S}^{N_{1}}_{i-d},...,\widetilde{S}^{N_{\tilde{s}-1}}_{i-d},\widetilde{S}^{N_{\tilde{s}+1}}_{i-d},...,
\widetilde{S}^{N_{k}}_{i-d})$, which implies the $\tilde{s}$-th component of the private
message $W_{i,p}$ is only related with the $\tilde{s}$-th component of $\widetilde{U}_{i}$, $\widetilde{S}_{i}$, $\widetilde{Z}_{i}$,
$\widetilde{U}_{i-d}$, $\widetilde{S}_{i-d}$ and $\widetilde{Z}_{i-d}$, and (e) is from the Markov chain $W_{i,\tilde{s},p,1}\rightarrow
(\widetilde{Z}^{N_{\tilde{s}}}_{i},\widetilde{S}^{N_{\tilde{s}}}_{i},
\widetilde{U}^{N_{\tilde{s}}}_{i})\rightarrow (\widetilde{Z}^{N_{\tilde{s}}}_{i-d},\widetilde{S}^{N_{\tilde{s}}}_{i-d},\widetilde{U}^{N_{\tilde{s}}}_{i-d})$.

Now it remains for us to bound the conditional entropies $H(W_{i,\tilde{s},p,1}|\widetilde{Z}^{N_{\tilde{s}}}_{i},\widetilde{S}^{N_{\tilde{s}}}_{i},
\widetilde{U}^{N_{\tilde{s}}}_{i})$ and $H(W_{i,\tilde{s},p,2}|W_{i,\tilde{s},p,1},\widetilde{Z}^{N_{\tilde{s}}}_{i},\\ \widetilde{S}^{N_{\tilde{s}}}_{i},
\widetilde{U}^{N_{\tilde{s}}}_{i},
\widetilde{Z}^{N_{\tilde{s}}}_{i-d},\widetilde{S}^{N_{\tilde{s}}}_{i-d},\widetilde{U}^{N_{\tilde{s}}}_{i-d})$
in (\ref{deadgame1}), see the followings.

The conditional entropy $H(W_{i,\tilde{s},p,1}|\widetilde{Z}^{N_{\tilde{s}}}_{i},\widetilde{S}^{N_{\tilde{s}}}_{i},
\widetilde{U}^{N_{\tilde{s}}}_{i})$ can be bounded by
\begin{eqnarray}\label{deadgame2}
&&H(W_{i,\tilde{s},p,1}|\widetilde{Z}^{N_{\tilde{s}}}_{i},\widetilde{S}^{N_{\tilde{s}}}_{i},\widetilde{U}^{N_{\tilde{s}}}_{i})\geq
H(W_{i,\tilde{s},p,1}|\widetilde{Z}^{N_{\tilde{s}}}_{i},\widetilde{S}^{N_{\tilde{s}}}_{i},\widetilde{U}^{N_{\tilde{s}}}_{i},\tilde{S}=\tilde{s})\nonumber\\
&&=H(W_{i,\tilde{s},p,1},\widetilde{Z}^{N_{\tilde{s}}}_{i},\widetilde{S}^{N_{\tilde{s}}}_{i},\widetilde{U}^{N_{\tilde{s}}}_{i},\tilde{S}=\tilde{s})-
H(\widetilde{Z}^{N_{\tilde{s}}}_{i},\widetilde{S}^{N_{\tilde{s}}}_{i},\widetilde{U}^{N_{\tilde{s}}}_{i},\tilde{S}=\tilde{s})\nonumber\\
&&=H(\widetilde{V}^{N_{\tilde{s}}}_{i},W_{i,\tilde{s},p,1},\widetilde{Z}^{N_{\tilde{s}}}_{i},\widetilde{S}^{N_{\tilde{s}}}_{i},\widetilde{U}^{N_{\tilde{s}}}_{i},\tilde{S}=\tilde{s})
-H(\widetilde{V}^{N_{\tilde{s}}}_{i}|W_{i,\tilde{s},p,1},\widetilde{Z}^{N_{\tilde{s}}}_{i},\widetilde{S}^{N_{\tilde{s}}}_{i},\widetilde{U}^{N_{\tilde{s}}}_{i},\tilde{S}=\tilde{s})
-H(\widetilde{Z}^{N_{\tilde{s}}}_{i},\widetilde{S}^{N_{\tilde{s}}}_{i},\widetilde{U}^{N_{\tilde{s}}}_{i},\tilde{S}=\tilde{s})\nonumber\\
&&\stackrel{(f)}=H(\widetilde{Z}^{N_{\tilde{s}}}_{i}|\widetilde{V}^{N_{\tilde{s}}}_{i},\widetilde{S}^{N_{\tilde{s}}}_{i},\widetilde{U}^{N_{\tilde{s}}}_{i},\tilde{S}=\tilde{s})
+H(\widetilde{V}^{N_{\tilde{s}}}_{i},W_{i,\tilde{s},p,1},\widetilde{S}^{N_{\tilde{s}}}_{i},\widetilde{U}^{N_{\tilde{s}}}_{i},\tilde{S}=\tilde{s})\nonumber\\
&&-H(\widetilde{V}^{N_{\tilde{s}}}_{i}|W_{i,\tilde{s},p,1},\widetilde{Z}^{N_{\tilde{s}}}_{i},\widetilde{S}^{N_{\tilde{s}}}_{i},\widetilde{U}^{N_{\tilde{s}}}_{i},\tilde{S}=\tilde{s})
-H(\widetilde{Z}^{N_{\tilde{s}}}_{i},\widetilde{S}^{N_{\tilde{s}}}_{i},\widetilde{U}^{N_{\tilde{s}}}_{i},\tilde{S}=\tilde{s})\nonumber\\
&&\stackrel{(g)}=N_{\tilde{s}}H(Z|V,U,S,\tilde{S}=\tilde{s})+H(\widetilde{V}^{N_{\tilde{s}}}_{i}|\widetilde{S}^{N_{\tilde{s}}}_{i},\widetilde{U}^{N_{\tilde{s}}}_{i},\tilde{S}=\tilde{s})
-H(\widetilde{Z}^{N_{\tilde{s}}}_{i}|\widetilde{S}^{N_{\tilde{s}}}_{i},\widetilde{U}^{N_{\tilde{s}}}_{i},\tilde{S}=\tilde{s})\nonumber\\
&&-H(\widetilde{V}^{N_{\tilde{s}}}_{i}|W_{i,\tilde{s},p,1},\widetilde{Z}^{N_{\tilde{s}}}_{i},\widetilde{S}^{N_{\tilde{s}}}_{i},
\widetilde{U}^{N_{\tilde{s}}}_{i},\tilde{S}=\tilde{s})\nonumber\\
&&\geq N_{\tilde{s}}H(Z|V,U,S,\tilde{S}=\tilde{s})+H(\widetilde{V}^{N_{\tilde{s}}}_{i}|\widetilde{S}^{N_{\tilde{s}}}_{i},\widetilde{U}^{N_{\tilde{s}}}_{i},\tilde{S}=\tilde{s})
-N_{\tilde{s}}H(Z|U,S,\tilde{S}=\tilde{s})\nonumber\\
&&-H(\widetilde{V}^{N_{\tilde{s}}}_{i}|W_{i,\tilde{s},p,1},\widetilde{Z}^{N_{\tilde{s}}}_{i},\widetilde{S}^{N_{\tilde{s}}}_{i},
\widetilde{U}^{N_{\tilde{s}}}_{i},\tilde{S}=\tilde{s})\nonumber\\
&&\stackrel{(h)}\geq N_{\tilde{s}}I(V;Y|U,S,\tilde{S}=\tilde{s})-1-N_{\tilde{s}}I(V;Z|U,S,\tilde{S}=\tilde{s})
-H(\widetilde{V}^{N_{\tilde{s}}}_{i}|W_{i,\tilde{s},p,1},\widetilde{Z}^{N_{\tilde{s}}}_{i},\widetilde{S}^{N_{\tilde{s}}}_{i},
\widetilde{U}^{N_{\tilde{s}}}_{i},\tilde{S}=\tilde{s})\nonumber\\
&&\stackrel{(i)}\geq N_{\tilde{s}}I(V;Y|U,S,\tilde{S}=\tilde{s})-1-N_{\tilde{s}}I(V;Z|U,S,\tilde{S}=\tilde{s})
-N_{\tilde{s}}\epsilon_{1},
\end{eqnarray}
where (f) is from the fact that $H(W_{i,\tilde{s},p,1}|\widetilde{V}^{N_{\tilde{s}}}_{i})=0$, (g) is also from
$H(W_{i,\tilde{s},p,1}|\widetilde{V}^{N_{\tilde{s}}}_{i})=0$ and
the fact that
the channel is a DMC with transition probability $P_{Y,Z|X,S}(y,z|x,s)$, and for each $\tilde{s}$,
$X^{N_{\tilde{s}}}$ is i.i.d. generated according to a new DMC with transition probability $P_{X|U,V,\tilde{S}}(x|u,v,\tilde{s})$, thus we have
$H(\widetilde{Z}^{N_{\tilde{s}}}_{i}|\widetilde{V}^{N_{\tilde{s}}}_{i},\widetilde{S}^{N_{\tilde{s}}}_{i},\widetilde{U}^{N_{\tilde{s}}}_{i},\tilde{S}=\tilde{s})
=N_{\tilde{s}}H(Z|V,U,S,\tilde{S}=\tilde{s})$, (h) is from the fact that for given $\tilde{s}$, $\widetilde{u}^{N_{\tilde{s}}}_{i}$
and $\widetilde{s}^{N_{\tilde{s}}}_{i}$, $\widetilde{V}^{N_{\tilde{s}}}_{i}$ has $A_{\tilde{s}}\cdot B_{\tilde{s}}$
possible values, using a similar lemma in \cite{CK}, we have
\begin{eqnarray}\label{rotk1}
&&H(\widetilde{V}^{N_{\tilde{s}}}_{i}|\widetilde{S}^{N_{\tilde{s}}}_{i},\widetilde{U}^{N_{\tilde{s}}}_{i},\tilde{S}=\tilde{s})\geq
\log A_{\tilde{s}}+\log B_{\tilde{s}}-1\stackrel{(1)}=N_{\tilde{s}}I(V;Y|U,S,\tilde{S}=\tilde{s})-1,
\end{eqnarray}
where (1) is from (\ref{swjtu.3}) and (\ref{swjtu.4}), and (i) is from the fact that given $\tilde{s}$,
$w_{i,\tilde{s},p,1}$, $\widetilde{z}^{N_{\tilde{s}}}_{i}$, $\widetilde{s}^{N_{\tilde{s}}}_{i}$ and
$\widetilde{u}^{N_{\tilde{s}}}_{i}$, the eavesdropper's decoding error probability of $\widetilde{v}^{N_{\tilde{s}}}_{i}$ tends
to zero (see (\ref{dead4.3})), then, by using Fano's inequality, we have
\begin{eqnarray}\label{rotk2.1}
&&\frac{1}{N_{\tilde{s}}}H(\widetilde{V}^{N_{\tilde{s}}}_{i}|W_{i,\tilde{s},p,1},\widetilde{Z}^{N_{\tilde{s}}}_{i},\widetilde{S}^{N_{\tilde{s}}}_{i},
\widetilde{U}^{N_{\tilde{s}}}_{i},\tilde{S}=\tilde{s})\leq \epsilon_{1},
\end{eqnarray}
where $\epsilon_{1}\rightarrow 0$ as $N_{\tilde{s}}\rightarrow \infty$.

The conditional entropy $H(W_{i,\tilde{s},p,2}|W_{i,\tilde{s},p,1},\widetilde{Z}^{N_{\tilde{s}}}_{i},\widetilde{S}^{N_{\tilde{s}}}_{i},
\widetilde{U}^{N_{\tilde{s}}}_{i},
\widetilde{Z}^{N_{\tilde{s}}}_{i-d},\widetilde{S}^{N_{\tilde{s}}}_{i-d},\widetilde{U}^{N_{\tilde{s}}}_{i-d})$
can be bounded by
\begin{eqnarray}\label{rotk3}
&&H(W_{i,\tilde{s},p,2}|W_{i,\tilde{s},p,1},\widetilde{Z}^{N_{\tilde{s}}}_{i},\widetilde{S}^{N_{\tilde{s}}}_{i},\widetilde{U}^{N_{\tilde{s}}}_{i},
\widetilde{Z}^{N_{\tilde{s}}}_{i-d},\widetilde{S}^{N_{\tilde{s}}}_{i-d},\widetilde{U}^{N_{\tilde{s}}}_{i-d})\nonumber\\
&&\geq H(W_{i,\tilde{s},p,2}|W_{i,\tilde{s},p,1},\widetilde{Z}^{N_{\tilde{s}}}_{i},\widetilde{S}^{N_{\tilde{s}}}_{i},\widetilde{U}^{N_{\tilde{s}}}_{i},
\widetilde{Z}^{N_{\tilde{s}}}_{i-d},\widetilde{S}^{N_{\tilde{s}}}_{i-d},\widetilde{U}^{N_{\tilde{s}}}_{i-d},W^{*}_{i,\tilde{s},p,2}\oplus K_{i,\tilde{s}}^{*},
\tilde{S}=\tilde{s},\widetilde{V}^{N_{\tilde{s}}}_{i},\widetilde{V}^{N_{\tilde{s}}}_{i-d})\nonumber\\
&&\stackrel{(j)}=H(W_{i,\tilde{s},p,2}|
\widetilde{Z}^{N_{\tilde{s}}}_{i-d},\widetilde{S}^{N_{\tilde{s}}}_{i-d},\widetilde{U}^{N_{\tilde{s}}}_{i-d},W^{*}_{i,\tilde{s},p,2}\oplus K_{i,\tilde{s}}^{*},
\tilde{S}=\tilde{s},\widetilde{V}^{N_{\tilde{s}}}_{i-d})\nonumber\\
&&\stackrel{(k)}=H(W_{i,\tilde{s},p,2}|
\widetilde{Z}^{N_{\tilde{s}}}_{i-d},\widetilde{S}^{N_{\tilde{s}}}_{i-d},W^{*}_{i,\tilde{s},p,2}\oplus K_{i,\tilde{s}}^{*},
\tilde{S}=\tilde{s},\widetilde{V}^{N_{\tilde{s}}}_{i-d})\nonumber\\
&&=H(K_{i,\tilde{s}}^{*}|
\widetilde{Z}^{N_{\tilde{s}}}_{i-d},\widetilde{S}^{N_{\tilde{s}}}_{i-d},W^{*}_{i,\tilde{s},p,2}\oplus K_{i,\tilde{s}}^{*},
\tilde{S}=\tilde{s},\widetilde{V}^{N_{\tilde{s}}}_{i-d})\nonumber\\
&&\stackrel{(l)}=H(K_{i,\tilde{s}}^{*}|
\widetilde{Z}^{N_{\tilde{s}}}_{i-d},\widetilde{S}^{N_{\tilde{s}}}_{i-d},\widetilde{V}^{N_{\tilde{s}}}_{i-d},\tilde{S}=\tilde{s})\nonumber\\
&&\stackrel{(m)}\geq N_{\tilde{s}}R_{f}(\tilde{s})-\log(1+\delta),
\end{eqnarray}
where (j) is from the Markov chain $W_{i,\tilde{s},p,2}\rightarrow (\widetilde{Z}^{N_{\tilde{s}}}_{i-d},\widetilde{S}^{N_{\tilde{s}}}_{i-d},
\widetilde{U}^{N_{\tilde{s}}}_{i-d},W^{*}_{i,\tilde{s},p,2}\oplus K_{i,\tilde{s}}^{*},
\tilde{S}=\tilde{s},\widetilde{V}^{N_{\tilde{s}}}_{i-d})\rightarrow (W_{i,\tilde{s},p,1},\widetilde{Z}^{N_{\tilde{s}}}_{i},
\widetilde{S}^{N_{\tilde{s}}}_{i},\\ \widetilde{U}^{N_{\tilde{s}}}_{i},\widetilde{V}^{N_{\tilde{s}}}_{i})$, (k) is from the fact that
$H(\widetilde{U}^{N_{\tilde{s}}}_{i-d}|\widetilde{V}^{N_{\tilde{s}}}_{i-d})=0$, (l) is from the Markov chain
$W^{*}_{i,\tilde{s},p,2}\oplus K_{i,\tilde{s}}^{*}\rightarrow
(\widetilde{Z}^{N_{\tilde{s}}}_{i-d},\widetilde{S}^{N_{\tilde{s}}}_{i-d},\widetilde{V}^{N_{\tilde{s}}}_{i-d},\tilde{S}=\tilde{s})
\rightarrow K_{i,\tilde{s}}^{*}$, and (m) is from (\ref{newbee2}).

Substituting (\ref{deadgame2}) and (\ref{rotk3}) into (\ref{deadgame1}), we have
\begin{eqnarray}\label{rotk4}
\Delta&\geq&\frac{1}{nN}\sum_{i=2d+1}^{n}\sum_{\tilde{s}=1}^{k}[N_{\tilde{s}}I(V;Y|U,S,\tilde{S}=\tilde{s})-1-N_{\tilde{s}}I(V;Z|U,S,\tilde{S}=\tilde{s})
-N_{\tilde{s}}\epsilon_{1}+N_{\tilde{s}}R_{f}(\tilde{s})-\log(1+\delta)]\nonumber\\
&=&\frac{1}{nN}\sum_{i=2d+1}^{n}\sum_{\tilde{s}=1}^{k}[N_{\tilde{s}}(I(V;Y|U,S,\tilde{S}=\tilde{s})-I(V;Z|U,S,\tilde{S}=\tilde{s})+R_{f}(\tilde{s})-\epsilon_{1})
-1-\log(1+\delta)]\nonumber\\
&\stackrel{(n)}=&\frac{1}{nN}\sum_{i=2d+1}^{n}\sum_{\tilde{s}=1}^{k}[N(\pi(\tilde{s})-\epsilon^{'})(I(V;Y|U,S,\tilde{S}=\tilde{s})-I(V;Z|U,S,\tilde{S}=\tilde{s})
+R_{f}(\tilde{s})-\epsilon_{1})-1-\log(1+\delta)]\nonumber\\
&=&\frac{n-2d}{nN}\sum_{\tilde{s}=1}^{k}[N\pi(\tilde{s})(I(V;Y|U,S,\tilde{S}=\tilde{s})-I(V;Z|U,S,\tilde{S}=\tilde{s})+R_{f}(\tilde{s}))
-N\pi(\tilde{s})\epsilon_{1}\nonumber\\
&&-N\epsilon^{'}(I(V;Y|U,S,\tilde{S}=\tilde{s})-I(V;Z|U,S,\tilde{S}=\tilde{s})+R_{f}(\tilde{s}))+N\epsilon^{'}\epsilon_{1}
-1-\log(1+\delta)]\nonumber\\
&\stackrel{(o)}=&I(V;Y|U,S,\tilde{S})-I(V;Z|U,S,\tilde{S})+R_{f}
-\frac{2d}{n}(I(V;Y|U,S,\tilde{S})-I(V;Z|U,S,\tilde{S})+R_{f})-\frac{n-2d}{n}\epsilon_{1}\sum_{\tilde{s}=1}^{k}\pi(\tilde{s})\nonumber\\
&&-\frac{n-2d}{n}\epsilon^{'}\sum_{\tilde{s}=1}^{k}(I(V;Y|U,S,\tilde{S}=\tilde{s})-I(V;Z|U,S,\tilde{S}=\tilde{s})+R_{f}(\tilde{s}))\nonumber\\
&&+\frac{n-2d}{n}k(\epsilon^{'}\epsilon_{1}-\frac{1+\log(1+\delta)}{N}),
\end{eqnarray}
where (n) is from (\ref{swjtu.1}), and (o) is from (\ref{dead1}). Thus, choosing sufficiently large $n$ and $N$ (here note that
$\epsilon^{'}$ and $\epsilon_{1}$ tend to zero while $N\rightarrow \infty$),
$\Delta\geq I(V;Y|U,S,\tilde{S})-I(V;Z|U,S,\tilde{S})+R_{f}-\epsilon$ is proved.

\subsubsection*{Equivocation analysis for case 2}\label{sub-x5.2}

For the case 2, (\ref{final-2.x}) implies that the private message $W_{i,p,1}=(W_{i,1,p,1},...,W_{i,k,p,1})$ of block $i$ is a constant, and thus
the conditional entropy $H(W_{i,\tilde{s},p,1}|\widetilde{Z}^{N_{\tilde{s}}}_{i},\widetilde{S}^{N_{\tilde{s}}}_{i},
\widetilde{U}^{N_{\tilde{s}}}_{i})$ of (\ref{deadgame1}) satisfies
\begin{eqnarray}\label{rotk5}
&&H(W_{i,\tilde{s},p,1}|\widetilde{Z}^{N_{\tilde{s}}}_{i},\widetilde{S}^{N_{\tilde{s}}}_{i},
\widetilde{U}^{N_{\tilde{s}}}_{i})=0.
\end{eqnarray}
Moreover, using (\ref{newbee3}), the last step of (\ref{rotk3}) can be re-written by
\begin{eqnarray}\label{rotk6}
&&H(W_{i,\tilde{s},p,2}|W_{i,\tilde{s},p,1},\widetilde{Z}^{N_{\tilde{s}}}_{i},\widetilde{S}^{N_{\tilde{s}}}_{i},\widetilde{U}^{N_{\tilde{s}}}_{i},
\widetilde{Z}^{N_{\tilde{s}}}_{i-d},\widetilde{S}^{N_{\tilde{s}}}_{i-d},\widetilde{U}^{N_{\tilde{s}}}_{i-d})\nonumber\\
&&\geq N_{\tilde{s}}R^{*}_{f}(\tilde{s})-\log(1+\delta).
\end{eqnarray}
Substituting (\ref{rotk5}) and (\ref{rotk6}) into (\ref{deadgame1}), we have
\begin{eqnarray}\label{rotk4}
\Delta&\geq&\frac{1}{nN}\sum_{i=2d+1}^{n}\sum_{\tilde{s}=1}^{k}(N_{\tilde{s}}R^{*}_{f}(\tilde{s})-\log(1+\delta))\nonumber\\
&=&\frac{1}{nN}\sum_{i=2d+1}^{n}\sum_{\tilde{s}=1}^{k}(N(\pi(\tilde{s})-\epsilon^{'})R^{*}_{f}(\tilde{s})-\log(1+\delta))\nonumber\\
&=&\frac{n-2d}{nN}(N\sum_{\tilde{s}=1}^{k}\pi(\tilde{s})R^{*}_{f}(\tilde{s})-N\epsilon^{'}\sum_{\tilde{s}=1}^{k}R^{*}_{f}(\tilde{s})-k\log(1+\delta))\nonumber\\
&\stackrel{(1)}=&\frac{n-2d}{n}R^{*}_{f}-\frac{n-2d}{n}\epsilon^{'}\sum_{\tilde{s}=1}^{k}R^{*}_{f}(\tilde{s})-\frac{n-2d}{n}\frac{\log(1+\delta)}{N}k,
\end{eqnarray}
where (1) is from (\ref{dead2}). Thus, choosing sufficiently large $n$ and $N$ (here note that
$\epsilon^{'}$ tends to zero while $N\rightarrow \infty$),
$\Delta\geq R^{*}_{f}-\epsilon$ is proved.

Thus, the achievability proof of $\mathcal{R}^{fi\diamond}$ for both cases are completed.
Finally, using Fourier-Motzkin elimination to eliminate $R_{c}$ and $R_{p}$ from $\mathcal{R}^{fi\diamond}$,
$\mathcal{R}^{fi}$ is obtained.
The proof of Theorem \ref{T1} is completed.

\section{Proof of Theorem \ref{T1.1}\label{appen2.xx}}

Since $R_{e}\leq R$ is obvious, we only need to prove the
inequalities $R\leq I(V;Y|S,\tilde{S})$ and $R_{e}\leq H(Y|Z,U,S,\tilde{S})$.
Define the auxiliary random variables $U$, $V$, $X$, $S$, $\tilde{S}$, $Y$ and $Z$ the same as those in (\ref{jmds1}).
Then it is easy to see that the proof of $R\leq I(V;Y|S,\tilde{S})$ is exactly the same as that in (\ref{jmds2}). Now it
remains to show $R_{e}\leq H(Y|Z,U,S,\tilde{S})$, see the followings.

By using (\ref{e210}) and (\ref{e211}), we have
\begin{eqnarray}\label{jmds3.xx}
R_{e}-\epsilon&\stackrel{(1)}\leq&\frac{1}{N}H(W|Z^{N},S^{N})\nonumber\\
&=&\frac{1}{N}(H(W|Z^{N},S^{N})-H(W|Z^{N},S^{N},Y^{N})+H(W|Z^{N},S^{N},Y^{N}))\nonumber\\
&\stackrel{(2)}\leq&\frac{1}{N}I(W;Y^{N}|Z^{N},S^{N})+\frac{\delta(P_{e})}{N}\nonumber\\
&\leq&\frac{1}{N}H(Y^{N}|Z^{N},S^{N})+\frac{\delta(P_{e})}{N}\nonumber\\
&=&\frac{1}{N}\sum_{i=1}^{N}H(Y_{i}|Y^{i-1},Z^{N},S^{N})+\frac{\delta(P_{e})}{N}\nonumber\\
&\stackrel{(3)}\leq&\frac{1}{N}\sum_{i=1}^{N}H(Y_{i}|Y^{i-1},Z_{i+1}^{N},S^{N},Z_{i},S_{i},S_{i-d})+\frac{\delta(P_{e})}{N}\nonumber\\
&\stackrel{(4)}=&H(Y|U,Z,S,\tilde{S})+\frac{\delta(P_{e})}{N}\nonumber\\
&\stackrel{(5)}\leq&H(Y|U,Z,S,\tilde{S})+\frac{\delta(\epsilon)}{N},
\end{eqnarray}
where (1) from (\ref{e211}), and (2) is from the Fano's inequality, (3) is from the
fact that $S_{i}$ and $S_{i-d}$ (here $S_{i-d}=const$ when $i\leq d$) are included in $S^{N}$, (4)
is from the definitions in (\ref{jmds1}) and the fact that $J$ is a random variable (uniformly distributed
over $\{1,2,...,N\}$), and it is independent of $Y^{N}$, $Z^{N}$, $W$ and $S^{N}$, and (5) is from
$\delta(P_{e})$ is increasing while $P_{e}$ is increasing, and $P_{e}\leq \epsilon$.

Letting $\epsilon\rightarrow 0$, $R_{e}\leq H(Y|Z,U,S,\tilde{S})$ is proved, and the proof of Theorem \ref{T1.1} is completed.

\section{Proof of (\ref{giveup1.rmb})\label{appengiveup1.xx}}

\subsection{Achievability proof of (\ref{giveup1.rmb})\label{appengiveup1.xx.1}}

Replacing $V^{N}$ by $X^{N}$, and letting $W_{c}$, $U^{N}$ be constants, the achievability of $\mathcal{R}^{fi*}$ is along the lines of
the proof of Theorem \ref{T1} for case 1, where
\begin{eqnarray*}
&&\mathcal{R}^{fi*}=\{(R, R_{e}): 0\leq R_{e}\leq R,\\
&&R\leq I(X;Y|S,\tilde{S}),\\
&&R_{e}\leq I(X;Y|S,\tilde{S})-I(X;Z|S,\tilde{S})+H(Y|X,Z,S,\tilde{S})\}.
\end{eqnarray*}
Here note that since $Z$ is a degraded version of $Y$,
\begin{eqnarray*}
&&I(X;Y|S,\tilde{S})-I(X;Z|S,\tilde{S})+H(Y|X,Z,S,\tilde{S})\nonumber\\
&&=H(X|S,\tilde{S})-H(X|S,\tilde{S},Y)-H(X|S,\tilde{S})+H(X|S,\tilde{S},Z)+H(Y|X,Z,S,\tilde{S})\nonumber\\
&&\stackrel{(1)}=H(X|S,\tilde{S},Z)-H(X|S,\tilde{S},Y,Z)+H(Y|X,Z,S,\tilde{S})\nonumber\\
&&=I(X;Y|S,\tilde{S},Z)+H(Y|X,Z,S,\tilde{S})\nonumber\\
&&=H(Y|S,\tilde{S},Z),
\end{eqnarray*}
where (1) is from the Markov chain $X\rightarrow (S,\tilde{S},Y)\rightarrow Z$. Thus, it is easy to see that $\mathcal{R}^{fi*}=\mathcal{R}^{f*}$,
and the achievability of (\ref{giveup1.rmb}) is completed.

\subsection{Converse proof of (\ref{giveup1.rmb})\label{appengiveup1.xx.2}}

Since $R_{e}\leq R$ is obvious and the proof of $R\leq I(X;Y|S,\tilde{S})$ is exactly the same as that in Appendix \ref{appengiveup1} (see (\ref{jmds2})),
it remains to show that $R_{e}\leq H(Y|S,\tilde{S},Z)$, see the followings.

Note that
\begin{eqnarray}\label{b1}
R_{e}-\epsilon&\stackrel{(1)}\leq& \frac{H(W|Z^{N},S^{N})}{N}\nonumber\\
&=&\frac{1}{N}(H(W|Z^{N},S^{N})-H(W|Z^{N},S^{N},Y^{N})+H(W|Z^{N},S^{N},Y^{N}))\nonumber\\
&\stackrel{(2)}\leq&\frac{1}{N}(I(W;Y^{N}|Z^{N},S^{N})+\delta(P_{e}))\nonumber\\
&\leq&\frac{1}{N}(H(Y^{N}|Z^{N},S^{N})+\delta(P_{e}))\nonumber\\
&\stackrel{(3)}=&\frac{1}{N}\sum_{i=1}^{N}H(Y_{i}|Y^{i-1},Z^{N},S^{N},S_{i},S_{i-d})+\frac{\delta(P_{e})}{N}\nonumber\\
&\leq&\frac{1}{N}\sum_{i=1}^{N}H(Y_{i}|Z_{i},S_{i},S_{i-d})+\frac{\delta(P_{e})}{N}\nonumber\\
&\stackrel{(4)}=&\frac{1}{N}\sum_{i=1}^{N}H(Y_{i}|Z_{i},S_{i},S_{i-d},J=i)+\frac{\delta(P_{e})}{N}\nonumber\\
&\stackrel{(5)}=&H(Y_{J}|Z_{J},S_{J},S_{J-d},J)+\frac{\delta(P_{e})}{N}\nonumber\\
&\stackrel{(6)}\leq&H(Y_{J}|Z_{J},S_{J},S_{J-d})+\frac{\delta(\epsilon)}{N}\nonumber\\
&\stackrel{(7)}=&H(Y|Z,S,\tilde{S})+\frac{\delta(\epsilon)}{N},
\end{eqnarray}
where (1) is from (\ref{e211}), (2) is from Fano's inequality, (3) is from the
fact that $S_{i}$ and $S_{i-d}$ (here $S_{i-d}=const$ when $i\leq d$) are included in $S^{N}$,
(4) and (5) are from the fact that $J$ is a random variable (uniformly distributed
over $\{1,2,...,N\}$), and it is independent of $Y^{N}$, $Z^{N}$, $W$ and $S^{N}$,
(6) is from $P_{e}\leq \epsilon$ and
$\delta(P_{e})$ is increasing while $P_{e}$ is increasing, and (7) is from the definitions in (\ref{jmds1}).

Letting $\epsilon\rightarrow 0$,
$R_{e}\leq H(Y|Z,S,\tilde{S})$ is proved.
The converse and entire proof of (\ref{giveup1.rmb}) is completed.


\begin{thebibliography}{99}

\bibitem{wang} H. S. Wang and N. Moayeri, ``Finite-state markov channel-A useful
model for radio communication channels,'' {\sl IEEE Trans. Veh. Technol},
vol. 44, pp. 163-171, 1995.

\bibitem{zhang} Q. Zhang and S. Kassam, ``Finite-state Markov model for Rayleigh fading channels,''
{\sl IEEE Trans. Commun}, vol. 47, no. 11, pp. 1688-1692, 1999.

\bibitem{god} A. J. Goldsmith and P. P. Varaiya, ``Capacity, mutual information, and
coding for finite-state Markov channels,'' {\sl IEEE Trans. Inf. Theory}, vol. IT-42, pp. 868-886, 1996.

\bibitem{coverx} T. M. Cover and J. A. Thomas,
{\sl Elements of Information Theory}. New York, NY: Wiley-Interscience, 1991.

\bibitem{coverz} \textcolor[rgb]{1.00,0.00,0.00}{T. M. Cover and C. S. K. Leung,
``An achievable rate region for the multiple-access channel with feedback,"
{\sl IEEE Trans. Inf. Theory}, vol. IT-27, no. 3, pp. 292-298, 1981.}

\bibitem{CG1} \textcolor[rgb]{1.00,0.00,0.00}{T. M. Cover and A. El Gamal, ``Capacity theorems for the relay channel,''
{\sl IEEE Trans. Inf. Theory}, vol. IT-25, pp. 572-584, 1979.}

\bibitem{vis} H. Viswanathan, ``Capacity of Markov channels with receiver CSI
and delayed feedback,'' {\sl IEEE Trans. Inf. Theory}, vol. IT-45, no. 2, pp.
761-771, 1999.

\bibitem{bash} U. Basher, A. Shirazi and H. H. Permuter, ``Capacity region of finite state multiple-access
channels with delayed state information at the Transmitters,'' {\sl IEEE Trans. Inf. Theory}, vol. IT-58, no. 6, pp.
3430-3452, 2012.

\bibitem{chenjun} J. Chen and T. Berger, ``The capacity of finite-state Markov
channels with feedback,'' {\sl IEEE Trans. Inf. Theory}, vol. IT-51, pp.
780-789, 2005.

\bibitem{per1} H. H. Permuter and T. Weissman, ``Capacity region of the finite-state
multiple access channel with and without feedback,'' {\sl IEEE Trans. Inf. Theory}, vol. IT-55, no. 6, 2009.

\bibitem{per2} H. H. Permuter, T. Weissman, and A. J. Goldsmith, ``Finite state
channels with time-invariant deterministic feedback,'' {\sl IEEE Trans. Inf. Theory}, vol. IT-55, pp. 644-662, 2009.

\bibitem{como} G. Como and S. Y\"{u}ksel, ``On the capacity of finite state multiple access
channels with asymmetric partial state feedback,'' in {\sl WiOPT¡¯09:
Proc. 7th Int. Conf.Modeling and Optimization inMobile, Ad Hoc, and
Wireless Networks, IEEE Press}, Piscataway, NJ, 2009, pp. 589-594.

\bibitem{god2} A. J. Goldsmith and P. P. Varaiya, ``Capacity of fading channels
with channel side information,'' {\sl IEEE Trans. Inf. Theory}, vol. IT-43, pp. 1986-1992, 1997.


\bibitem{Wy} A. D. Wyner, ``The wire-tap channel,''
{\sl The Bell System Technical Journal}, vol. 54, no. 8, pp.
1355-1387, 1975.

\bibitem{CH} S. K. Leung-Yan-Cheong, M. E. Hellman,
``The Gaussian wire-tap channel,'' {\sl IEEE Trans. Inf. Theory}, vol. IT-24, no. 4, pp. 451-456, July 1978.


\bibitem{CK} I. Csisz$\acute{a}$r and J. K\"{o}rner, ``Broadcast channels with confidential messages,'' {\sl IEEE Trans.
Inf. Theory}, vol. IT-24, no. 3, pp. 339-348, May 1978.

\bibitem{LPS} Y. Liang, H. V. Poor and S. Shamai, ``Secure communication over fading channels,''
{\sl IEEE Trans. Inf. Theory}, vol. IT-54, pp. 2470-2492, 2008.


\bibitem{LMSY} R. Liu, I. Maric, P. Spasojevic and R. D. Yates, ``Discrete memoryless
interference and broadcast channels with confidential messages: secrecy
rate regions,'' {\sl IEEE Trans. Inf. Theory}, vol. IT-54, no. 6, pp. 2493-2507, Jun.
2008.


\bibitem{XCC} J. Xu, Y. Cao, and B. Chen, ``Capacity bounds for broadcast channels with confidential messages,''
{\sl IEEE Trans. Inf. Theory}, vol. IT-55, no. 6, pp. 4529-4542.
2009.

\bibitem{LP} Y. Liang and H. V. Poor, ``Multiple-access channels with confidential
messages,'' {\sl IEEE Trans. Inf. Theory}, vol. IT-54, no. 3, pp. 976-1002,
Mar. 2008.

\bibitem{TY2} E. Tekin and A. Yener, ``The Gaussian multiple access wire-tap channel,''
{\sl IEEE Trans. Inf. Theory}, vol. IT-54, no. 12, pp. 5747-5755, Dec. 2008.


\bibitem{TY1} E. Tekin and A. Yener, ``The general Gaussian multiple access and
two-way wire-tap channels: Achievable rates and cooperative jamming,''
{\sl IEEE Trans. Inf. Theory}, vol. IT-54, no. 6, pp. 2735-2751, June 2008.

\bibitem{WB} M. Wiese and H. Boche, ``An Achievable Region for the Wiretap Multiple-Access Channel with Common Message,''
{\sl Proceedings of 2012 IEEE International Symposium on Information Theory}, 2012.


\bibitem{YA} M. H. Yassaee and M. R. Aref, ``Multiple access wiretap channels with strong secrecy,''
{\sl Proceedings of IEEE Information Theory Workshop}, 2010.

\bibitem{XDD} P. Xu, Z. Ding, and X. Dai, ``Rate Regions for Multiple Access Channel With Conference and Secrecy Constraints,''
{\sl IEEE Trans. Inf. Forensics and Security}, vol. 8, no. 12, pp. 1961-1974, 2013.

\bibitem{AZV1} Z. H. Awan, A. Zaidi and L. Vandendorpe, ``Multi-access Channel with Partially Cooperating Encoders and Security Constraints,''
{\sl IEEE Trans. Inf. Forensics and Security}, Vol. 8, No. 7, pp. 1243-1254, Jul. 2013.

\bibitem{tang} X. Tang, R. Liu, P. Spasojevi$\acute{c}$ and H. V. Poor, ``Interference assisted secret communication,''
{\sl IEEE Trans. Inf. Theory}, vol. IT-57, no. 5, pp. 3153-3167,
May 2011.

\bibitem{LP2} Y. Liang, A. Somekh-Baruch, H. V. Poor, S. Shamai, and S. Verdu, ``Capacity of cognitive interference channels with and without secrecy,''
{\sl IEEE Trans. Inf. Theory}, vol. IT-55, pp. 604-619,
2009.

\bibitem{LG} L. Lai and H. El Gamal, ``The relay-eavesdropper channel: cooperation
for secrecy,'' {\sl IEEE Trans. Inf. Theory}, vol. IT-54, no. 9, pp. 4005-4019,
Sep. 2008.

\bibitem{daima} B. Dai and Z. Ma, ``Multiple-access relay wiretap channel,'' {\sl IEEE Trans. Inf. Forensics and Security},
vol. 10, no. 9, pp. 1835-1849, Sep.
2015.

\bibitem{Oo} Y. Oohama, ``Coding for relay channels with confidential messages,'' in
{\sl Proceedings of IEEE Information Theory Workshop}, Australia, 2001.

\bibitem{daima2} B. Dai, L. Yu and Z. Ma, ``Relay broadcast channel with confidential messages,''
{\sl IEEE Trans. Inf. Forensics and Security}, vol. 11, no. 2, pp. 410-425, 2016.


\bibitem{EU1} E. Ekrem and S. Ulukus, ``Secrecy in cooperative relay broadcast channels,''
{\sl IEEE Trans. Inf. Theory}, vol. IT-57, pp. 137-155, 2011.


\bibitem{MVL} C. Mitrpant, A. J. Han Vinck and Y. Luo,
``An Achievable Region for the Gaussian Wiretap Channel with Side
Information,'' {\sl IEEE Trans. Inf. Theory}, vol. IT-52, no. 5, pp.
2181-2190, 2006.


\bibitem{Ch} Y. Chen, A. J. Han Vinck, ``Wiretap channel with side information,''
{\sl IEEE Trans. Inf. Theory}, vol. IT-54, no. 1, pp. 395-402, January 2008.

\bibitem{ha} M. El Halabi, T. Liu, C. N. Georghiades and S. Shamai, ``Secret writing on dirty paper: a deterministic view,''
{\sl IEEE Trans. Inf. Theory}, vol. IT-58, no. 6, pp. 3419-3429,
June 2012.

\bibitem{CG} Y. K. Chia and A. El Gamal, ``Wiretap channel with causal state
information,'' {\sl IEEE Trans. Inf. Theory}, vol. 58, no. 5, pp. 2838-2849,
May 2012.

\bibitem{dai1} B. Dai, Z. Ma and X. Fang, ``Feedback Enhances the Security of State-Dependent
Degraded Broadcast Channels With Confidential Messages,''
{\sl IEEE Trans. Inf. Forensics and Security}, Vol. 10, No. 7, pp. 1529-1542, 2015.

\bibitem{san1} M. Bloch and J. N. Lanema, ``On the secrecy capacity of arbitrary wiretap
channels,'' in {\sl Proc. of 46th Allerton Conference on Communication,
Control and Computing}, Monticello, IL, September 2008.

\bibitem{san} Y. Sankarasubramaniam, A. Thangaraj and K. Viswanathan, ``Finite-state wiretap channels: secrecy under
memory constraints,'' in {\sl Proc. 2009 IEEE Information Theory Workshop}, Taormina, Italy, 2009, pp. 115-119.


\bibitem{mush} M. Mushkin and I. Bar-David, ``Capacity and coding for the Gilbert-Elliott channels,''
{\sl IEEE Trans. Inf. Theory}, vol. 35, pp. 1277-1290,
1989.

\bibitem{AC} R. Ahlswede and N. Cai, ``Transmission, Identification and Common Randomness Capacities for
Wire-Tap Channels with Secure Feedback from the Decoder,'' book
chapter in {\sl General Theory of Information Transfer and
Combinatorics}, LNCS 4123,  pp. 258-275, Berlin: Springer-Verlag,
2006.


\bibitem{lall} S. Lall, ``Advanced topics in computation for control,''
Lecture notes for Engr210b, Stanford University, Fall, 2004.

\end{thebibliography}
\end{document}